\documentclass[paper]{JHEP3}
\pdfoutput=1
\usepackage{amsmath,amssymb,amsthm,amscd,graphicx}
\usepackage{psfrag}
\input epsf.sty
\newtheorem{theorem}{Theorem}[section]

\theoremstyle{definition}

\newtheorem{example}[theorem]{Example}

\newcommand{\CC}{{\cal C}}

\newcommand{\CG}{{\cal G}}
\newcommand{\CH}{{\cal H}}
\newcommand{\CI}{{\cal I}}

\newcommand{\CL}{{\cal L}}

\newcommand{\CN}{{\cal N}}
\newcommand{\CO}{{\cal O}}

\newcommand{\CQ}{{\cal Q}}

\newcommand{\CW}{{\cal W}}

\def\IZ{{\mathbb Z}}
\def\IR{{\mathbb R}}
\def\IC{{\mathbb C}}
\def\IP{{\mathbb P}}

\def\IS{{\mathbb S}}

\newcommand{\tr}{{\rm Tr}}
\newcommand{\re}{{\rm e}}
\newcommand{\ri}{{\rm i}}
\newcommand{\rd}{{\rm d}}
\def\Ds{D \hskip -7pt / \hskip 2pt}
\def\sDs{D \hskip -5.5pt / \hskip 3.5 pt}

\newcommand{\be}{\begin{equation}}
\newcommand{\ee}{\end{equation}}
\newcommand{\ba}{\begin{aligned}}
\newcommand{\ea}{\end{aligned}}
\newcommand{\ben}{\begin{eqnarray}\displaystyle}
\newcommand{\een}{\end{eqnarray}}

\newcommand{\sectiono}[1]{\section{#1}\setcounter{equation}{0}}

\newdimen\tableauside\tableauside=1.0ex
\newdimen\tableaurule\tableaurule=0.4pt
\newdimen\tableaustep
\def\phantomhrule#1{\hbox{\vbox to0pt{\hrule height\tableaurule width#1\vss}}}
\def\phantomvrule#1{\vbox{\hbox to0pt{\vrule width\tableaurule height#1\hss}}}
\def\sqr{\vbox{%
  \phantomhrule\tableaustep
  \hbox{\phantomvrule\tableaustep\kern\tableaustep\phantomvrule\tableaustep}%
  \hbox{\vbox{\phantomhrule\tableauside}\kern-\tableaurule}}}
\def\squares#1{\hbox{\count0=#1\noindent\loop\sqr
  \advance\count0 by-1 \ifnum\count0>0\repeat}}
\def\tableau#1{\vcenter{\offinterlineskip
  \tableaustep=\tableauside\advance\tableaustep by-\tableaurule
  \kern\normallineskip\hbox
    {\kern\normallineskip\vbox
      {\gettableau#1 0 }%
     \kern\normallineskip\kern\tableaurule}%
  \kern\normallineskip\kern\tableaurule}}
\def\gettableau#1{\ifnum#1=0\let\next=\null\else
\squares{#1}\let\next=\gettableau\fi\next}

\newcommand{\figref}[1]{Fig.~\protect\ref{#1}}
\title{\huge{Lectures on localization and matrix models in supersymmetric Chern--Simons--matter theories}}

\author{
Marcos Mari\~no \\
D\'epartement de Physique Th\'eorique et Section de Math\'ematiques,\\
Universit\'e de Gen\`eve, Gen\`eve, CH-1211 Switzerland
\\
\email{marcos.marino@unige.ch}
}

\abstract{In these lectures I give a pedagogical presentation of some of the recent progress in supersymmetric Chern--Simons--matter theories, 
coming from the use of localization and matrix model techniques. The goal is to provide
 a simple derivation of the exact interpolating function for the free energy 
of ABJM theory on the three-sphere, which implies in particular the $N^{3/2}$ behavior at strong coupling. I explain in detail 
part of the background needed to understand this derivation, like holographic renormalization, localization of path integrals, and large $N$ techniques in matrix models. 
}    

\begin{document}

\sectiono{Introduction}
The AdS/CFT correspondence postulates a remarkable equivalence between certain gauge theories and string/M-theory on backgrounds involving 
AdS spaces. In these lectures we will look at the correspondence in the case of AdS$_4$/CFT$_3$, focusing on ABJM theory \cite{abjm}. One of the 
mysterious consequences of this correspondence is that, at strong coupling, the number of degrees of freedom of the CFT$_3$ should scale as $N^{3/2}$ \cite{kt}. 
This was finally established at the gauge theory level in \cite{dmp}. 

The derivation of the $N^{3/2}$ behavior 
in \cite{dmp} is based on various ingredients. The first one, due to \cite{kapustin}, is the fact that certain path integrals in superconformal Chern--Simons--matter 
theories can be reduced to 
matrix integrals, by using the localization techniques of \cite{pestun}. Localization techniques have a long story in supersymmetric and topological 
QFTs, and the applications 
of \cite{pestun,kapustin} to superconformal field theories provide a powerful technique to analyze certain observables in terms of matrix models. One of these observables is, in the three-dimensional case, the free energy on the three-sphere. In \cite{dmp} it was pointed out that this quantity is a good measure of the number of degrees of freedom in a CFT$_3$. The planar limit of this free energy can be calculated at weak 't Hooft coupling by using perturbation theory. If the superconformal field theory has an AdS dual, its value at strong coupling should be given by the regularized gravity action in AdS$_4$. In the case of ABJM theory one obtains in this way 
\be
\label{introgoal}
-\lim_{N \to \infty} {1\over N^2} F_{\rm ABJM} (\IS^3) \approx \begin{cases} -\log (2 \pi \lambda) +{3\over 2} +2 \log(2) , &\lambda \rightarrow 0, \\
\\ {\pi {\sqrt{2}} \over 3 {\sqrt{\lambda}}},  &\lambda \rightarrow \infty.
\end{cases}
\ee
Here, 
\be
\label{abjmthooft}
\lambda={N\over k}
\ee
is the 't Hooft parameter of ABJM theory. In \cite{dmp} it was shown, from the matrix model representation of \cite{kapustin}, that the planar free energy of ABJM theory on the three-sphere can be computed explicitly at all values of the 't Hooft coupling, and it gives a non-trivial interpolating function between the perturbative, weak coupling 
result, and the strong coupling result quoted in (\ref{introgoal}), confirming in this way the prediction of the large $N$ AdS dual. This gives a new, powerful test of the AdS/CFT correspondence which has been generalized to many other superconformal Chern--Simons--matter theories with AdS duals.

The goal of these lectures is to provide a simple derivation of the exact planar free energy of ABJM theory, first 
presented in \cite{dmp}. To begin with, we explain how to obtain the weak coupling result in field theory and the strong coupling result in AdS supergravity in (\ref{introgoal}). 
Then we show how the recent progress in \cite{kapustin,dmp} makes possible to obtain the exact function of the 't Hooft coupling interpolating 
between these two results. To do this, we present the localization technique of \cite{kapustin} and the matrix model techniques of \cite{hy,mp,dmp}. The derivation of the interpolating function in \cite{dmp} used results from the special geometry of Calabi--Yau manifolds, but one can obtain it by using only standard ideas in large $N$ 
matrix models. This is what we do here. 

The purpose of these lectures is mainly {\it pedagogical}. We try to provide detailed accounts of the background knowledge underlying the calculations in 
\cite{kapustin,dmp}. For example, we explain in some 
detail how to perform perturbative calculations in Chern--Simons theory and how to 
renormalize the gravity action in AdS spaces. We don't provide however a full overview
of the large $N$ string/gauge duality of \cite{abjm}, which can be found in other reviews like 
\cite{klebrev}. Also, we don't review the very recent developments based on localization and 
matrix models in supersymmetric Chern--Simons--matter theories, but in the last section we summarize some 
of the most interesting results which have been obtained. Given that these developments 
are still taking place, it is probably still premature to cover them appropriately. 

The organization of these lectures is as follows. Section 2 introduces Euclidean supersymmetric 
Chern--Simons--matter theories on the three-sphere and their 
classical properties. In section 3 we analyze perturbative Chern--Simons theory in some detail and in some generality. In section 4 we calculate 
the free energy of Chern--Simons--matter theories on 
$\IS^3$ at one-loop, and we derive the weak coupling result for ABJM theory quoted above in (\ref{introgoal}). 
Next, in section 5, we look at the free energy at strong coupling by using the 
AdS dual, and we explain the basics of holographic renormalization of the gravitational action. 
In section 6 we review the localization computation of \cite{kapustin} (incorporating 
some simplifications in \cite{hama}), which leads to a matrix model formulation of the free energy of ABJM theory. 
In section 7 we review some well-established techniques to 
analyze matrix integrals at large $N$, which are then used to solve the ABJM matrix model in section 8. This makes it possible to derive the interpolating function for 
the planar free energy and, in particular, to verify that its strong coupling limit matches the AdS prediction. Finally, in section 9 we give a brief summary of some recent developments. An Appendix collects many useful facts about harmonic analysis in $\IS^3$ which play a key r\^ole in the calculations.

\sectiono{Supersymmetric Chern--Simons--matter theories}

In this section we will introduce the basic building blocks of supersymmetric Chern--Simons--matter theories. We will work in Euclidean space, and we will put the theories 
on the three-sphere, since we are eventually interested in computing the free energy of the gauge theory in this curved space. In this section we will closely follow the presentation of \cite{hama}. 

\subsection{Conventions}

Our conventions for Euclidean spinors follow essentially \cite{wb}. In Euclidean space, the fermions $\psi$ and $\bar \psi$ are independent and they transform in the same representation of the Lorentz group. Their index structure is 
\be
\psi^{\alpha}, \quad \bar \psi^{\alpha}. 
\ee
We will take $\gamma_\mu$ to be the Pauli matrices, which are hermitian, and 
\be
\label{dgamma}
\gamma_{\mu \nu} = \frac{1}{2} [\gamma_\mu, \gamma_\nu] =\ri \epsilon_{\mu \nu \rho} \gamma^\rho.
\ee
We introduce the usual symplectic product through the antisymmetric matrix
\be
C_{\alpha \beta}=\begin{pmatrix} 0 & C \\-C &0\end{pmatrix}. 
\ee
In \cite{wb} we have $C=-1$ and the matrix is denoted by $\epsilon_{\alpha \beta}$. The product is 
\be
\bar \epsilon  \lambda =\bar\epsilon^\alpha C_{\alpha \beta} \lambda^\beta.
\ee
Notice that 
\be
 \bar \epsilon \gamma^\mu \lambda=\bar \epsilon^{\beta} C_{\beta \gamma} \left(\gamma^\mu\right)^{\gamma}_{~\alpha}\lambda^{\alpha}.
 \ee
 It is easy to check that 
\be
\bar \epsilon  \lambda =\lambda\bar\epsilon, \quad \bar \epsilon \gamma^\mu \lambda =-\lambda \gamma^\mu \bar\epsilon,
\ee
and in particular
\be
\label{ssign}
\left( \gamma^\mu \bar \epsilon\right) \lambda=-\bar \epsilon \gamma^\mu \lambda.
\ee
We also have the following Fierz identities
\be
\label{fierzone}
\bar \epsilon \left( \epsilon \psi\right) + \epsilon \left( \bar \epsilon \psi\right) +\left( \bar \epsilon \epsilon\right) \psi=0
\ee
and 
\be\label{fierztwo}
 \epsilon \left( \bar \epsilon \psi\right) +2 \left( \bar \epsilon \epsilon\right) \psi +\left( \bar \epsilon \gamma_\mu \psi\right) \gamma^\mu \epsilon=0. 
 \ee

\subsection{Vector multiplet and supersymmetric Chern--Simons theory}

We first start with theories based on vector multiplets. The three dimensional Euclidean ${\cal N}=2$ vector
superfield $V$ has the following content
\be
V: \qquad A_\mu, \,\, \sigma,\,\, \lambda, \,\, \bar\lambda, \,\, D,
\ee
where $A_\mu$ is a gauge field, $\sigma$ is an auxiliary
scalar field, $\lambda$, $\bar\lambda$ are two-component complex Dirac spinors, and
$D$ is an auxiliary scalar.  This is just the dimensional reduction of the $\mathcal{N}=1$ 
vector multiplet in $4$ dimensions, and $\sigma$ is the reduction of the fourth component of $A_\mu$.  All fields are valued in the Lie algebra $\mathfrak{g}$ of the gauge group $G$. For $G=U(N)$ our convention is that $\mathfrak{g}$ are Hermitian matrices. It follows that the gauge covariant derivative is given by
\be
 \partial_\mu + \ri [A_\mu, . ]
\ee
while the gauge field strength is 
\be
F_{\mu \nu}=\partial_\mu A_\nu -\partial_\nu A_\mu + \ri [A_\mu, A_\nu]. 
\ee

The transformations of the fields are generated by two independent complex spinors $\epsilon$, $\bar \epsilon$. They are given by,
\be\ba
 \delta A_\mu &=
{ \ri \over  2} (\bar\epsilon\gamma_\mu\lambda-\bar\lambda\gamma_\mu\epsilon), \\
 \delta\sigma &=
 {1\over 2} (\bar\epsilon\lambda-\bar\lambda\epsilon),
  \\
 \delta\lambda &=
-{1 \over 2} \gamma^{\mu\nu}\epsilon F_{\mu\nu}-D\epsilon
      +\ri\gamma^\mu\epsilon D_\mu\sigma
 +{2\ri \over 3}\sigma\gamma^\mu D_\mu\epsilon,
  \\
 \delta\bar\lambda &=
- {1 \over 2}\gamma^{\mu\nu}\bar\epsilon F_{\mu\nu}+D\bar\epsilon
                 -\ri\gamma^\mu\bar\epsilon D_\mu\sigma
 -{2\ri \over 3}\sigma\gamma^\mu D_\mu\bar\epsilon,
  \\
 \delta D &=
 -{\ri \over 2} \bar\epsilon\gamma^\mu D_\mu\lambda
 -{\ri \over 2} D_\mu\bar\lambda\gamma^\mu\epsilon
 +{\ri \over 2}[\bar\epsilon\lambda,\sigma]
 +{\ri \over 2}[\bar\lambda\epsilon,\sigma]
 -{\ri \over 6}(D_\mu\bar\epsilon\gamma^\mu\lambda
         +\bar\lambda\gamma^\mu D_\mu\epsilon),
\label{trvec}
\ea
\ee
and we split naturally
\be
\delta =\delta_{\epsilon} +\delta_{\bar \epsilon}. 
\ee
Here we follow the conventions of \cite{hama}, but we change the sign of the gauge connection: $A_\mu \rightarrow -A_\mu$. The derivative $D_\mu$ is covariant 
with respect to both the 
gauge field and the spin connection. On all the
fields, except $D$, the commutator $[\delta_\epsilon,\delta_{\bar\epsilon}]$
becomes a sum of translation, gauge transformation, Lorentz rotation,
dilation and R-rotation:
\be
\label{sqtrans}
\ba
 ~[\delta_\epsilon,\delta_{\bar\epsilon}]A_\mu =&
\ri  v^\nu\partial_\nu A_\mu + \ri \partial_\mu v^\nu A_\nu
 -D_\mu\Lambda,
  \\
 ~[\delta_\epsilon,\delta_{\bar\epsilon}]\sigma =&
 \ri v^\mu\partial_\mu\sigma+\ri[\Lambda,\sigma]+\rho\sigma,
  \\
 ~[\delta_\epsilon,\delta_{\bar\epsilon}]\lambda =&
 \ri v^\mu \partial_\mu\lambda+{\ri \over 4}\Theta_{\mu\nu}\gamma^{\mu\nu}\lambda
 +\ri[\Lambda,\lambda]+{3 \over 2} \rho\lambda
 +\alpha\lambda,
  \\
 ~[\delta_\epsilon,\delta_{\bar\epsilon}]\bar\lambda =&
\ri  v^\mu \partial_\mu\bar\lambda
 +{\ri \over 4}\Theta_{\mu\nu}\gamma^{\mu\nu}\bar\lambda
 +\ri[\Lambda,\bar\lambda]+{3 \over 2}\rho\bar\lambda
 -\alpha\bar\lambda,
  \\
 ~[\delta_\epsilon,\delta_{\bar\epsilon}]D =&
\ri v^\mu\partial_\mu D+\ri[\Lambda,D]+2\rho D  +{1 \over 3}\sigma(\bar\epsilon\gamma^\mu\gamma^\nu D_\mu D_\nu\epsilon
                -\epsilon\gamma^\mu\gamma^\nu D_\mu D_\nu\bar\epsilon),
\ea
\ee
where
\be
\ba
 v^\mu =& \bar\epsilon\gamma^\mu\epsilon,
  \\
 \Theta^{\mu\nu} =& D^{[\mu}v^{\nu]}+v^\lambda\omega_\lambda^{\mu\nu},
  \\
 \Lambda =& v^\mu \ri A_\mu +\sigma\bar\epsilon\epsilon,
  \\
 \rho =& \tfrac \ri3
 (\bar\epsilon\gamma^\mu D_\mu\epsilon
 +D_\mu\bar\epsilon\gamma^\mu\epsilon),
  \\
 \alpha =& \tfrac
  \ri3(D_\mu\bar\epsilon\gamma^\mu\epsilon-\bar\epsilon\gamma^\mu D_\mu\epsilon).
\label{sypar}
\ea\ee
Here, $\omega_\lambda^{\mu\nu}$ is the spin connection. As a check, let us calculate the commutator acting on $\sigma$. We have, 
\be
\ba
~[\delta_\epsilon, \delta_{\bar \epsilon} ] \sigma &= \delta_\epsilon\left( {1\over 2} \bar \epsilon \lambda \right) -\delta_{\bar \epsilon} \left( -{1\over 2} \bar \lambda \epsilon\right) \\
&={1\over 2} \bar \epsilon \left( -{1\over 2} \gamma^{\mu \nu} \epsilon F_{\mu \nu} - D \epsilon +\ri \gamma^\mu \epsilon D_\mu \sigma\right) +{ \ri \over 3} \bar\epsilon 
\gamma^\mu \left( D_\mu \epsilon \right) \sigma \\
&+{1\over 2}  \left( -{1\over 2} \gamma^{\mu \nu} \bar \epsilon F_{\mu \nu} + D \epsilon -\ri \gamma^\mu \bar \epsilon D_\mu \sigma\right) \epsilon-{ \ri \over 3} \gamma^\mu \left( D_\mu  \bar\epsilon \right)  \epsilon \sigma \\
& = \ri \bar \epsilon \gamma^\mu \epsilon D_\mu \sigma  + \rho \sigma, 
\ea
\ee
where we have used (\ref{ssign}). 

In order for the supersymmetry algebra to close, the last term in the
right hand side of $[\delta_\epsilon,\delta_{\bar\epsilon}]D$ must 
vanish. This is the case if the Killing spinors satisfy
\begin{equation}
 \gamma^\mu\gamma^\nu D_\mu D_\nu\epsilon = h\epsilon, \qquad  \gamma^\mu\gamma^\nu D_\mu D_\nu\bar \epsilon = h\bar \epsilon
\label{Kil2}
\end{equation}
for some scalar function $h$. 
A sufficient condition for this is to have
\be
\label{skil}
D_\mu\epsilon={\ri \over 2r} \gamma_\mu \epsilon, \qquad D_\mu \bar \epsilon={\ri \over 2r} \gamma_\mu \bar \epsilon
\ee
and
\be
\label{hsimple}
h=-{9 \over 4 r^2}
\ee
where $r$ is the radius of the three-sphere. 
This condition is satisfied by one of the Killing spinors on the three-sphere (the one which is constant in the left-invariant frame). Notice that, with this choice, $\rho$ in (\ref{sypar}) vanishes. 

The (Euclidean) SUSY Chern--Simons (CS) action, in flat space, is given by
\be
\ba
\label{susycs}
 S_{\rm SCS} &=  -\int \rd^3 x \, {\rm Tr} \left(A\wedge \rd A+{2 \ri \over
3}A^3- \bar \lambda \lambda+2D\sigma\right)\\
&=-\int \rd^3 x \, {\rm Tr} \left( \epsilon^{\mu \nu \rho} \left(A_\mu \partial_\nu A_\rho + {2 \ri \over 3} A_\mu A_\nu A_\rho \right) - \bar \lambda \lambda+2D\sigma\right).
\ea
\ee
Here $\tr$ denotes the trace in the fundamental representation. The part of the action involving the gauge connection $A$ is the standard, bosonic CS action 
in three dimensions. This action was first considered from the point of view of QFT in \cite{djt}, where the total action for a non-abelian gauge field was the sum of the standard Yang--Mills action and the CS action. In \cite{witten}, the CS action was considered by itself and shown to lead to a topological gauge theory. 

We can check that the supersymmetric CS action is invariant under the supersymmetry generated by $\delta_\epsilon$ (the proof for $\delta_{\bar \epsilon}$ is similar).The 
supersymmetric variation of the integrand of (\ref{susycs}) is
\be
\label{deltaLCS}
\ba
&\left( 2\delta A_\mu \partial_\nu A_\rho + 2\ri \delta A_\mu A_\nu A_\rho\right) \epsilon^{\mu \nu \rho} -\bar \lambda \delta \lambda +2 (\delta D)\sigma 
+2 D \delta \sigma\\
 &=-\ri \bar \lambda \gamma_\mu \epsilon \partial_\nu A_\rho \epsilon^{\mu \nu \rho} + \bar\lambda \gamma_\mu \epsilon A_\nu A_\rho \epsilon^{\mu \nu \rho} \\
&-\bar\lambda \left( -{1\over 2} \gamma^{\mu \nu} F_{\mu \nu} -D +\ri \gamma^\mu D_\mu \sigma \right) \epsilon -{2 \ri \over 3} \bar \lambda \gamma^\mu D_\mu \epsilon \sigma \\
&-\ri \left( D_\mu \bar\lambda\right) \gamma^\mu \sigma \epsilon + \ri [\bar \lambda \epsilon, \sigma]\sigma-{\ri \over 3} \bar\lambda \gamma^\mu D_\mu \epsilon \sigma -\bar\lambda \epsilon D.
\ea 
\ee
The terms involving $D$ cancel on the nose. Let us consider the terms involving the gauge field. After using (\ref{dgamma}) we find
\be
{1\over 2} \bar\lambda  \gamma^{\mu \nu} F_{\mu \nu} \epsilon=\ri  \bar\lambda  \gamma_\rho \epsilon \epsilon^{\mu \nu \rho}  \partial_\mu A_\nu - \bar\lambda  \gamma_\rho \epsilon \epsilon^{\mu \nu \rho} A_\mu A_\nu
\ee
which cancel the first two terms in (\ref{deltaLCS}). Let us now look at the remaining terms. The covariant derivative of $\bar\lambda$ is
\be
D_\mu \bar\lambda =\partial_\mu \bar \lambda +{\ri \over 2r} \gamma_\mu \bar \lambda+\ri [A_\mu, \bar \lambda].
\ee
If we integrate by parts the term involving the derivative of $\lambda$ we find in total 
\be
\ba
&\ri \bar\lambda \gamma^\mu \epsilon  \partial_\mu \sigma + \ri \bar\lambda \gamma^\mu\partial_\mu \epsilon  \sigma+ {1\over 2r} \left( \gamma^\mu\bar\lambda \right) \gamma_\mu \epsilon  + 
[A_\mu, \bar \lambda]\gamma^\mu \epsilon \sigma\\
&=\ri \bar\lambda \gamma^\mu \epsilon  \partial_\mu \sigma + \ri \bar\lambda \gamma^\mu D_\mu \epsilon  \sigma  + 
[A_\mu, \bar \lambda]\gamma^\mu \epsilon \sigma,
\ea 
\ee
where we used that
\be
 \left( \gamma^\mu\bar\lambda \right) \gamma_\mu \epsilon =-\bar\lambda \gamma^\mu \gamma_\mu \epsilon.
 \ee
The derivative of $\sigma$ cancels against the corresponding term in the covariant derivative of $\sigma$. Putting all together, we find
\be
\ri \bar\lambda \gamma^\mu \left( D_\mu \epsilon  \right)\sigma -\ri  \bar\lambda  \gamma^\mu \left(D_\mu \epsilon \right) \sigma +[A_\mu, \bar\lambda] \gamma^\mu \epsilon \sigma + \bar\lambda \gamma^\mu \epsilon [A_\mu, \sigma] + \ri [\bar \lambda \epsilon, \sigma]\sigma.
\ee
The last three terms cancel due to the cyclic property of the trace. This proves the invariance of the supersymmetric CS theory. 

In the path integral, 
the supersymmetric CS action enters in the form 
\be
\label{cspi}
\exp \left( {\ri k \over 4 \pi} S_{\rm SCS} \right)
\ee
where $k$ plays the role of the inverse coupling constant and it is referred to as the level of the CS theory. In a consistent quantum theory, $k$ must be an integer \cite{djt}. 
This is due to the 
fact that the Chern--Simons action for the connection $A$ is not invariant under large gauge transformations, but changes by an 
integer times $8 \pi^2$. The quantization of $k$ guarantees that (\ref{cspi}) remains invariant. 

Of course, there is another Lagrangian for vector multiplets, namely the Yang--Mills Lagrangian, 
\be
\label{yml}
 {\cal L}_\text{YM} = \text{Tr}\left[
  {1\over 4} F_{\mu\nu}F^{\mu\nu}+{1\over 2} D_\mu\sigma D^\mu\sigma
 +{1\over 2} \left(D+{\sigma \over r}\right)^2
 +{\ri \over 2} \bar\lambda\gamma^\mu D_\mu\lambda
 +{\ri \over 2} \bar\lambda[\sigma,\lambda]
 -{1\over 4r} \bar\lambda\lambda
 \right].
\ee
In the flat space limit $r\rightarrow \infty$, this becomes the standard (Euclidean) super Yang--Mills theory in three dimensions. 
The Lagrangian (\ref{yml}) is not only invariant under the SUSY transformations (\ref{trvec}), but it can be written as a superderivative, 
\be
\label{YMder}
 \bar\epsilon\epsilon\,{\cal L}_\text{YM} =
 \delta_{\bar\epsilon}\delta_\epsilon\text{Tr}\Big(
  {1\over 2} \bar\lambda\lambda-2D\sigma \Big).
  \ee
 This will be important later on.  

\subsection{Supersymmetric matter multiplets}

We will now add supersymmetric matter, i.e. a chiral multiplet $\Phi$ in a representation $R$ of the gauge group. Its components are 
\be
\Phi: \qquad \phi,\,\, \bar \phi, \,\, \psi, \, \, \bar \psi, \,\, F,\,\, \bar F. 
\ee
The
supersymmetry transformations are
\be
\ba
 \delta\phi =& \bar\epsilon\psi,  \\
 \delta\bar\phi =& \epsilon\bar\psi,  \\
 \delta\psi =& \ri\gamma^\mu\epsilon D_\mu\phi +\ri\epsilon\sigma\phi
 +{2\Delta \ri \over 3}\gamma^\mu D_\mu\epsilon\phi+\bar\epsilon F,
  \\
 \delta\bar\psi =& \ri\gamma^\mu\bar\epsilon D_\mu\bar\phi
 +\ri\bar\phi\sigma\bar\epsilon+{2\Delta \ri \over 3}\bar\phi\gamma^\mu D_\mu\bar\epsilon
 +\bar F\epsilon,
  \\
 \delta F =&
 \epsilon(\ri\gamma^\mu D_\mu\psi-\ri\sigma\psi-\ri\lambda\phi)
 +{\ri \over 3}(2\Delta-1)D_\mu\epsilon\gamma^\mu\psi,
  \\
 \delta\bar F =&
 \bar\epsilon(\ri\gamma^\mu D_\mu\bar\psi-\ri\bar\psi\sigma+\ri\bar\phi\bar\lambda)
 +{\ri \over 3}(2\Delta-1)D_\mu\bar\epsilon\gamma^\mu\bar\psi,
\label{dncm}
\ea
\ee
where $\Delta$ is the possible anomalous dimension of $\phi$. For theories with $\CN\ge 3$ supersymmetry, the field has the canonical dimension 
\be
\Delta={1\over 2}, 
\ee
but in general this is not the case. 

The commutators of these transformations are given by
\begin{eqnarray}
~[\delta_\epsilon,\delta_{\bar\epsilon}]\phi &=&
 \ri v^\mu\partial_\mu\phi+\ri\Lambda\phi+\Delta\rho\phi-\Delta\alpha\phi,
 \nonumber \\
~[\delta_\epsilon,\delta_{\bar\epsilon}]\bar\phi &=&
 \ri v^\mu\partial_\mu\bar\phi-\ri\bar\phi\Lambda+\Delta\rho\bar\phi
 +\Delta\alpha\bar\phi,
 \nonumber \\
~[\delta_\epsilon,\delta_{\bar\epsilon}]\psi &=&
 \ri v^\mu \partial_\mu\psi
 +\tfrac14\Theta_{\mu\nu}\gamma^{\mu\nu}\psi
 +\ri\Lambda\psi+\left(\Delta+{1\over 2}\right)\rho\psi+(1-\Delta)\alpha\psi,
 \nonumber \\
~[\delta_\epsilon,\delta_{\bar\epsilon}]\bar\psi &=&
 \ri v^\mu \partial_\mu\bar\psi
 +\tfrac14\Theta_{\mu\nu}\gamma^{\mu\nu}\bar\psi
 -\ri\bar\psi\Lambda+\left(\Delta+{1\over 2} \right)\rho\bar\psi
 +(\Delta-1)\alpha\bar\psi,
 \nonumber \\
~[\delta_\epsilon,\delta_{\bar\epsilon}]F &=&
 \ri v^\mu\partial_\mu F+\ri\Lambda F+(\Delta+1)\rho F+(2-\Delta)\alpha F,
 \nonumber \\
~[\delta_\epsilon,\delta_{\bar\epsilon}]\bar F &=&
 \ri^v\mu\partial_\mu\bar F-\ri\bar F\Lambda+(\Delta+1)\rho\bar F
 +(\Delta-2)\alpha\bar F.
\end{eqnarray}
The lowest components of the superfields are assigned the dimension $\Delta$ and R-charge
$\mp \Delta$. The supersymmetry algebra closes off-shell when the Killing spinors
$\epsilon,\bar\epsilon$ satisfy (\ref{Kil2}) and $h$ is given by (\ref{hsimple}). As a check, we compute
\be
\ba
 ~[\delta_\epsilon, \delta_{\bar \epsilon}] \phi &=\delta_\epsilon \left(\bar \epsilon\psi\right)\\ 
 &=\bar \epsilon\left(\ri \gamma^\mu \epsilon D_\mu  \phi +\ri \epsilon \sigma \phi + {2 \ri \Delta \over 3}\gamma^\mu \left( D_\mu \epsilon \right) \phi \right) =\ri v^\mu  D_\mu \phi +\ri \sigma \bar \epsilon \epsilon +{2\ri \Delta \over 3} \left( \bar \epsilon  \gamma^\mu D_\mu \epsilon \right),\ea
 \ee
which is the wished-for result. 
 %
 %
%
%
%
%

Let us now consider supersymmetric Lagrangians for the matter hypermultiplet. If the fields have their canonical dimensions, the Lagrangian
\be
 {\cal L} = D_\mu\bar\phi D^\mu\phi -\ri\bar\psi\gamma^\mu D_\mu\psi
 +{3 \over 4 r^2}\bar\phi\phi+\ri\bar\psi\sigma\psi 
 +\ri\bar\psi\lambda\phi-\ri\bar\phi\bar\lambda\psi
 +\ri\bar\phi D\phi+\bar\phi\sigma^2\phi+\bar FF 
\label{Lcm}
\ee
is invariant under supersymmetry if the Killing spinors
$\epsilon,\bar\epsilon$ satisfy (\ref{Kil2}), with $h$ given in (\ref{hsimple}). The quadratic part of the Lagrangian for $\phi$ gives indeed the standard conformal 
coupling for a scalar field. We recall that the action for a massless scalar field in a curved space of $n$ dimensions contains a coupling to the curvature $R$ given by
\be
S=\int \rd^n x {\sqrt{g}} \left(  g^{\mu \nu} \partial_\mu \phi \partial_\nu \phi +\xi R \phi^2 \right), 
\ee
where $\xi$ is a constant. This action is conformally invariant when 
\be
\xi={1\over 4} {n-2\over n-1}. 
\ee
If the spacetime is an $n$-sphere of radius $r$, the curvature is
\be
R={n(n-1) \over r^2},
\ee
and the conformal coupling of the scalar leads to an effective mass term of the form 
\be
{n(n-2)\over 4 r^2}  \phi^2 
\ee
which in $n=3$ dimensions gives the quadratic term for $\phi$ in (\ref{Lcm}). 

If the fields have non-canonical dimensions, the Lagrangian 
\be
\ba
 {\cal L}_\text{mat} =&
  D_\mu\bar\phi D^\mu\phi
 +\bar\phi\sigma^2\phi
 +{\ri(2\Delta-1)\over r} \bar\phi\sigma\phi
 +{\Delta (2-\Delta)\over r^2} \bar\phi\phi
 +\ri\bar\phi D\phi
 +\bar FF \\ &
 -\ri\bar\psi\gamma^\mu D_\mu\psi
 +\ri\bar\psi\sigma\psi
 - {2\Delta-1 \over 2r}\bar\psi\psi
 +\ri\bar\psi\lambda\phi
 -\ri\bar\phi\bar\lambda\psi 
\label{Lncm}
\ea
\ee
is supersymmetric, provided the parameters
$\epsilon,\bar\epsilon$ satisfy the Killing spinor conditions (\ref{skil}). The Lagrangian (\ref{Lncm}) is not only invariant under the 
supersymmetries $\delta_{\epsilon, \bar \epsilon}$, but it can be written as a total superderivative, 
\be
\label{matterder}
 \bar\epsilon\epsilon \, \CL_\text{mat} =
 \delta_{\bar\epsilon}\delta_\epsilon\left(
  \bar\psi\psi-2\ri\bar\phi\sigma\phi+{2(\Delta-1) \over r}\bar\phi\phi
 \right).
 \ee
\vskip1.5cm
\subsection{ABJM theory}

\FIGURE{
\includegraphics[height=4cm]{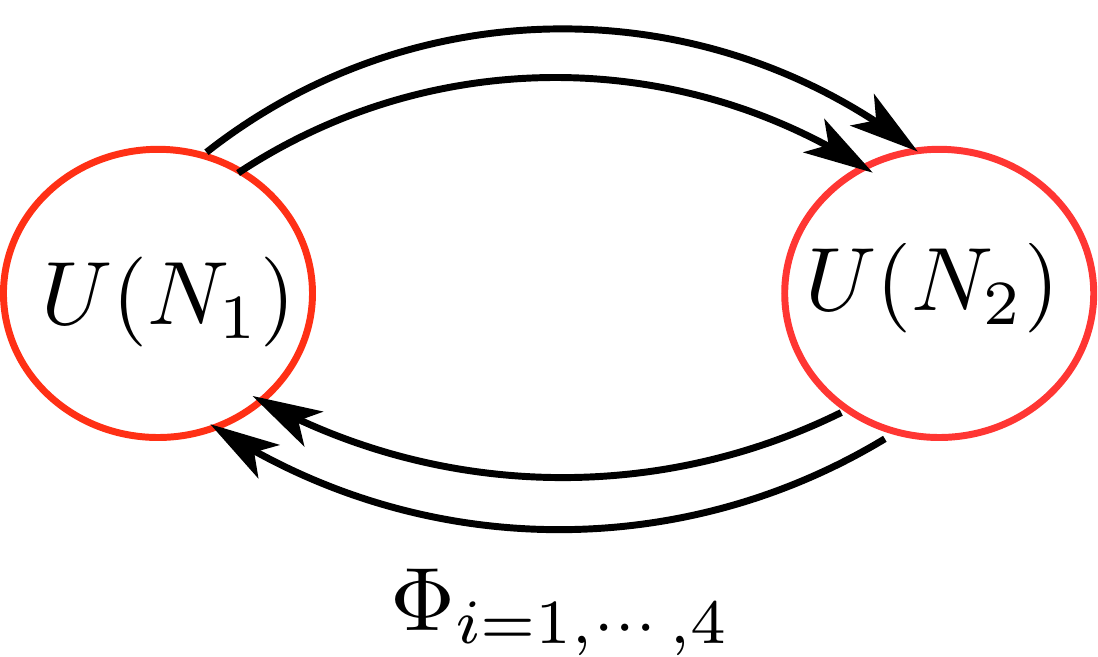}
\caption{The quiver for ABJ(M) theory. The two nodes represent the $U(N_{1,2})$ Chern--Simons theories (with opposite levels) and the arrows between 
the nodes represent the four matter multiplets in the bifundamental representation.}
\label{quiver}
}

The theory proposed in \cite{abjm,abj} to describe $N$ M2 branes is a particular example of a supersymmetric Chern--Simons theory. It consist of 
two copies of Chern--Simons theory with gauge groups $U(N_1)$, $U(N_2)$, and opposite levels $k$, $-k$. In addition, we have four matter supermultiplets 
$\Phi_i$, $i=1,\cdots, 4$, in the bifundamental representation of the gauge group $U(N_1) \times U(N_2)$. This theory can be represented as a quiver, with two nodes representing the 
Chern--Simons theories, and four edges between the nodes representing the matter supermultiplets, see \figref{quiver}. In addition, there is a superpotential involving the matter fields, which after integrating out the auxiliary fields in the Chern--Simons--matter system, reads (on $\IR^3$)
\be
\label{superpot}
W={4 \pi \over k} \tr\left(\Phi_1 \Phi_2^{\dagger} \Phi_3 \Phi_4^{\dagger} - \Phi_1 \Phi_4^{\dagger} \Phi_3 \Phi_2^{\dagger}\right), 
\ee
where we have used the standard superspace notation for $\CN=1$ supermultiplets \cite{wb}.

\sectiono{A brief review of Chern--Simons theory}
Since one crucial ingredient in the theories we are considering is Chern--Simons theory, we review here some results concerning 
the perturbative structure of this theory on general three-manifolds. These results were first obtained in the seminal paper by Witten \cite{witten} and then extended 
and refined in various directions in \cite{fg,jeffrey, rozansky,rozanskyrev,adams, adamstwo}. Chern--Simons perturbation theory on general three-manifolds is a important subject in itself, hence we will try to give a general presentation which might be useful in other 
contexts. This will require a rather formal development, and the reader interested in the result for the one-loop contribution might want to skip some of the derivations in the next two subsections. 

\subsection{Perturbative approach}
In this section, we will denote the bosonic Chern--Simons action by 
\be
S=-{1\over 4\pi} \int_M {\rm Tr} \Bigl( A\wedge \rd A + {2 \ri \over 3} A
\wedge A \wedge A \Bigr)
\label{csactbis}
\ee
where we use the conventions appropriate for Hermitian connections, and we included the factor $1/4 \pi$ in the action for notational 
convenience. The group of gauge transformations $\CG$ acts on the gauge connections as follows, 
\be
A \rightarrow A^U=U A U^{-1} -\ri U \, \rd U^{-1}, \qquad U \in \CG. 
\ee
We will assume that the theory is defined on a compact 
three-manifold $M$. The partition function is defined as
\begin{equation}
Z (M)={1\over {\rm vol}(\CG)} \int [{\cal D} A]  {\rm e}^{\ri k S}
\label{partcs}
\end{equation}
where we recall that $k \in \IZ$. 

There are many different approaches to the calculation 
of (\ref{partcs}), but the obvious strategy is to use perturbation theory. Notice that, since the theory is defined on a compact manifold, 
there are no IR divergences and we just have to deal with UV divergences, as in standard QFT. Once these are treated appropriately, the 
partition function (\ref{partcs}) is a well-defined observable. In perturbation theory we evaluate (\ref{partcs}) by expanding around saddle--points. These are
flat connections, which are in
one-to-one correspondence with group homomorphisms
\be
\pi_1(M) \rightarrow G
\ee
modulo conjugation. For example, if $M=\IS^3/{\IZ}_p$ is the lens space $L(p,1)$, one has
$\pi_1(L(p,1))={\IZ}_p$, and flat connections are labelled by homomorphisms
${\IZ}_p \rightarrow G$. Let us assume that
these are a discrete set of points (this happens, for example, if $M$ is a
rational homology sphere, since in that case $\pi_1(M)$ is a finite group). 
We will label the flat connections with an index $c$, and a flat connection 
will be denoted by $A^{(c)}$. Each flat connection leads to a covariant derivative 
\be
\label{cd}
\rd_{A^{(c)}}=\rd + \ri [A^{(c)}, \cdot],
\ee
and flatness implies that
\be
\rd_{A^{(c)}}^2=\ri \, F_{A^{(c)}}=0.
\ee
Therefore, the covariant derivative leads to a complex
\be
\label{eqcomplex}
0\rightarrow \Omega^0(M,{\bf g}) \xrightarrow{\rd_{A^{(c)}}} \Omega^1(M,{\bf g}) \xrightarrow{\rd_{A^{(c)}}}  \Omega^2(M,{\bf g}) \xrightarrow{\rd_{A^{(c)}}}  \Omega^3(M,{\bf g}).
\ee
\FIGURE[ht]{
\includegraphics[height=3cm]{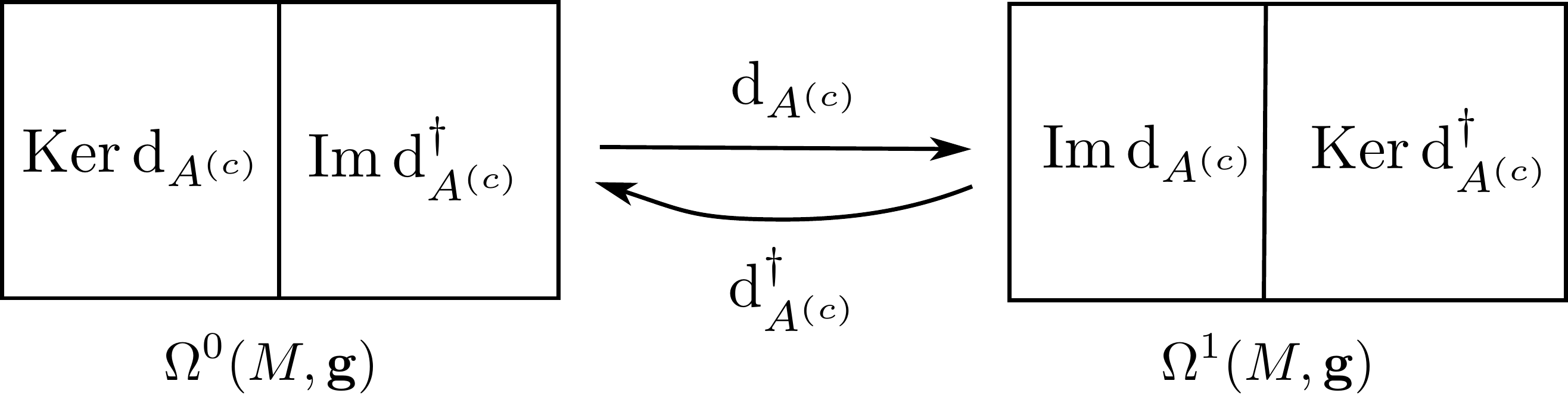} 
\caption{The standard elliptic decomposition of $\Omega^{0}(M, {\bf g})$ and $\Omega^{1}(M, {\bf g})$.} 
\label{complex}}
The first two terms in this complex have a natural interpretation in the context of gauge theories: $\Omega^0(M, {\bf g})$ is the Lie algebra of the group of gauge transformations, and we can write a gauge transformation as
\be
U=\re^{\ri \phi}, \qquad \phi \in \Omega^0(M, {\bf g}).
\ee
The elements of $\Omega^0(M, {\bf g})$ generate infinitesimal gauge transformations, 
\be
\label{igt}
\delta A =-\rd_A \phi.
\ee
The second term, $\Omega^1(M, {\bf g})$, can be identified with the tangent space to the space of gauge connections. The first map in the complex (\ref{eqcomplex}) is 
interpreted as (minus) an infinitesimal gauge transformation in the background of $A^{(c)}$. 

We recall that the space of ${\bf g}$-valued forms on $M$ has a natural inner product given by 
\be
\langle a,b \rangle =\int_M {\rm Tr}\, (a \wedge *b),
\ee
where $*$ is the Hodge operator. 
With respect to this product we can define an adjoint operator on ${\bf g}$-valued $p$-forms in the same way that is done for the usual 
de Rham operator, 
\be
\rd^\dagger_{A^{(c)}} =(-1)^{3(1+p)+1} *\rd_{A^{(c)}}*. 
\ee
We then have the orthogonal decompositions (see \figref{complex}) 
\be
\label{ortdec}
\ba
\Omega^0(M,{\bf g})=&{\rm Ker}\, \rd_{A^{(c)}} \oplus {\rm Im}\, \rd_{A^{(c)}}^{\dagger},\\
\Omega^1(M,{\bf g})=&{\rm Ker}\, \rd_{A^{(c)}}^{\dagger} \oplus{\rm Im}\,  \rd_{A^{(c)}}.
\ea
\ee
These decompositions are easily proved. For the first one, for example, we just note that 
\be
a \in {\rm Ker}\, \rd_{A^{(c)}} \Rightarrow \langle \rd_{A^{(c)}} a , \phi\rangle =\langle a, \rd_{A^{(c)}}^{\dagger}\phi\rangle=0, \qquad \forall \phi
\ee
therefore
\be
\left( {\rm Ker}\, \rd_{A^{(c)}}  \right)^{\perp}={\rm Im}\,  \rd_{A^{(c)}}^{\dagger}.
\ee
%
%
%
One also has the analogue of the Laplace--Beltrami operator acting on $p$-forms
\be
\Delta^{p}_{A^{(c)}}=\rd^\dagger_{A^{(c)}} \rd_{A^{(c)}}+\rd_{A^{(c)}}\rd^\dagger_{A^{(c)}}.
\ee

In the following we will assume that
\be
\label{cohocond}
H^1(M, {\bf g})=0.
\ee
This means that the connection $A^{(c)}$ is {\it isolated}. However, we will consider the possibility that $A^{(c)}$ has a non-trivial isotropy group $\CH_c$. 
We recall that the isotropy group of a connection $A^{(c)}$ is the subgroup of gauge transformations which leave $A^{(c)}$ invariant, 
\be
\CH_c=\{\phi \in {\cal G}| \phi(A^{(c)})=A^{(c)}\}.
\label{isotropia}
\ee
The Lie algebra of this group is given by zero-forms annihilated by the covariant derivative (\ref{cd}),
\be
\label{lieiso}
{\rm Lie}(\CH_c)=H^0(M,  {\bf g})={\rm Ker}\, \rd_{A^{(c)}},
\ee
which is in general non-trivial. A connection is {\it irreducible} if its isotropy group 
is equal to the center of the group. In particular, if $A^{(c)}$ is irreducible one has 
\be
H^0(M,  {\bf g})=0. 
\ee
It can be shown that the isotropy group $\CH_c$ consists of constant 
gauge transformations that leave $A^{(c)}$ invariant, 
\be
\phi A^{(c)}\phi^{-1}=A^{(c)}. 
\ee
They are in one-to-one correspondence with a subgroup of $G$ which we will denote by $H_c$. 

In the semiclassical approximation, $Z(M)$ is written as a
sum of terms associated to saddle-points:
\begin{equation}
Z (M)= \sum_c Z^{(c)}(M),
\end{equation}
where $c$ labels the different flat connections $A^{(c)}$ on $M$.
Each of the $Z^{(c)}(M)$ will be a perturbative series in $1/k$ of the form
\begin{equation}
\label{perts}
Z^{(c)}(M)=Z^{(c)}_{\rm 1-loop}(M) \exp \Biggl\{ \sum_{\ell=1}^\infty
S^{(c)}_\ell  k^{-\ell}
\Biggr\},
\end{equation}
where $S^{(c)}_\ell$ is the $(\ell+1)$-loop contribution around the flat connection $A^{(c)}$. In order to 
derive this expansion, we split the connection into a ``background", which is the 
flat connection $A^{(c)}$, plus a ``fluctuation" $B$:
\be
A=A^{(c)} + B.
\ee
Expanding around this, we find
\be
\label{acsplit}
S(A)=S(A^{(c)}) + S(B), 
\ee
where
\be
S(B)=-{1\over 4\pi} \int_M {\rm Tr}\, (B\wedge \rd_{A^{(c)}} B + {2\ri \over 3} B^3).
\ee
The first term in (\ref{acsplit}) is the classical Chern--Simons invariant of the connection $A^{(c)}$. Since Chern--Simons theory is a gauge theory, in order 
to proceed we have to fix the gauge. We will follow the detailed analysis of
\cite{adams}. Our gauge choice will be the standard, covariant gauge,
\be
\label{gaugecond}
g_{A^{(c)}}(B)=\rd_{A^{(c)}}^{\dagger} B=0
\ee
where $g_{A^{(c)}}$ is the gauge fixing function. We recall that in the standard Fadeev--Popov (FP) gauge fixing one first defines 
\be
\label{deltaminusone}
\Delta^{-1}_{A^{(c)}}\left( B \right)=\int {\cal D} U\, \delta\left( g_{ A^{(c)}} \left(B^U \right)\right), 
\ee
and then inserts into the path integral 
\be
1=\left[ \int {\cal D} U\, \delta\left( g_{ A^{(c)}} \left(B^U \right)\right) \right] \Delta_{A^{(c)}}\left( B \right).
\ee
The key new ingredient here is the presence of a non-trivial isotropy group $\CH_c$ for the flat connection $A^{(c)}$. When there is a non-trivial isotropy group, 
the gauge-fixing condition does not fix the gauge completely, since 
\be
\label{residual}
g_{A^{(c)}}(B^{\phi})=\phi g_{A^{(c)}}(B) \phi^{-1}, \qquad \phi \in \CH_c,
\ee
i.e. the basic assumption that $g(A)=0$ only cuts the gauge orbit once is not true, and there is a residual symmetry given by the 
isotropy group. Another way to see this is that the standard FP determinant vanishes due to 
zero modes. In fact, the standard 
calculation of (\ref{deltaminusone}) (which is valid if the isotropy group of $A^{(c)}$ is trivial) gives
\be
\label{FPdet}
\Delta^{-1}_{A^{(c)}}\left( B \right)=\left| {\rm det} {\delta g_{A^{(c)}}(B^U) \over \delta U}\right|^{-1}=\biggl| {\rm det} \, \rd_{A^{(c)}}^{\dagger} \rd_A \biggr|^{-1}.
\ee
But when $\CH_c\not=0$, the operator $\rd_{A^{(c)}}$ has zero modes due to the nonvanishing of (\ref{lieiso}), and the FP procedure is ill-defined. 
The correct way to proceed in the calculation of (\ref{deltaminusone}) is to split the integration over 
the gauge group into two pieces. The first piece is the integration over the isotropy group. Due to (\ref{residual}), the integrand does not depend on it, and we obtain a factor of ${\rm Vol}(\CH_c)$. The second piece gives an integration over the remaining part of the gauge transformations, which has as its Lie algebra
\be
({\rm Ker}\, \rd_{A^{(c)}})^{\perp}.
\ee
The integration over this piece leads to the standard FP determinant (\ref{FPdet}) but with the zero modes removed. We then find,  
\be
\Delta^{-1}_{A^{(c)}}\left( B \right)={\rm Vol}(\CH_c) \biggl| {\rm det} \, \rd_{A^{(c)}}^{\dagger} \rd_A \biggr|^{-1}_{({\rm Ker}\, \rd_{A^{(c)}})^{\perp}}
\ee
This phenomenon was first observed by Rozansky in \cite{rozansky}, and developed in this language in \cite{adams}. As usual, the determinant appearing here 
can be written as a path integral over ghost fields, with action
\be
S_{\rm ghosts}(C,{\overline C},B) = \langle {\overline C},  \rd_{A^{(c)}}^{\dagger} \rd_A C\rangle,
\ee
where $C, \overline C$ are Grassmannian fields taking values in 
\be
\label{ghostres}
({\rm Ker}\, \rd_{A^{(c)}})^{\perp}.
\ee
The action for the ghosts can be divided into a kinetic term plus an interaction term between
the ghost fields and the fluctuation $B$:
\be
S_{\rm ghosts}(C,{\overline C},B) =
\langle {\overline C}, \Delta^{0}_{A^{(c)}}C \rangle + \ri \langle{\overline C}, \rd_{A^{(c)} }^{\dagger}[B,C]\rangle.
\ee
The modified FP gauge--fixing leads then to the path integral
\be
\ba
Z^{(c)}(M)&= \re^{ \ri k S(A^{(c)})}  \int_{\Omega^1(M,{\bf g})} {\cal {D}}B \, \re^{\ri k S(B)}\Delta_{A^{(c)} }(B)
\delta\left(\rd_{A^{(c)}}^{\dagger} B \right)\\
&={\re^{ \ri k S(A^{(c)})} \over {\rm Vol}(\CH_c)}
 \int_{\Omega^1(M,{\bf g})}{\cal {D}}B \,  \delta\left(\rd_{A^{(c)}}^{\dagger} B \right) \int_{({\rm Ker}\, \rd_{A^{(c)}})^{\perp}} {\cal {D}}C {\cal {D}}{\overline C} \, 
 \re^{ \ri k S(B) -S_{\rm ghosts}(C,{\overline C},B)}.
 \ea
 \ee
Finally, we analyze the delta constraint on $B$. Due to the decomposition of $\Omega^1(M, {\bf g})$ in (\ref{ortdec}), we can write
\be
\label{cvar}
B=\rd_{A^{(c)}} \phi + B', 
\ee
where
\be
\phi \in \left( {\rm Ker} \, \rd_{A^{(c)}} \right)^{\perp}, \quad B'\in {\rm Ker}\, \rd_{A^{(c)}} ^{\dagger}.
\ee
The presence of the operator $\rd_{A^{(c)}}$ in the change of variables (\ref{cvar}) leads to a non-trivial Jacobian. Indeed, we have
\be
\Vert B\Vert^2=\langle \phi, \Delta^{0}_{A^{(c)}} \phi \rangle+ \Vert B'\Vert^2, 
\ee
 and the measure in the functional integral becomes
\be
\label{jacobiB}
{\cal D}B =\left({\rm det}'\, \Delta^{0}_{A^{(c)}} \right)^{1\over 2} {\cal D}\phi\, {\cal D} B',
\ee
where the $'$ indicates, as usual, that we are removing zero modes. Notice that the operator in (\ref{jacobiB}) is positive--definite, so the square root of its determinant is well--defined. 
We also have that
\be
\delta\left(\rd_{A^{(c)}}^{\dagger} B \right)=\delta\left(\Delta^{0}_{A^{(c)}} \phi\right)=\left({\rm det}'\, \Delta^{0}_{A^{(c)}} \right)^{-1} \delta(\phi),
\ee
which is a straightforward generalization of the standard formula
\be
\delta(ax)={1\over |a|} \delta (x).
\ee
We conclude that the delta function, together with the Jacobian in (\ref{jacobiB}), lead to the the following factor in the path integral:
\be
\label{detdelta}
\left({\rm det}'\, \Delta^{0}_{A^{(c)}} \right)^{-{1\over 2}}.
\ee
In addition, the delta function sets $\phi=0$. It only remains the  integration over $B'$, which we relabel $B' \rightarrow B$. 
%
%
%
The final result for the gauge-fixed path integral is then 
\be
\label{finalgf}
Z^{(c)}(M)={\re^{ \ri k S(A^{(c)})} \over {\rm Vol}(\CH_c)} \left({\rm det}'\, \Delta^{0}_{A^{(c)}} \right)^{-{1\over 2}}
 \int_{{\rm Ker}\, \rd_{A^{(c)}}^{\dagger}} {\cal {D}}B
 \int_{({\rm Ker}\, \rd_{A^{(c)}})^{\perp}} {\cal {D}}C {\cal {D}}{\overline C} \, 
 \re^{ \ri k S(B) -S_{\rm ghosts}(C,{\overline C},B)}.
 \ee
 This is starting point to perform gauge-fixed perturbation theory in Chern--Simons theory. 
 
\subsection{The one--loop contribution}

We now consider the one-loop contribution of a saddle-point to the path integral. This has been studied in many papers \cite{fg,jeffrey,rozansky, 
rozanskyrev}. We will follow the detailed presentation in \cite{adamstwo}. 

Before proceeding, we should specify what is the regularization method that we will use to define 
the functional determinants appearing in our calculation. A natural 
and useful regularization for quantum field theories in curved space is zeta-function regularization \cite{hawkingzeta}. We recall that the zeta function of a 
self-adjoint operator $T$ with eigenvalues $\lambda_n >0$ is defined as
\be
\zeta_T(s)=\sum_n \lambda_n ^{-s}.
\ee
Under appropriate conditions, this defines a meromorphic function on the complex $s$-plane which is regular at $s=0$. Since
\be
-\zeta'_T(0)=\sum_n \log \lambda_n
\ee
we can define the determinant of $T$ as
\be
\label{zetadet}
{\rm det}(T)=\re^{-\zeta'_T(0)}. 
\ee
This is the regularization we will adopt in the following. It has the added advantage that, when used in Chern--Simons theory, it leads to natural 
topological invariants like the Ray--Singer torsion, as we will see. 

The main ingredients in the one-loop contribution to the path integral (\ref{finalgf}) 
are the determinants of the operators appearing in the kinetic terms for $B$, $C$ and $\overline C$. Putting together the determinant (\ref{detdelta}) 
and the determinant coming from the ghost fields, we obtain
 \be
 \label{ghdelta}
\left( {\rm det}'\, \Delta^0_{A^{(c)}} \right)^{1/2}
\ee
since the ghosts are restricted to (\ref{ghostres}) and their determinant is also primed. The operator appearing in the kinetic term for $B$ is $\ri Q/2$, where
\be
Q= -{k \over 2 \pi} *\rd_{A^{(c)}}
\ee
is a self-adjoint operator acting on $\Omega^1(M, {\bf g})$ which has to be restricted to 
\be
\label{restric}
{\rm Ker}\, \rd_{A^{(c)}}^ {\dagger}=({\rm Im}\, \rd_{A^{(c)}})^\perp
\ee
due to the gauge fixing. Notice that, 
if (\ref{cohocond}) holds, one has 
\be
H^1(M, {\bf g}) =0\Rightarrow {\rm Ker}\, \rd_{A^{(c)}}={\rm Im}\, \rd_{A^{(c)}},
\ee
and due to the restriction to (\ref{restric}), $Q$ has no zero modes. However, the operator $Q$ is {\it not} positive definite, and one has to be careful in order to define its determinant. We will now do this following the discussion in \cite{witten, atiyah}. 
A natural definition takes as its starting point the trivial 
Gaussian integral
\be
\int_{-\infty}^{\infty}{\rd x}\, \exp\left( \ri {\lambda x^2 \over 2} \right)={\sqrt{2 \pi \over |\lambda|}} \exp \left( {\ri \pi \over 4} {\rm sign}\, \lambda\right). 
\ee
If we want to have a natural generalization of this, the integration over $B$ should be
\be
\label{Qdet}
\exp \left( {\ri \pi \over 4} {\rm sign}(Q) \right)  \left| {\rm det}\left( { Q \over 2 \pi} \right)\right|^{-1/2}.
\ee
In order to compute the determinant in absolute value, we can consider the square of the operator $-*\rd_{A^{(c)}}$, which is given by 
\be
\label{sqop}
*\rd_{A^{(c)}}*\rd_{A^{(c)}}=\rd^{\dagger}_{A^{(c)}} \rd_{A^{(c)}}, 
\ee
and it is positive definite when restricted to (\ref{restric}). It is the Laplacian on one-forms, restricted to (\ref{restric}). We then define 
\be
 \left| {\rm det}{ Q \over 2 \pi} \right|^2 = {\rm det}'\left[ \left( {k \over 4 \pi^2}\right)^2 \rd_{A^{(c)}}^{\dagger}\rd_{A^{(c)}} \right]
 \ee
where we have subtracted the zero--modes (i.e., we consider the restriction to (\ref{restric})). 
In order to take into account the sign in (\ref{Qdet}), we need the $\eta$ invariant of the operator $-*\rd_{A^{(c)}}$. We recall that the $\eta$ invariant is defined as 
\be
\eta_T(s)=\sum_j {1\over \left(\lambda^+_j\right)^s} -\sum_j {1\over \left(-\lambda^-_j\right)^s}
\ee
where $\lambda^{\pm}_j$ are the strictly positive (negative, respectively) eigenvalues of $T$. The regularized difference of eigenvalues is then $\eta_T(0)$. 
In our case, this gives
\be
\eta(A^{(c)})\equiv \eta_{-*\rd_{A^{(c)}}}( 0).
\ee
Finally, we have to take into account that for each eigenvalue of the operator (\ref{sqop}) we have a factor of 
\be
\label{factork}
\biggl( {k \over 4\pi^2}\biggr)^{-1/2}
\ee
appearing in the final answer. The regularized number of eigenvalues of the operator is simply 
\be
\label{zeta0}
 \zeta(A^{(c)}) \equiv \zeta_{\rd_{A^{(c)}}^{\dagger}\rd_{A^{(c)}}} (0),
\ee
restricted again to (\ref{restric}). Putting all together we obtain, 
\be
\label{quotfirst}
 \left({\rm det}\, {\ri Q \over 2 \pi}  \right)^{-{1\over 2}}=\biggl( {k \over 4\pi^2}\biggr)^{-\zeta(A^{(c)})/2} \exp \biggl( {\ri \pi\over 4} \eta(A^{(c)})\biggr) 
 \left({\rm det}'\, \rd_{A^{(c)}}^{\dagger}\rd_{A^{(c)}}\right)^{-{1\over 4}}.
 \ee
 Here we assumed that $k>0$. For a negative level $-k<0$ the answer is still given by (\ref{quotfirst}), but the phase involving the eta invariant 
has the opposite sign. We can now combine this result with (\ref{ghdelta}). The quotient of the determinants of the Laplacians gives 
 the square root of the so-called {\it Ray--Singer torsion} of the flat connection $A^{(c)}$ \cite{rs},
 \be
 \label{quotdet}
{ ({\rm det}'\, \Delta^0_{A^{(c)}})^{1\over 2} \over  \left({\rm det} ' \, \rd_{A^{(c)}}^{\dagger}\rd_{A^{(c)}}\right)^{1\over 4} }={\sqrt { \tau'_R(M, A^{(c)})}}.
\ee
This was first observed by Schwarz in the Abelian theory \cite{schwarz}. When the connection $A^{(c)}$ is isolated and irreducible, this quotient 
is a topological invariant of $M$, but in general it is not. However, for a reducible and isolated flat connection, the dependence 
on the metric is just given by an overall factor, equal to the volume of the manifold $M$:
\be
\label{mftorsion}
\tau'_R(M, A^{(c)})=\left({\rm vol}(M)\right)^{{\rm dim}(\CH_c)} \tau_R(M, A^{(c)}).
\ee
where $ \tau_R(M, A^{(c)})$ is now metric-indepedendent. For an explanation of this fact, see for example Appendix B in \cite{fw}. However, 
this volume, which is a metric-dependent factor, cancels in the final answer for the one-loop path integral \cite{adamstwo}. The isotropy group $\CH_c$ is the space of constant zero forms, taking values in a subgroup $H_c \subset G$ of the gauge group. Each generator of its Lie algebra has a norm given by its norm as an element of ${\bf g}$, times 
\be
\left( \int_M *1\right)^{1/2}=\left({\rm vol}(M)\right)^{1/2}. 
\ee
Therefore, 
\be
{\rm vol}(\CH_c)=\left({\rm vol}(M)\right)^{{\rm dim}(\CH_c)/2}{\rm vol}(H_c), 
\ee
and 
\be
{ {\sqrt {\tau'_R(M, A^{(c)})}} \over {\rm vol}(\CH_c)}={ {\sqrt {\tau_R(M, A^{(c)})}} \over {\rm vol}(H_c)}
\ee
which does not depend on the metric of $M$. Finally, in order to write down the answer, we take into account that, for isolated flat connections, $\zeta(A^{(c)})$ can be evaluated as \cite{asen}
\be
\label{zeta0ex}
 \zeta(A^{(c)})= {\rm dim}\, H^0(M, {\bf g}).
 \ee
 Putting everything together, we find for the one--loop contribution to the path integral
\begin{equation}
\label{asympt}
Z^{(c)}_{\rm 1-loop}(M)=
{ 1\over {\rm vol} (H_c)} \left( {k \over 4\pi^2} \right)^{-{1\over 2} {\rm dim}\, H^0(M, {\bf g})}{\rm e}^{\ri k S(A^{(c)})
+{\ri \pi \over 4} \eta(A^{(c)})} {\sqrt {\tau_R(M, A^{(c)})}}.
\end{equation}
As noticed above, this expression is valid for $k>0$. For a negative level $-k<0$, the phase involving the gravitational $\eta$ invariant has the opposite sign. 
It was pointed out in \cite{witten} that this phase can be written in a more suggestive form by using the Atiyah--Patodi--Singer theorem, which says that
\be
\eta(A^{(c)})=\eta(0) +{4 y \over \pi}  S(A^{(c)}).
\ee
Here $y$ is the dual Coxeter number of $G$ (for $U(N)$, $y=N$), and $\eta(0)$ is the eta invariant of the trivial connection. Let us denote by
\be
d_G={\rm dim}(G),
\ee
the dimension of the gauge group. The operator involved in the calculation of $\eta(0)$ is just 
$d_G$ copies of the ``gravitational" operator $-*\rd$, which is only coupled to the metric. We can then write
\be
\eta(0)=d_G  \, \eta_{\rm grav}, 
\ee
where $\eta_{\rm grav}$ is the ``gravitational" eta invariant of $-*\rd$ . We then find, 
\be
\label{asfin}
Z^{(c)}_{\rm 1-loop}(M)=
{ 1\over {\rm vol} (H_c)} \left( {k \over 4\pi^2} \right)^{-{1\over 2} {\rm dim}\, H^0(M, {\bf g})}{\rm e}^{\ri \left( k +y \right) S(A^{(c)})
+{\ri \pi \over 4} d_G \eta_{\rm grav}} {\sqrt {\tau_R(M, A^{(c)})}}.
\end{equation}
This formula exhibits a one--loop renormalization of $k$ 
\be
\label{kshift}
k \rightarrow k+y
\ee
which is simply a shift by the dual Coxeter number \cite{witten}. However, different regularizations of the theory seem to 
lead to different shifts \cite{asorey}. 

When $A^{(c)}=0$ is the trivial flat connection, one has that $H_c=G$, where $G$ is the gauge group, and the cohomology twisted by $A^{(c)}$ reduces to the ordinary cohomology. The Ray--Singer torsion is the torsion $\tau_R(M)$ of the ordinary de Rham differential, to the power $d_G$. We then obtain, for the trivial connection, 
\be
\label{trivial}
Z_{\rm 1-loop}(M)=
{ 1\over {\rm vol} (G)} \left( {k \over 4\pi^2} \right)^{-d_G/ 2} {\rm e}^{ {\ri \pi   \over 4} d_G \eta_{\rm grav}}  \left(\tau_R(M)\right)^{{d_G/2}}.
\end{equation}

 As explained in detail in \cite{witten}, the phase in (\ref{asfin}) and (\ref{trivial}) involving the $\eta$ invariant is metric-dependent. In constructing a topological 
 field theory out of Chern--Simons gauge theory, as in \cite{witten}, one wants to preserve 
 topological invariance, and an appropriate counterterm has to be added to the action. The counterterm involves the gravitational Chern--Simons action $S(\omega)$, where $\omega$ is the spin connection. However, this action is ambiguous, and it depends on a trivialization of the tangent bundle to $M$. Such a choice of trivialization is called a {\it framing} of the three-manifold. The difference between two trivializations 
 can be encoded in an integer $s$, and when one changes the trivialization the gravitational Chern--Simons action changes as 
 \be
 S(\omega) \rightarrow S(\omega) + 2 \pi s,
 \ee
 similarly to the gauge Chern--Simons action. According to the Atiyah--Patodi--Singer theorem, the combination 
\be
{1\over 4} \eta_{\rm grav} + {1\over 12} {S(\omega) \over 2 \pi}
\ee
is a topological invariant. It depends on the choice of framing of $M$, but not on its metric. Therefore, if we modify (\ref{trivial}) by including in the phase an appropriate multiple of the gravitational Chern--Simons action, 
\be
\exp\left(  {\ri \pi   \over 4} d_G \eta_{\rm grav}\right) \rightarrow \exp\left[   \ri \pi d_G  \left( { \eta_{\rm grav} \over  4} 
+ {1\over 12} {S(\omega) \over 2 \pi} \right)\right],
\ee
the resulting one-loop partition function is a topological invariant of the framed three-manifold $M$. If we change the framing of $M$ by $s$ units, the above factor induces a 
change in the partition function which at one-loop is of the form 
\be
Z (M)\rightarrow \exp\left( 2\pi \ri s \cdot{d_G\over 24}\right) Z(M). 
\ee

One of the most beautiful results of \cite{witten} is that Chern--Simons theory is exactly solvable on any three-manifold $M$, and its partition function can 
be computed exactly as a function of $k$, for any gauge group $G$, by using current algebra in two dimensions. In particular, one can compute the exact change of the partition function under a change of framing, and one finds
\be
\label{npchf}
Z (M)\rightarrow \exp\left( 2\pi \ri s \cdot{c\over 24}\right) Z(M), 
\ee
where
\be
c={k d_G \over k+y}.
\ee
A detailed explanation of the exact solution of CS theory would take us too far, and we refer the reader to the original paper \cite{witten} or to the presentation in \cite{mmrev} for more details. We will however list later on the relevant results when $M=\IS^3$.

\sectiono{The free energy at weak coupling}

The partition function of a CFT on $\IS^3$ should encode information about the number of degrees of freedom of the theory, in the sense that 
at weak coupling it should scale as the number $\CN$ of elementary constituents. This follows simply from the factorization property of the partition function in the 
absence of interactions:
\be
Z(\IS^3, \CN) \approx \left(Z(\IS^3, 1)\right)^{\CN}. 
\ee
For example, a gauge theory with gauge group $U(N)$ has at weak coupling $N^2$ degrees of freedom, and 
we should expect the free energy on the three-sphere to scale in this regime as
\be
F(\IS^3) \sim \CO(N^2).
\ee
Of course, at strong coupling this is not necessarily the case. 

In this section we will compute the partition function on $\IS^3$ of supersymmetric Chern--Simons--matter theories in the weak coupling approximation, i.e. at one loop. 
First, we will do the computation in Chern--Simons theory, and then we will 
consider the much simpler case of supersymmetric matter multiplets. 

\subsection{Chern--Simons theory on $\IS^3$}

In the previous section we presented the general procedure to calculate the one-loop contribution of Chern--Simons theory on any three-manifold, around 
an isolated flat connection. This procedure can be made very concrete when the manifold is $\IS^3$. In this case there is only one flat connection, 
the trivial one $A^{(c)}=0$, and we can use (\ref{trivial}). Therefore, we just have to compute the Ray--Singer torsion $\tau(\IS^3)$ for the standard de Rham differential, i.e. 
the quotient of determinants appearing in (\ref{quotdet}) with $A^{(c)}=0$ (a similar calculation was made in Appendix A of \cite{fw}). 

We endow $\IS^3$ with its standard metric (the one induced by its standard embedding in $\IR^4$ with Euclidean metric), and 
we choose the radius $r=1$ (it is easy to verify explicitly that the final result is independent of $r$). The determinant of the scalar Laplacian on the sphere can be computed very explicitly, since its eigenvalues are known to be (see the Appendix)
\be
\lambda_n=n(n+2), \qquad n=0,1,\cdots
\ee
where $n$ is related to $j$ in (\ref{eigenlaplace}) by $n=2j$. The degeneracy of this eigenvalue is 
\be
d_n=(n+1)^2. 
\ee
Removing the zero eigenvalue just means that we remove $n=0$ from the spectrum. 
To calculate the determinant we must calculate the zeta function, 
\be
\zeta_{\Delta^{0}}(s)=\sum_{n=1}^{\infty} {d_n \over \lambda_n^s}=\sum_{n=1}^\infty {(n+1)^2 \over \left( n(n+2)\right)^s}=\sum_{m=2}^{\infty} {m^2 \over \left( m^2-1\right)^s}.
\ee
This zeta function can not be written in closed form, but its derivative at $s=0$ is easy to calculate. 
The calculation can be done in many ways, and general results for the determinant of Laplacians on $\IS^m$ can be found in for example \cite{rocky,wy}. 
We will follow 
a simple procedure inspired by \cite{noc}. We split
\be
{m^2 \over \left( m^2-1\right)^s}={1\over m^{2(s-1)}} + {s \over m^{2s}} +R(m,s), 
\ee
where
\be
R(m,s)={m^2 \over \left( m^2-1\right)^s}-{1\over m^{2(s-1)} }- {s \over m^{2s}}
\ee
which decreases as $m^{-2s -2}$ for large $m$, and therefore leads to a convergent series for all $s\ge -1/2$ which is moreover uniformly convergent. Therefore, it 
is possible to exchange sums with derivatives. 
The derivative of $R(m,s)$ at $s=0$ can be calculated as
\be
{\rd R (m,s) \over \rd s}\bigg|_{s=0}=-1 -m^2 \log \left(1-{1\over m^2}\right).
\ee
The sum of this series can be explicitly calculated by using the Hurwitz zeta function, and one finds
\be
-\sum_{m=2}^{\infty} \left[1 + m^2 \log \left(1-{1\over m^2}\right)\right] ={3\over 2}-\log(\pi). 
\ee
We then obtain
\be
\zeta_{\Delta^{0}}(s)=\zeta(2s-2)-1+s\left( \zeta(2s)-1\right) + \sum_{m=2}^{\infty} R(m,s), 
\ee
where $\zeta(s)$ is Riemann's zeta function, and 
\be
-\zeta'_{\Delta^{0}}(0)=\log(\pi) -2 \zeta'(-2). 
\ee
The final result can be expressed in terms of $\zeta(3)$, since
\be
\zeta'(-2)=-{\zeta(3) \over 4 \pi^2}. 
\ee
We conclude that the determinant of the scalar Laplacian on $\IS^3$ is given by 
\be
\log\, {\rm det}' \, \Delta^{0}=\log(\pi) + {\zeta(3) \over 2 \pi^2}. 
\ee

We now compute the determinant in the denominator of (\ref{quotdet}). We must consider the space of one-forms on $\IS^3$, and restrict to the ones that are not 
in the image of $\rd$. These forms are precisely the vector spherical harmonics, whose properties are reviewed in the Appendix. The eigenvalues of the operator $\rd^{\dagger} \rd$ are given in (\ref{lapvsh}), and they read
\be
\label{vsheigen}
\lambda_n=(n+1)^2, \qquad n=1, 2, \cdots, 
\ee
with degeneracies
\be
\label{vshdeg}
d_n=2n (n+2). 
\ee
The zeta function associated to this Laplacian (restricted to the vector spherical harmonics) is 
\be
\zeta_{\Delta^{1}}(s)=\sum_{n=1}^\infty {2n(n+2) \over \left( n+1\right)^{2s}}=2 \sum_{m=1}^{\infty} {m^2-1 \over m^{2s}}=2\zeta(2s-2) -2\zeta(2s), 
\ee
and
\be
\log\, {\rm det}' \, \Delta^{1} =-4\zeta'(-2) -2\log (2 \pi)= -2\log (2 \pi) +{\zeta(3) \over \pi^2}.
\ee
Here we have used that
\be
\zeta'(0)=-{1\over 2} \log (2 \pi). 
\ee
We conclude that
\be
\log \tau'_R(\IS^3)=\log\, {\rm det}' \, \Delta^{(0)} -{1\over 2}  \log\, {\rm det}' \, \Delta^{(1)} =\log (2 \pi^2). 
\ee
This is in agreement with the calculation of the analytic torsion for general spheres in for example \cite{wyou}. In view of (\ref{mftorsion}), and since
\be
{\rm vol}(\IS^3)=2 \pi^2,
\ee
we find
\be
\tau_R(\IS^3)=1.
\ee
One can also calculate the invariant (\ref{zeta0}) directly, since this is $d_G$ times
\be
\zeta_{\Delta^{1}}(0)=-2\zeta(0)=1, 
\ee
and it agrees with (\ref{zeta0ex}). Since the eigenvalues of $*\rd$ on the vector spherical harmonics come in pairs with the same absolute value but 
opposite signs (see (\ref{dvh})), $\eta_{\rm grav}=0$. We conclude that, for $k>0$, 
\be
\label{olcs}
Z_{\rm 1-loop}(\IS^3)=
 {1\over {\rm vol} (G)} \left(  {k \over 4\pi^2} \right)^{-{d_G\over 2} }.
\end{equation}
In particular, for $G=U(N)$ we have
\be
\label{olcsun}
Z_{\rm 1-loop}(\IS^3)=
 {1\over {\rm vol} (U(N))} \left(  {k \over 4\pi^2} \right)^{-{N^2\over 2}}
\end{equation}
The volume of $U(N)$ is given by
\be
\label{volun}
{\rm vol}(U(N))={ (2\pi)^{ {1 \over 2}N(N+1)} \over G_2(N+1)},
\ee
where $G_2(z)$ is the Barnes function, defined by
\be
\label{gaussianvalue}
G_2 (z+1)=\Gamma (z) G_2(z), \,\,\,\,\, G_2(1)=1.
\ee
Notice that 
\be
G_2(N+1)=(N-1)! (N-2)! \cdots 2! 1!. 
\ee

As we mentioned before, the partition function of Chern--Simons theory on $\IS^3$ can be computed exactly for any gauge group. In the case of $U(N)$, the answer 
is, for $k>0$ \cite{witten}
\be\label{csun}
Z(\IS^3)={1\over (k+N)^{N/2}} \prod_{j=1}^{N-1} \left( 2 \sin {\pi j \over k+N}\right)^{N-j}. 
\ee
Here an explicit choice of framing has been made, but one can compute the partition function for any choice of framing by simply using (\ref{npchf}). 
The expansion for large $k$ should reproduce the perturbative result, and in particular the leading term should agree with the result (\ref{olcsun}). Indeed, we 
have for $k$ large 
\be
Z(\IS^3)\approx k^{-N/2} \prod_{j=1}^{N-1} \left( {2 \pi j \over k}\right)^{N-j}=\left(2\pi\right)^{{1\over 2} N (N-1)}k^{-N^2/2} G_2(N+1),
\ee
which is exactly (\ref{olcsun}). 

\subsection{Matter fields}

The supersymmetric multiplet contains a conformally coupled complex scalar and a fermion, both in a representation $R$ of the gauge group. The 
partition function at one loop is just given by the quotient of functional determinants
\be
Z^{\rm matter}_{\rm 1-loop}=\left( {{\rm det} \left(-\ri \Ds \right) \over {\rm det} \, \Delta_c  } \right)^{d_R}
\ee
where $d_R$ is the dimension of the representation, and 
\be
\Delta_{c}=\Delta^0 +{3\over 4}
\ee
is the conformal Laplacian (we have set again $r=1$). We now compute these determinants. 

The eigenvalues of the conformal Laplacian are simply 
\be
n(n+2) +{3\over 4}, \qquad n=0,1, \cdots, 
\ee
with the same multiplicity as the standard Laplacian, namely $(n+1)^2$. We then have 
\be
\zeta_{\Delta_c}(s)=\sum_{n=0}{ (n+1)^2 \over \left( n(n+2) + {3\over 4} \right)^s}=\sum_{m=1}^{\infty} {m^2 \over \left(m^2-{1\over 4} \right)^s}. 
\ee
As in the case of the standard Laplacian, we split
\be
{m^2 \over \left(m^2-{1\over 4} \right)^s}={1\over m^{2(s-1)}} + {s \over 4 m^{2s}} +R_c(m,s), 
\ee
where
\be
R_c(m,s)={m^2 \over\left(m^2-{1\over 4} \right)^s}-{1\over m^{2(s-1)} }- {s \over 4 m^{2s}}.
\ee
The derivative of $R_c(m,s)$ at $s=0$ is
\be
{\rd R_c (m,s) \over \rd s}\bigg|_{s=0}=-{1\over 4} - m^2 \log \left(1-{1\over4 m^2}\right).
\ee
The sum of this series can be explicitly calculated as
\be
-\sum_{m=1}^{\infty} \left[{1\over 4}  + m^2 \log \left(1-{1\over 4m^2}\right)\right] ={1\over 8} -{1\over 4} \log (2) +{7\zeta(3) \over 8 \pi^2}. 
\ee
We then find
\be
\zeta_{\Delta_c}(s)=\zeta(2s-2)+{s\over 4}  \zeta(2s) + \sum_{m=1}^{\infty} R_c(m,s), 
\ee
and we conclude that the determinant of the conformal Laplacian on $\IS^3$ is given by 
\be
\log\, {\rm det} \, \Delta_c=-\zeta'_{\Delta_c}(0)={1\over 4} \log (2) -{3\zeta(3) \over 8 \pi^2}.
\ee
This is in agreement with the result quoted in the Erratum to \cite{dowker}\footnote{Beware: the arXiv version of this paper gives a wrong result for this determinant.}. 

Let us now consider the determinant (in absolute value) for the spinor field. We have, using (\ref{spinoreigen}),  
\be
\zeta_{|\sDs|}(s)=2 \sum_{n=1}^{\infty} {n(n+1) \over \left( n+{1\over 2} \right)^s}.
\ee
After a small manipulation we can write it as 
\be
\label{fermionzeta}
\ba
\zeta_{|\sDs|}(s)=& 2\cdot 2^{s-2}\left\{ \sum_{m\ge1} {1\over (2m+1)^{s-2}} -\sum_{m\ge1} {1\over (2m+1)^s} \right\}\\
=&2 \left( 2^{s-2}-1\right) \zeta(s-2) -{1\over 2} \left(2^s-1\right) \zeta(s), 
\ea
\ee
where we have used that 
\be
\sum_{m\ge0} {1\over (2m+1)^{s}}=\sum_{n\ge 1}{1\over n^s}-\sum_{k\ge 1}{1\over (2k)^s}=\left(1 -2^{-s}\right) \zeta(s).
\ee
The regularized number of negative eigenvalues of this operator is given by $\zeta_{|\sDs|}(0)/2$ and it vanishes, so the determinant 
of the Dirac operator equals its absolute value. We deduce
\be
\log \, {\rm det}\, \left(-\ri \Ds\right)=-\zeta'_{|\sDs|}(0)=-{3\over 8 \pi^2} \zeta(3) -{1\over 4} \log \, 2.
\ee
Combining the conformal scalar determinant with the spinor determinant we obtain, 
\be
\label{freelogz}
\log \, {\rm det}\, \left(-\ri \Ds\right)-\log\, {\rm det}\, \Delta_c=-{1\over 2} \log \, 2. 
\ee
This can be seen directly at the level of eigenvalues. The quotient of determinants is 
\be
\label{freequot}
\prod_{m=1}^{\infty} {\left( m+{1\over 2}\right)^{m(m+1)}\left( m-{1\over 2}\right)^{m(m-1)} \over \left( m^2-{1\over 4} \right)^{m^2}}=\prod_{m=1}^{\infty} {\left( m+{1\over 2}\right)^{m} \over \left( m-{1\over 2} \right)^{m}}
\ee
and its regularization leads directly to the result above (see Appendix A of \cite{dmp}). We conclude that
\be
Z^{\rm matter}_{\rm 1-loop}=2^{-d_R/2}. 
\ee

\subsection{ABJM theory at weak coupling}
We can now calculate the free energy on $\IS^3$ of ABJM theory. We will restrict ourselves to the ``ABJM slice" where the two gauge groups have the same rank, i.e. the 
theory originally considered in \cite{abjm}. We have two copies of CS theory with gauge group $U(N)$ and opposite levels $k$, $-k$, together with four chiral multiplets in 
the bifundamental representation of $U(N)\times U(N)$. 
Keeping the first term (one-loop) in perturbation theory we find, at one-loop, 
\be
F_{\rm ABJM}(\IS^3)\approx-N^2  \log \left( {k \over 4 \pi^2} \right) -2 \log\left( {\rm vol}(U(N))\right) -2 N^2 \log (2)
\ee
where the first two terms come from the CS theories, and the last term comes from the supersymmetric matter. Here we assume $k>0$. Notice that the theory with opposite level $-k$ gives the same contribution as the theory with level $k$. In order to obtain the planar limit of this quantity, we 
have to expand the volume of $U(N)$ at large $N$. 
Using the asymptotic expansion of 
the Barnes function
\ben
\log\, G_2 (N+1)&=& {N^2\over 2} \log \, N -{1\over 12} \log \, N -{3\over 4} N^2 +{1\over 2} N \, \log \, 2\pi + \zeta'(-1)
\nonumber\\
&+& \sum_{g=2}^{\infty} {B_{2g} \over 2g(2g-2)}N^{2-2g},
\een
where $B_{2g}$ are the Bernoulli numbers, we finally obtain the weakly coupled, planar result
\be
\label{fabjmweak}
F_{\rm ABJM}(\IS^3)\approx N^2\left\{ \log (2 \pi \lambda) -{3\over 2} -2 \log(2)\right\}. 
\ee

\sectiono{Strong coupling and AdS duals}
Some SCFTs in three dimensions have AdS duals given by M-theory/string theory on backgrounds of the form
\be
\label{xdual}
X=\text{AdS}_4\times X_{6,7},  
\ee
where $X_{6,7}$ is a six-dimensional or seven-dimensional compactification manifold, depending on whether we consider a superstring or an M-theory dual, respectively. 
One of the consequences of the AdS/CFT duality is that the partition function of the Euclidean gauge theory 
on $\IS^3$ should be equal to the partition function of the Euclidean version of M-theory/string theory on the dual AdS background \cite{wittenads}, i.e. 
\be
Z_{\rm CFT}(\IS^3)=Z(X).
\ee
In the large $N$ limit, we can compute the r.h.s. in classical (i.e. genus zero) string theory, and at strong coupling it is sufficient to consider the supergravity approximation. This means that the partition function of the strongly coupled gauge theory on $\IS^3$ in the planar limit should be given by 
\be
Z_{\rm CFT}(\IS^3) \approx \re^{-I(\text{AdS}_4)} 
\ee
where $I$ is the classical gravity action evaluated on the AdS$_4$ metric. This gives a prediction for the strongly coupled behavior of the gauge theory. 
However, the gravitational action on AdS$_4$ is typically divergent, and it has to be regularized in order to obtain finite results. I will now review 
the method of holographic renormalization and work out two examples: the (related) example of the Casimir energy of $\CN=4$ SYM on $\IR \times \IS^3$, and 
the case of main interest for us, namely the free energy of ABJM theory on $\IS^3$. 

\subsection{Holographic renormalization}

The gravitational action in an Euclidean space with boundary has two contributions. The first contribution is the bulk term, given by the Einstein--Hilbert action
\be
I_{\rm bulk}=-{1\over 16 \pi G_N} \int_{M} \rd^{n+1} x \, {\sqrt{G}} \left( R -2 \Lambda\right)
\ee
where $G_N$ is Newton's constant in $n+1$ dimensions and $G$ is the $(n+1)$-dimensional metric. The second contribution is the surface term \cite{gh}
\be
I_{\rm surf}=-{1\over 8 \pi G_N} \int_{\partial M}   K |\gamma|^{1/2} \rd^n x, 
\ee
where $\partial M$ is the boundary of spacetime, $\gamma$ is the metric induced by $G$ on the boundary, and $K$ is the extrinsic curvature of 
the boundary. $K$ satisfies the useful 
relation (see for example \cite{poisson})
\be
{\sqrt{\gamma}} K =\CL_n {\sqrt{\gamma}}
\ee
where $n$ is the normal unit vector to $\partial M$, and $\CL_n$ is the Lie derivative along this vector. 
Both actions, when computed on an AdS background, diverge due to the non-compactness of the space. For example, after using Einstein's equations, the bulk action of an AdS space of radius $L$ can be written as
\be
\label{onshellEH}
I_{\rm bulk}={n \over 8 \pi G_N L^2} \int \rd^{n+1} x {\sqrt{G}} 
\ee
which is proportional to the volume of space-time, and it is divergent. 

In order to use the AdS/CFT correspondence, we have to regularize the gravitational action in an appropriate way. The procedure which has 
emerged in studies of the AdS/CFT correspondence is to introduce a set of {\it universal counterterms}, 
depending only on the induced metric on the boundary, which lead to finite values of the gravitational action, energy-momentum tensor, etc. This procedure gives values for the gravitational quantities in agreement with the corresponding quantities computed in the CFT side, and we will adopt it here. It is sometimes called ``holographic renormalization" and it has been developed in for example \cite{hs,bk,ejm,dhss}. We now present the basics of holographic renormalization in AdS. Useful reviews, focused on AdS$_5$, can be found in for example \cite{skenderis,kiritsis}. 

An asymptotically AdS metric in $n+1$ dimensions with radius $L$ and cosmological constant 
\be
\Lambda=-{n(n-1) \over 2 L^2}
\ee
can be written near its boundary at $u=0$ as 
\be
\rd s^2 =L^2 \left[ {\rd u^2 \over u^2} + {1 \over u^2} g_{ij}(u^2,x) \rd x^i \rd x^j \right].
\ee
The metric $g_{ij}(u^2,x)$ can be expanded in a power series in $u$ near $u=0$,
\be
\label{metricps}
g_{ij}(u^2, x)=g^{(0)}_{ij}(x)+ u^2 g^{(2)}_{ij}(x) + u^4 \left[ g^{(4)}(x)+ \log(u^2) h^{(4)}(x)\right]+\cdots
\ee
where the first term, $g^{(0)}_{ij}$, is the metric of the CFT on the boundary. 
The coefficients $g^{(2n)}_{ij}$ appearing here can be solved recursively in terms of $g^{(0)}_{ij}$ by plugging (\ref{metricps}) in Einstein's equations. One finds, for example \cite{dhss}\footnote{The sign in the curvature is opposite to the conventions in \cite{dhss}, which give a positive curvature to AdS.}
\be
\label{ntlmetric}
g^{(2)}_{ij}=-{1\over n-2} \left(R_{ij} -{1\over 2(n-1)} R g^{(0)}_{ij} \right),
\ee
where $R_{ij}$ and $R$ are the Ricci tensor and curvature of $g^{(0)}_{ij}$. 
The resulting metric is then used to compute the gravitational action with a cut-off at $u=\epsilon$ which regulates 
the divergences, 
\be
I_{\epsilon}=-{1\over 16 \pi G_N} \int_{M_\epsilon} \rd^{n+1} x \, {\sqrt{G}} \left( R+{n(n-1) \over L^2} \right) -{1\over 8 \pi G_N} \int_{\partial M_\epsilon}  K |\gamma|^{1/2} \rd^n x. 
\label{gaction}
\ee
Here, $M_\epsilon$ is the manifold with $u\ge \epsilon$ and a boundary $\partial M_\epsilon$ at $u=\epsilon$. To calculate the boundary term, we consider the normal vector to the hypersurfaces of constant $u$, 
\be
n^u=-{u\over L}.
\ee
The minus sign is due to the fact that the boundary is at $u=0$, so that the normal vector points towards the origin. The induced metric is 
\be
\label{inducedmetric}
\gamma_{ij} \rd x^i \rd x^j={L^2  \over u^2} g_{ij}(u^2,x) \rd x^i \rd x^j,
\ee
with element of volume 
\be
{\sqrt{\gamma}}=\left( {L\over u}\right)^n {\sqrt{g}}.
\ee
The intrinsic curvature of the hypersurface at constant $u$ is then 
\be
{\sqrt{\gamma}} K =\CL_n {\sqrt{\gamma}}=-{u \over L} \partial_u \left[ \left( {L\over u}\right)^n {\sqrt{g}} \right]={ n L^{n-1} \over u^n} \left( 1- {1\over n} u \partial_u \right) {\sqrt{g}}.
\ee
We then find, 
\be
I_{\epsilon}={n L^{n-1} \over 8 \pi G_N } \int \rd^{n} x \int_{\epsilon}  {\rd u \over u^{n+1}} {\sqrt{g}} -{n L^{n-1} \over 8 \pi G_N \epsilon^n }  \int \rd^{n} x   \left( 1- {1\over n} u \partial_u \right) {\sqrt{g}}\Bigl|_{u=\epsilon}.
\ee
The singularity structure of this regulated action is \cite{hs,dhss}
\be
\label{Ieps}
I_{\epsilon}={L^{n-1} \over 16 \pi G_N} \int \rd^n x {\sqrt{g^{(0)}}}\left( \epsilon^{-n} a_{(0)} + \epsilon^{-n+2} a_{(2)} +\cdots -2 
\log(\epsilon) a_{(n)}\right) 
+\CO(\epsilon^0).
\ee
The logarithmic divergence appears only when $n$ is even. In order to regularize power-type divergences in $n=3$ and $n=4$, it suffices 
to calculate the first two coefficients, $a_{(0)}$ and $a_{(2)}$. Let us now calculate these coefficients (the next two are computed in \cite{dhss}). We first expand, 
\be
{\rm det} \, g ={\rm det} \, g^{(0)} \left( 1+ u^2 \tr \left( g^{(0)-1} g^{(2)} \right) +\cdots\right), 
\ee
so that
\be
\label{gexp}
\ba
{\sqrt{g(u^2,x)}}&={\sqrt{g^{(0)}}}  \left( 1+ {u^2 \over 2}  \tr \left( g^{(0)-1} g^{(2)} \right) +\cdots\right),\\
 \left( 1- {1\over n} u \partial_u \right) {\sqrt{g(u^2,x)}}&={\sqrt{g^{(0)}}}  \left(1 + {n-2\over 2n} u^2 \tr \left( g^{(0)-1} g^{(2)} \right) +\cdots\right). 
 \ea
\ee
The regulated Einstein--Hilbert action gives
\be
{n L^{n-1} \over 8 \pi G_N } \int \rd^{n} x \, {\sqrt{g^{(0)}}} \left[ {1\over n \epsilon^n} +{1\over 2(n-2) \epsilon^{n-2}}  \tr \left( g^{(0)-1} g^{(2)} \right) +\cdots\right],
\ee
while the regulated Gibbons--Hawking term gives
\be
-{n L^{n-1} \over 8 \pi G_N } \int \rd^{n} x \, {\sqrt{g^{(0)}}} \left[ {1\over \epsilon^n} +{n-2\over 2n \epsilon^{n-2}}  \tr \left( g^{(0)-1} g^{(2)} \right) +\cdots\right].
\ee
In total, we find
\be
I_\epsilon={ L^{n-1} \over 16 \pi G_N } \int \rd^{n} x \, {\sqrt{g^{(0)}}} \left[ {2(1-n) \over  \epsilon^n} -{n^2 -5n+4 \over (n-2) \epsilon^{n-2}}  \tr \left( g^{(0)-1} g^{(2)} \right) +\cdots\right],
\ee
and we deduce, 
\be
\ba
a_{(0)}&=2(1-n),\\
a_{(2)}&=-{(n-4)(n-1) \over n-2} \tr \left( g^{(0)-1} g^{(2)} \right), \quad n\not=2. 
\ea
\ee
The counterterm action is obtained by using a gravitational analogue of the minimal subtraction scheme, and it is given by minus the divergent part of $I_\epsilon$, 
\be
\ba
I_{\rm ct}={ L^{n-1} \over 16 \pi G_N } \int \rd^{n} x \, {\sqrt{g^{(0)}}} \biggl[ &{2(n-1) \over  \epsilon^n} +{(n-4)(n-1) \over (n-2) \epsilon^{n-2}}  \tr \left( g^{(0)-1} g^{(2)} \right) +\cdots\\
&+2 \log(\epsilon) a_{(n)} \biggr].\ea
\ee
As pointed out in \cite{bk}, we should re-write this in terms of the induced metric in the boundary (\ref{inducedmetric}), evaluated at $u=\epsilon$. From (\ref{gexp}) we deduce
\be
{\sqrt{g^{(0)}}} =\left( {\epsilon \over L}\right)^n \left(1 - {\epsilon^2 \over 2}  \tr \left( g^{(0)-1} g^{(2)} \right)  +\CO(\epsilon^4)\right){\sqrt{\gamma}}.
\ee
On the other hand, from (\ref{ntlmetric}) we obtain
\be
\ba
\tr \left( g^{(0)-1} g^{(2)} \right)& =-{1\over n-2} \left( g^{(0)}_{ij}R^{ij}-{1\over 2(n-1)} R g^{(0)}_{ij} g^{(0) ij}\right)=-{1\over 2(n-1)} R \\
&=-{L^2\over 2 (n-1) \epsilon^2} R[\gamma]+\cdots,
\ea
\ee
where in the first line the Ricci tensor and curvature are computed for $g^{(0)}_{ij}$, while in the second line the curvature is computed for the induced metric $\gamma$. 
Plugging these results into the counterterm action we find
\be
\label{ctI}
\ba
I_{\rm ct}&={ 1 \over 16 \pi G_N  L }  \int \rd^{n} x \,  {\sqrt{\gamma}} \left(1 + {L^2 \over 4 (n-1) } R[\gamma] +\cdots\right)\\
& \qquad  \qquad \qquad \times \left[ 2(n-1) -{n-4 \over 2(n-2) } L^2 R[\gamma]   +2 \log (\epsilon) a_{(n)} [\gamma] +\cdots \right]\\
&={ 1 \over 8 \pi G_N  }  \int \rd^{n} x \,  {\sqrt{\gamma}} \left( 2 \log (\epsilon) a_{(n)} [\gamma] + {n-1 \over L} +{L \over 2(n-2)} R[\gamma]+\cdots\right),
\ea
\ee
which is the result written down in \cite{dhss,ejm} (for Euclidean signature). This is the counterterm action which is relevant for AdS in four and five dimensions, and the dots denote higher order counterterms (in the Riemann tensor of the induced metric) which are 
needed for higher dimensional spaces \cite{bk,ejm,dhss}. The total, regularized gravitational action is then 
\be
\label{reggrav}
I=I_{\rm bulk}+I_{\rm surf}+ I_{\rm ct}
\ee
and it yields a finite result by construction. The removal of these IR divergences in the gravitational theory is dual to the removal of UV divergences in the 
CFT theory. It can be verified in examples that the gravity answers obtained by holographic renormalization match the answers obtained in CFT on a curved background 
after using zeta-function regularization \cite{bk}. In the next subsection we work out a beautiful example of this matching closely related to the techniques developed here, 
namely the Casimir energy for $\CN=4$ super Yang--Mills on $\IR \times \IS^3$, which was first derived in \cite{bk}.

\subsection{Example 1: Casimir energy in AdS$_5$}
Let us consider an $n$-dimensional CFT on the manifold $ \IS^{n-1} \times \IS^1_\beta$, where the one-circle has length $\beta$, and with periodic boundary conditions for the fermions. The supersymmetric partition function on this manifold is given by
\be
Z_{\rm CFT} \left(\IS^{n-1} \times \IS^1_\beta \right) =\tr \left[ (-1)^F \re^{-\beta H(\IS^{n-1})}\right] \approx \re^{-\beta E_0}
\ee
where in the last step we have considered the large $\beta$ limit, and $E_0$ is the energy of the ground state, i.e. the Casimir energy on $\IS^{n-1}$. The AdS/CFT correspondence implies that this partition function can be obtained 
by computing the partition function of a superstring/M-theory on a manifold of the form (\ref{xdual}), where now the AdS space has the boundary $\IS^{n-1} \times \IS^1_\beta$ \cite{wittenads}. In the SUGRA approximation, this can be computed by evaluating the regularized gravity action $I$ (\ref{reggrav}), and this should give 
the planar, strongly coupled limit of the CFT partition function, 
\be
Z_{\rm CFT} \left(\IS^{n-1} \times \IS^1_\beta  \right) \approx \re^{-I(\text{AdS}_{n+1})}. 
\ee
In order to calculate $I$, we need an Euclidean AdS metric which is asymptotic to $\IS^{n-1} \times \IS^1_\beta$ at the boundary. The relevant metric is given by \cite{ejm}
\be
\label{genadsk}
\rd s^2 =\left( 1+ {r^2 \over L^2} \right) \rd \tau^2 + {\rd r^2 \over  1 + r^2/L^2 } +r^2 \rd\Omega_{n-1}^2
\ee
where the boundary is now at $r\rightarrow \infty$. The sphere $\IS^{n-1}$ at the boundary has radius $L$. 
Here, $\rd \Omega_n^2$ is the metric on the unit $n$-sphere, and $\tau$ has length $\beta$. Let us 
compute the gravitational action for this theory, with a cutoff at the boundary $\partial M$ located at $r$. We will only include the counterterms presented in (\ref{ctI}), which are sufficient in three and four dimensions. At the end of the calculation we will take 
$r\rightarrow \infty$. Let us denote
\be
V(r)=1 +{r^2 \over L^2}.
\ee
The normal unit vector to the boundary is 
\be
n=V^{1/2}(r) {\partial \over \partial r},
\ee
while the induced metric is
\be
\gamma_{ij}\rd x^i \rd x^j=V(r) \rd \tau^2 + r^2 \rd \Omega_{n-1}^2.
\ee
The calculation of the different contributions to the total gravitational action is straightforward:
\be
\ba
8 \pi G_N I_{\rm bulk}&={1 \over L^2} r^{n} {\rm vol}(\IS^{n-1}) \beta ,\\
 8 \pi G_N I_{\rm surf}&=- {\rm vol}(\IS^{n-1})\beta r^{n-1} \left\{ {n-1\over r} V(r) + {1\over 2} V'(r)\right\},\\
8 \pi G_N I_{\rm ct}&={\rm vol}(\IS^{n-1})\beta r^{n-1} \left\{ {n-1\over L} + { (n-1)L \over 2 r^2 } \right\}  V^{1/2} (r). \ea
\ee
The total action is 
\be
\ba
&8 \pi G_N I \\
&=  {{\rm vol}(\IS^{n-1}) \beta\over L^2}  r^{n-1}  \left[  -{(n-1) L^2 \over r} \left( 1+{r^2 \over L^2} \right)  +(n-1) r \left(1 + { L^2 \over 2 r^2} \right) \left( 1+ {L^2 \over r^2}  \right)^{1/2}\right].
\ea
\ee
Expanding for $r\to \infty$ we find, for $n=3$, a vanishing action, while for $n=4$ (i.e. for AdS$_5$) we find
\be
8 \pi G_N I= {3  {\rm vol}(\IS^3) \beta L^2 \over 8} \Rightarrow I={3 \pi L^2  \beta  \over 32 G_N}.
\ee
In this approximation the Casimir energy is then
\be
\label{casgravity}
E_0={3 \pi L^2  \over 32 G_N}.
\ee
Using the standard AdS$_5$/CFT$_4$ dictionary for $\CN=4$ super Yang--Mills \cite{maldacena}
\be
N^2= {\pi L^3 \over 2 G_N},
\ee
we obtain the planar, strong coupling value of the Casimir energy of $\CN=4$ super Yang--Mills on $\IS^3$, 
\be
\label{casn4}
E={3 N^2 \over 16 L}. 
\ee

Using the technology developed in the previous section it is an easy exercise to compute the 
Casimir energy directly in QFT, at weak coupling. A massless scalar field in a four-dimensional 
curved space with Minkowski signature satisfies the wave equation
\be
\label{mscalar}
\left(g^{\mu \nu}D_\mu D_\nu + \xi R\right) \phi=0.
\ee
For a conformally coupled scalar in 4d, we have 
\be
\xi={1\over 6}. 
\ee
Let $\eta, {\bf x}$ be coordinates for $\IR \times\IS^3$. The metric can be written as
\be
\rd s^2= L^2 \left( \rd \eta^2 -\rd \Omega_3^2\right)
\ee
where $L$ is the radius of $\IS^3$, and 
$\rd\Omega_3^2$ is the element of volume on an $\IS^3$ of unit radius. The coordinate $\eta$ (which is dimensionless) is called the conformal time parameter 
(see \cite{bd}, p. 120), and it is related to the time coordinate by
\be
t=L \eta. 
\ee
We write the wavefunctions in factorized form 
\be
u_m (\eta, {\bf x})= \chi_m(\eta) S_{m-1\over 2} ({\bf x}), 
\ee
where $S_j(x)$ is a scalar spherical harmonic (\ref{sshar}), and it is an eigenfunction of the Laplacian,
\be
\Delta^0 S_{m-1\over 2} =\left(m^2-1\right) S_{m-1\over 2} ,  \quad m=1, 2, \cdots. 
\ee
These functions have degeneracy $m^2$. $u_m$ satisfies the wave equation (\ref{mscalar}), which after separation of variables 
leads to the following equation for $\chi_m(\eta)$:
\be
\partial_\eta^2\chi_m+ \left( m^2 -1+L^2 \xi R \right) \chi_m=0.
\ee
Here $R$ is the curvature of an $\IS^3$ of radius $L$. In the case of a 4d conformally coupled scalar, the above equation reads
\be
\partial_\eta^2\chi_m+  m^2 \chi_m=0
\ee
with the solution 
\be
\chi_m(\eta) \propto \re^{-\ri m t/L}. 
\ee
The Casimir energy for this conformally coupled scalar is given by
\be
E={1\over 2} \sum_{m=1}^{\infty} m^2\cdot (m/L)
\ee
where the first factor $m^2$ comes from the degeneracies. 
The above sum is of course divergent, but we can use zeta-function regularization to obtain 
\be
E(s)= {1\over 2L} \sum_{m=1}^{\infty} m^2 m^{-s}={1\over 2L} \zeta(s-2)
\ee
which can be analytically continued to $s=-1$. We find in this way
\be
E=E(-1)={1\over 2L} \zeta(-3)={1\over 240 L}. 
\ee

For Weyl spinors, the Casimir energy is obtained by summing over the modes of the spinor spherical harmonics, and with a negative sign due to Fermi 
statistics, i.e. 
\be
E_{\text{spinor}}=-{1\over 2L} \sum_{n=1}^\infty 2n (n+1) (n+1/2), 
\ee
where $2n(n+1)$ is the degeneracy of the eigenvalue $n+1/2$ of the Dirac operator. This can be again regularized by considering
\be
E_{\text{spinor}}(s)=-{1\over L}\sum_{n=1}^{\infty} {n(n+1) \over \left( n+{1\over 2} \right)^s}
\ee
and analytically continuing it for $s=-1$. Using the previous result (\ref{fermionzeta}) we deduce
\be
E_{\text{spinor}}(-1)=-{1\over L}\left(\left( 2^{-3}-1\right) \zeta(-3) -{1\over 4} \left(2^{-1}-1\right) \zeta(-1)\right)={17 \over 960 L}. 
\ee
Finally, for a gauge field, we have
\be
E_{\text{gauge}}={1\over 2L} \sum_{n=1}^\infty 2n(n+2)  (n+1), 
\ee
where $n+1$ is the square root of the energies of the modes (i.e. the square root of the eigenvalues (\ref{vsheigen}) of the Laplacian, after gauge fixing). 
This is regularized as
\be
E_{\text{gauge}}(s)={1\over L} \sum_{n=1}^\infty {n(n+2)  \over (n+1)^{s}}={1\over L}\sum_{m=1}^{\infty} {m^2-1 \over m^{s}}= {1\over L} \left( \zeta(s-2) -\zeta(s) \right), 
\ee
and one finds
\be
E_{\text{gauge}}(-1)={1\over L} \left( \zeta(-3) -\zeta(-1) \right)={11\over 120 L}. 
\ee
It follows that the Casimir energy on $\IS^3\times \IR$ for a QFT with $n_0$ conformally coupled real scalars, $n_{1/2}$ Weyl spinors and $n_1$ vector fields is 
\be
E={1\over 960 L}\left( 4 n_0 +17 n_{1/2} + 88 n_1\right). 
\ee
This was first shown in \cite{ford} by a cutoff regularization of the sum over modes. In the case of $\CN=4$ SYM with gauge group $U(N)$ we have 
\be
n_0=6 N^2, \quad n_{1/2}= 4 N^2, \quad n_1=N^2, 
\ee
and the Casimir energy obtained at weak coupling agrees with (\ref{casn4}), which is {\it a priori} the strong-coupling value. This was first observed in \cite{bk}, 
and it can be also regarded \cite{dhss,sk2} as a consequence of the agreement between the trace anomaly in both sides of 
the correspondence \cite{hs}. Indeed, it is known that, for 
conformal fields in conformally flat backgrounds like $\IS^{n-1} \times \IS^1$, the vev of the energy-momentum tensor is completely determined by the trace anomaly \cite{brownc}, which fixes then the value of the Casimir energy on $\IS^{n-1}$ \cite{capcos}. The weak and the strong coupling results are 
the same because these quantities are protected by a non-renormalization theorem and do not depend on the 't Hooft coupling $\lambda=g^2 N$. 
This will not be the case for the free energy on the sphere for 3d SCFTs, which we now compute at strong coupling by using the large $N$ AdS dual. 

\subsection{Example 2: free energy in AdS$_4$}

We are interested in studying CFTs on $\IS^n$. Therefore, in the AdS dual we need the Euclidean version of the AdS metric with that boundary, which can be written 
as \cite{ejm}
\be
\label{freemetric}
\rd s^2 = {\rd r^2 \over  1 + r^2/L^2 } +r^2 \rd\Omega_{n}^2
\ee
with the notations of the previous subsection. The boundary is again at $r\rightarrow \infty$. This metric can be also written as \cite{wittenads,ejm}
\be
\rd s^2 =L^2 \left(\rd \rho^2  +\sinh^2(\rho)\rd\Omega_{n}^2 \right). 
\ee
Let us compute the regularized gravitational action (\ref{reggrav}), again with a cutoff at the boundary $\partial M$ located at constant $r$. The element of volume of the 
metric $G$ given in (\ref{freemetric}) is 
\be
{\sqrt{G}}={r^n \over {\sqrt{1+ r^2/L^2}}} {\sqrt{g_{\IS^{n}}}}, 
\ee
where $g_{\IS^n}$ is the metric on an $n$-sphere of unit radius. The bulk action is just (\ref{onshellEH}), i.e. 
\be
I_{\rm bulk}={ n {\rm vol}(\IS^n) \over 8 \pi G_N L} \int_0^r \rd \rho { \rho^n \over {\sqrt{L^2 + \rho^2}}}.
\ee
To calculate the surface action, we notice that the unit normal vector to $\partial M$ is 
\be
n={\sqrt{1+r^2/L^2}}{\partial \over \partial r}
\ee
while the induced metric is 
\be
\gamma= r^2 g_{\IS^n}
\ee
which has the element of volume
\be
{\sqrt{\gamma}}=r^n {\sqrt{g_{\IS^{n}}}}
\ee
and scalar curvature
\be
R[\gamma]={R_{\IS^n} \over r^2} ={n(n-1) \over r^2}.
\ee
We then obtain 
\be
\CL_n {\sqrt{\gamma}}=n r^{n-1}{\sqrt{1+r^2/L^2}} {\sqrt{g_{\IS^{n}}}},
\ee
and the surface term is 
\be
I_{\rm surf}=-{n r^{n-1}\over 8 \pi G_N} {\sqrt{1+r^2/L^2}} {\rm vol}(\IS^n).
\ee
Finally, the first two counterterms are given by
\be
{{\rm vol}(\IS^n)\over 8 \pi G_N}\left[ {n-1\over L} r^n  + {L r^{n-2} n(n-1)\over 2(n-2)} \right]={{\rm vol}(\IS^n)\over 8 \pi G_N}{r^n (n-1) \over L} 
\left[ 1  + { n\over 2(n-2)} {L^2\over r^2} \right].
\ee
Putting everything together we obtain, 
\be
\ba
I={{\rm vol}(\IS^n)  \over 8 \pi G_N L } \biggl[ & n  L^n \int_0^{r/L} \rd u { u^n \over {\sqrt{1 + u^2}}}-n r^{n-1}{\sqrt{r^2+ L^2}} \\
&+ r^n (n-1) \left(1  + { n\over 2(n-2)} {L^2\over r^2} 
\right)\biggr].
\ea
\ee
We should now take the limit of this expression when $r\rightarrow \infty$. For $n=3$ we find
\be
I={{\rm vol}(\IS^3)  \over 8 \pi G_N L } \left( 2L^3 +\CO(r^{-1})\right), 
\ee
therefore we obtain a {\it finite} action given by
\be
\label{freeI}
I={\pi L^2 \over 2 G_N}.
\ee
This will give us the strong coupling prediction for the free energy on $\IS^3$ of supersymmetric Chern--Simons--matter theories with an AdS dual. 
%
%

\subsection{ABJM theory and its AdS dual}

In order to compute the free energy of ABJM theory at strong coupling we have to be more precise about the gauge/gravity dictionary. We will 
now write down this dictionary for supersymmetric Chern--Simons--matter theories, which was first established for ABJM theory in \cite{abjm}. 
We will not attempt here to review the derivation of the duality. A pedagogical introduction can be found in \cite{klebrev}. 

The AdS duals to the theories we will consider are given by M-theory on 
\be
\text{AdS}_4 \times X_7, 
\ee
where $X_7$ is a seven-dimensional manifold. In the case of ABJM theory, 
\be
X_7=\IS^7/\IZ_k. 
\ee
The eleven-dimensional metric and four-form flux are given by the Freund--Rubin background (see \cite{dnp} for a review)
\be
\label{metricflux}
\ba
\rd s_{11}^2 &= L_{X_7}^2\left( \frac{1}{4}\rd s_{\text{AdS}_4}^2+\rd s_{X_7}^2\right), \\
	F &= \frac{3}{8}L_{X_7}^3 \omega_{\text{AdS}_4},
\ea
\ee
where $ \omega_{\text{AdS}_4}$ is the volume form with unit radius. 
The radius $L_{X_7}$ is determined by the flux quantization condition
\be
(2\pi \ell_p)^6 Q=\int_{X_7} \star_{11} F =6 L_{X_7}^6 {\rm vol}(X_7).
\ee
In this equation, $\ell_p$ is the eleven-dimensional Planck length. The charge $Q$ is 
given, at large radius, by the number of M2 branes $N$, but it receives corrections \cite{bh,ahho}. In ABJM theory we have
\be
\label{qshift} 
Q=N-{1\over 24}\left( k-{1\over k}\right).
\ee
This extra term comes from the coupling 
\be
\int C_3 \wedge I_8 
\ee
in M-theory, which contributes to the charge of M2 branes. Here, $I_8$ is proportional to the Euler density in eight dimensions, and it satisfies
\be
\int_{M_8} I_8=-{\chi \over 24} 
\ee
where $M_8$ is a compact eight-manifold. In ABJM theory, the relevant eight-manifold is $\IC^4/\IZ_k$, with regularized Euler characteristic
\be
\chi \left( \IC^4/\IZ_k\right)=k-{1\over k}. 
\ee
This leads to the shift in (\ref{qshift}). 

One final ingredient that we will need is Newton's constant in four dimensions. It can be obtained from the Einstein--Hilbert action in eleven dimensions, which leads to its four-dimensional counterpart by standard Kaluza--Klein reduction, 
 \be
 \label{nconstant4d}
 {1\over 16 \pi  G_{11}} \int \rd^{11} x \, {\sqrt{g_{11}}} R_{11} \rightarrow {1\over 4} \cdot {L_{X_7}^7 \over 16 \pi  G_{11}} {\rm vol}\left(X_7\right) \int \rd^{4} x \, {\sqrt{g_4 }} R_{4}= {1 \over 16 \pi  G_{N}} 
  \int \rd^{4} x \, {\sqrt{g_4 }} R_{4},
  \ee
  where $G_{11}$, $G_N$ denote the eleven-dimensional and the four-dimensional Newton's constant, respectively, and the volume of $X_7$ is calculated for unit radius. In the resulting Einstein--Hilbert action in four dimensions, the metric and scalar curvature refer to an AdS$_4$ space of radius $L_{X_7}$, and not $L_{X_7}/2$ as in (\ref{metricflux}). This 
  is the source for the extra factor of $1/4$ in the second term. 
Recalling that 
\be
16 \pi G_{11}={1\over 2\pi} \left( 2 \pi \ell_p\right)^9, 
\ee
we obtain 
\be
\label{newnew}
\frac{1}{G_{N}}={2 {\sqrt{6}} \pi^2 Q^{3/2} \over 9 {\sqrt{{\rm vol}(X_7)}}}{1\over L_{X_7}^2}.
\ee
It follows that the regularized gravitational action (\ref{freeI}) is given by
\be
I={\pi L_{X_7}^2 \over 2 G_N} =Q^{3/2} {\sqrt{ 2 \pi^6 \over 27 {\rm vol}(X_7)}}.
\ee
In particular, for ABJM theory we have
\be
I={\pi\sqrt{2} \over 3} k^{1/2}  Q^{3/2}.
\ee
In the supergravity and planar approximation we can just set $Q=N$, and we find indeed that the planar free energy is given by 
\be
\label{fvol}
-{1\over N^2} F(\IS^3) \approx {\sqrt{ 2 \pi^6 \over 27 {\rm vol}(X_7)}}{1\over N^{1/2}}. 
\ee
In the case of ABJM theory we finally obtain the strong coupling result stated in (\ref{introgoal}).

\sectiono{Localization}

Localization is an ubiquitous technique in supersymmetric QFT which makes possible to reduce an infinite-dimensional path integral to a finite dimensional integral. It features prominently in Witten's topological quantum field theories of the cohomological type, where one can argue that the semiclassical approximation is exact, see \cite{bt,cmr} for reviews and a list of references. 

The basic idea of localization is the following. Let $\delta$ be a Grassmann-odd symmetry of a theory with action $S(\phi)$, where $\phi$ denotes the set of fields in the theory. We assume that the measure of the path integral is invariant under $\delta$ as well (i.e. $\delta$ is not anomalous), and that
\be
\delta^2 =\CL_B
\ee
where $\CL_B$ is a Grassmann-even symmetry. In a Lorentz-invariant, gauge invariant theory, $\CL_B$ could be a combination of a Lorentz and a gauge transformation. Consider now the perturbed partition function 
\be
Z(t)=\int {\cal D}\phi\, \re^{-S - t \delta V}, 
\ee
where $V$ is a Grassmann-odd operator which is invariant under $\CL_B$. It is easy to see that $Z(t)$ is independent of $t$, since
\be
{\rd Z \over \rd t}=-\int {\cal D}\phi\, \delta V\, \re^{-S - t \delta V}=-\int {\cal D}\phi \, \delta \left( V \re^{-S - t \delta V}\right)=0. 
\ee
Here we have used the fact that $\delta^2V=\CL_B V=0$. In the final step we have used the fact that $\delta$ is a symmetry of the path integral, in order to interpret the integrand as a total derivative. In some cases, the integral of the total derivative does not vanish due to boundary terms 
(a closely related example appears in section 11.3 of \cite{mw}), but if the integral decays sufficiently fast in field space one expects the perturbed 
partition function $Z(t)$ to be independent of $t$. This means that it can be computed at $t=0$ (where one recovers the original partition function) but also for other values of $t$, like $t\rightarrow \infty$. In this regime, simplifications typically occur. For 
example, if $\delta V$ has a positive definite bosonic part $(\delta V)_B$, the limit $t \rightarrow \infty$ localizes the integral to a submanifold of field space where
\be
(\delta V)_B=0. 
\ee
It turns out that, in many interesting examples, this submanifold is finite-dimensional. This leads to a ``collpase" of the path integral to a finite-dimensional 
integral. It is easy to see that this method also makes it possible to calculate the correlation functions of $\delta$-invariant operators. 

In order to see how the method of localization works, let us briefly review a beautiful and simple 
example, namely the field theoretical version of the Poincar\'e--Hopf theorem. 

\subsection{A simple example of localization}

The Poincar\'e--Hopf theorem has been worked out from the point of view of supersymmetric localization in many references, like for example \cite{bt,cmr,ll}. 
Let $X$ be a Riemannian manifold of even dimension $n$, with metric $g_{\mu \nu}$, vierbein $e^a_\mu$, and let $V_\mu$ be a vector field on $X$. We will consider the following ``supercoordinates" on the tangent bundle $TX$
\be
\left( x^\mu, \psi^\mu\right), \qquad \left(\bar \psi_\mu, B_\mu\right), 
\ee
where the first doublet represents supercoordinates on the base $X$, and the second doublet represents supercoordinates on the fiber.  
$\psi^\mu$ and $\bar \psi_\mu$ are Grasmann variables. The above supercoordinates are related by the Grasmannian symmetry
\be
\ba
\delta x^\mu&=\psi^\mu,\\
\delta\psi^\mu&=0,\ea  \qquad
\ba 
\delta\bar\psi_\mu&=B_\mu,\\
\delta B_\mu&= 0,\ea
\ee
which squares to zero, $\delta^2=0$. With these fields we construct the ``action" 
 \be
 \label{phaction}
 S(t) =\delta \Psi, \qquad \Psi={1\over 2} \bar\psi_\mu \left(B^\mu +2\ri t V^\mu
+\Gamma^\sigma_{\tau\nu}\bar\psi_\sigma\psi^\nu g^{\mu\tau} \right),
\ee
 and we define the partition function of the theory as
 \be
 Z_X(t)={1\over{(2\pi)^{n}}} \int_X \rd x \, \rd\psi \,  \rd\bar\psi \,  \rd B \, \re^{-S(t)}.
 \ee
Using that
\be
{\partial g^{\mu \sigma} \over \partial x^{\tau}} =-\Gamma^\mu_{\tau \sigma} g^{\tau \sigma} -\Gamma^\sigma_{\tau \nu} g^{\tau \mu}, 
\ee
one finds that, in the resulting theory, $B_\mu$ is a Gaussian field with mean value 
\be
B^\mu=-\ri t V^\mu - g^{\mu \tau} \Gamma_{\tau \nu}^{\sigma} \bar \psi_\sigma \psi^\nu. 
\ee
If we integrate it out, we obtain an overall factor 
\be
{(2\pi)^{n/2} \over {\sqrt{g}}}, 
\ee
and the action becomes
\be
{t^2 \over 2}  g_{\mu\nu} V^{\mu}V^{\nu}
-{1\over4}R^{\rho\sigma}{}_{\mu\nu}\bar\psi_\rho\bar\psi_\sigma
\psi^\mu\psi^\nu
 -\ri t \nabla_\mu V^\nu \bar\psi_\nu \psi^\mu.
 \ee
We can define orthonormal coordinates on the fiber by using the inverse vierbein, 
\be
\chi_a=E_a^\mu \bar \psi_\mu, 
\ee
so that the partition function reads
\be
Z_X(t)={1\over{(2\pi)^{n/2}}} \int_X \rd x \, \rd\psi \,  \rd\chi  \, \re^{-{t^2 \over 2}  g_{\mu\nu}V^\mu V^\nu
+{1\over4}R^{ab}{}_{\mu\nu}\chi_a \chi_b 
\psi^\mu\psi^\nu+\ri t \nabla_\mu V^\nu e^a_\nu \chi_a \psi^\mu }.
\ee

It is clear that this partition function should be independent of $t$, since the action can be written as 
\be
S(t)=S(0) + t \delta V, \qquad V=\ri \bar \psi_\mu V^\mu. 
\ee
We can then evaluate it in different regimes: $t\rightarrow 0$ or $t\rightarrow \infty$. The calculation when $t=0$ is very easy, since we just have
\be
Z_X(0)={1\over{(2\pi)^{n/2}}} \int_X \rd x \, \rd\psi \,  \rd\chi \, \re^{
{1\over4}R^{a b}{}_{\mu\nu}\chi_a \chi_b
\psi^\mu\psi^\nu}={1\over {(2\pi)^{n/2}}} \int_X \rd x  \, {\rm Pf}(R),
\ee
where we have integrated over the Grassmann variables $\chi_a$ to obtain the Pfaffian of the matrix $R^{ab}$. 
The resulting top form in the integrand, 
\be
e(X)={1\over {(2\pi)^{n/2}}}  {\rm Pf}(R)
\ee
is nothing but the Chern--Weil representative of the Euler class, therefore the evaluation at $t=0$ produces the Euler characteristic of $X$, 
\be
\label{chix}
Z_X(0)=\chi(X).
\ee

Let us now calculate the partition function in the limit $t\rightarrow \infty$. We will now assume that $V^\mu$ has isolated, simple zeroes 
$p_k$ where $V^\mu(p_k)=0$. These are the saddle--points of the ``path integral," so we can write 
$Z_X(t)$ as a sum over saddle--points $p_k$, and for each saddle--point we have to perform a perturbative expansion. 
Let $\xi^\mu$ be coordinates around the point $p_k$. We have the expansion, 
\be
V^\mu(x)=\sum_{n\ge 1} {1\over n!} \partial_{\mu_1} \cdots \partial_{\mu_n}V^\mu(p_k) \xi^{\mu_1} \cdots \xi^{\mu_n}.
\ee
After rescaling the variables as
\be
\xi \to  t^{-1}\xi, \qquad \psi\to t^{-1/2}\psi, \qquad \chi \to t^{-1/2} \chi, 
\ee
the theory becomes Gaussian in the limit $t\rightarrow \infty$, since higher order terms in the fluctuating fields $\xi, \psi, \chi$ contain at least a power $t^{-1/2}$. 
Interactions are suppressed, and the partition function is one-loop exact:
\be
\lim_{t \to \infty} Z_X (t) =  \sum_{p_k}{1\over{(2\pi)^{n/2}}} \int_X \rd \xi \, \rd\psi \, \rd\chi\, 
\re^{ -{1\over 2}  g_{\mu\nu}H^{(k)\mu}_\alpha  H^{(k)\nu}_\beta  \xi^\alpha \xi^\beta 
+\ri H_{\mu}^{(k)\nu} e^a_\nu
\chi_a
\psi^\mu }
\ee
where we denoted, 
\be
H^{(k)\mu}_\sigma
=\partial_\sigma V^\mu \big|_{p_k}.
\ee
Each term in this sum can now be computed as a product of a bosonic Gaussian integral, times a Grassmann integral, and we obtain 
\be
\label{vf}
\lim_{t \to \infty} Z_X (t)=\sum_{p_k}{1 \over{\sqrt{g}\vert  {\hbox{\rm
det}} ~H^{(k)}\vert}}{\rm det}(e^a_\mu) {\rm det}\, H^{(k)}
=\sum_{p_k}{{\hbox{\rm det}}~H^{(k)}
\over{\vert {\hbox{\rm det}}~H^{(k)}\vert}}.
\ee
The equality between (\ref{chix}) and (\ref{vf}) is the famous Poincar\'e--Hopf theorem. 

Conceptually, the localization analysis in \cite{kapustin} that we will review now 
is not very different from this example, although technically it is more complicated. The key common ingredient in the analysis of the $t \rightarrow \infty$ limit is that the localization locus becomes very simple, and all Feynman diagrams involving at least two loops are suppressed by a factor $t^{-1/2}$, so that the one-loop approximation is exact. 

\subsection{Localization in Chern--Simons--matter theories: gauge sector}

We are now ready to use the ideas of localization in supersymmetric Chern--Simons--matter theories on $\IS^3$, following 
\cite{kapustin}. The Grassmann-odd symmetry is 
simply $\CQ$, defined by $\delta_\epsilon=\epsilon \CQ$, where $\epsilon$ is the conformal Killing spinor satisfying (\ref{skil}). This symmetry satisfies $\CQ^2=0$, and then 
it is a suitable symmetry for localization. To localize in the gauge sector, we add to the CS-matter theory the term 
\be
-t S_{\rm YM},
\ee
which thanks to (\ref{YMder}) and (\ref{sqtrans}) is of the form $\CQ V$, and its bosonic part is positive definite. By the localization argument, the partition function 
of the theory (as well as the correlators of $\CQ$-invariant operators) does not depend on $t$, and we can take $t\rightarrow \infty$. This forces the fields to take the values 
that make the bosonic part of (\ref{yml}) to vanish. Since this is a sum of positive definite terms, they have to vanish separately. We then have the 
localizing locus, 
\be
F_{\mu \nu}=0, \qquad D_\mu \sigma=0, \qquad D+{\sigma \over r} =0. 
\ee
The first equation says that the gauge connection $A_\mu$ must be flat, but since we are on $\IS^3$ the only flat connection is $A_\mu=0$. 
Plugging this into the second equation, we obtain 
\be
\partial_\mu \sigma=0 \Rightarrow \sigma=\sigma_0, 
\ee
a constant. Finally, the third equation says that
\be
D=-{\sigma_0 \over r}. 
\ee
The localizing locus is indeed finite-dimensional: it is just the submanifold where $\sigma$ and $D$ are constant Hermitian matrices, and $A_\mu=0$.

Let us now calculate the path integral over the vector multiplet in the limit $t\to \infty$. 
We have to perform a gauge fixing, and we will choose the standard covariant gauge (\ref{gaugecond}) as in the case of Chern--Simons theory. The path integral to be calculated is 
\be
\label{intcal}
{1 \over {\rm Vol}(G)} ({\rm det}'\, \Delta^{0})^{-{1\over 2}}
 \int_{{\rm Ker}\, \rd ^{\dagger}} {\cal {D}}A
 \int_{({\rm Ker}\, \rd)^{\perp}} {\cal {D}}C {\cal {D}}{\overline C} \, 
 \re^{ {\ri k \over 4 \pi} S_{\rm SCS}-t S_{\rm YM}(A) -S_{\rm ghosts}(C,{\overline C}, A)},
 \ee
 where $C, {\overline C}$ are ghosts fields. As in the example of the Poincar\'e--Hopf theorem, we expand the fields around the localizing locus, and we set 
 \be
 \label{fixedfluc}
\ba
\sigma &= \sigma_0 + \frac{1}{\sqrt{t}} \sigma' ,\\ 
D &= -{ \sigma_0 \over r}  + \frac{1}{\sqrt{t}} D' ,\\ 
A, \, \, \lambda  &\rightarrow \frac{1}{\sqrt{t}} A, \,  \,\frac{1}{\sqrt{t}} \lambda,
\ea
\end{equation}
where the factors of $t$ are chosen to remove the overall factor of $t$ in the Yang--Mills action. In the Yang--Mills Lagrangian, only the terms which are quadratic in the fluctuations survive in this limit, namely, 
\be
\ba
{1\over 2} \int \sqrt{g} \, \rd^3 x \, \tr\biggl( &- A^\mu \Delta A_\mu - [A_\mu,\sigma_0]^2 + \partial_\mu \sigma' \partial^\mu \sigma' + ( D' + \sigma')^2 \\ 
& + {\ri }\bar  \lambda \gamma^\mu \nabla_\mu \lambda + {\ri }  \bar \lambda [ \sigma_0, \lambda] -  \frac{1}{2 }\bar  \lambda \lambda + \partial_\mu {\overline C} 
\partial^\mu C \biggr),
\ea
\ee
where we set $r=1$. We are then left with a Gaussian theory, but with non-trivial quadratic operators for the fluctuations. In the same way, when we expand (\ref{susycs}) around the fixed-point limit (\ref{fixedfluc}), we obtain
\be
{\ri k \over 4 \pi} S_{\rm SCS}={\ri k \over  2\pi}  \tr (\sigma_0^2) {\rm vol}(\IS^3) +\CO(t^{-1/2}), 
\ee
so only the first term survives in the $t \to \infty$ limit. 

Let us now calculate the path integral when $t\to \infty$. Like in the example of the Poincar\'e--Hopf theorem, we just have to compute the one--loop 
determinants. In this calculation we will only take into account the factors which depend explicitly on $\sigma_0$. The remaining, numerical factors (which 
might depend on $N$, but not on the coupling constant $k$) can be incorporated afterwards by comparing to the weak coupling results. 
The integral over the fluctuation $D'$ can be done immediately. It just eliminates the term $( D' + \sigma')^2$. The integral over $\sigma'$ and over the ghost field $C, \overline C$ gives 
\be
({\rm det}'\, \Delta^{0})^{{1\over 2}}
\ee
which cancels the overall factor in (\ref{intcal}). 

Before proceeding, we just note that due to gauge invariance we can 
diagonalize $\sigma_0$ so that it takes values in the Cartan subalgebra. This introduces the usual Vandermonde factor in the integral over $\sigma_0$, namely
\be
\label{genvander}
\prod_{\alpha>0}\left (\alpha(\sigma_0)\right)^2,
\ee
where $\alpha$ denote the roots of the Lie algebra ${\bf g}$, and $\alpha>0$ are the positive roots. 
Using the Cartan decomposition of ${\bf g}$, we can write $A_\mu$ as
\be
\label{adec}
A_\mu = \sum_\alpha A_\mu^\alpha X_\alpha + h_\mu
\ee
In this equation, $X_\alpha$ are representatives of the root spaces of $G$, normalized as
\be
\tr (X_\alpha X_\beta)=\delta_{\alpha+\beta},
\ee
where $\delta_{\alpha+\beta}$ is one if $\alpha+\beta=0$, and zero otherwise. In (\ref{adec}), $h_\mu$ is the component of $A_\mu$ along the Cartan subalgebra. Notice that this part of $A_\mu$ will only contribute a $\sigma_0$-independent factor to the one-loop determinant, so we will ignore it. We have
\be
[\sigma_0,A_\mu] = \sum_\alpha \alpha(\sigma_0) A_\mu^\alpha X_\alpha
\ee
and a similar equation for $\lambda$.  Plugging this into the action, we can now write it in terms of ordinary (as opposed to matrix valued) vectors and spinors
\begin{equation}
{1\over 2} \int \sqrt{g} \rd^3 x \, \sum_\alpha \left( g^{\mu \nu} A_\mu^{-\alpha} \left( - \Delta + \alpha(\sigma_0)^2 \right) A_{\nu}^{\alpha} + \bar \lambda^{-\alpha} \left( \ri \gamma^\mu \nabla_\mu  +\ri \alpha(\sigma_0) - \frac{1}{2} \right) \lambda^\alpha \right).
\end{equation}
We now have to calculate 
the determinants of the above operators. The integration over the fluctuations of the gauge field is restricted, as in the Chern--Simons case, to the 
vector spherical harmonics. Using the results (\ref{vsheigen}), (\ref{vshdeg}), we find that the bosonic part of the determinant is:
\begin{equation}
{\rm det}({\text{bosons}})= \prod_{\alpha}\prod_{n=1}^\infty \left( (n+1)^2 + \alpha(\sigma_0)^2 \right)^{2n(n+2)}.
\end{equation}
For the gaugino, we can use (\ref{spinoreigen}) to write the fermion determinant as:
\begin{equation}
{\rm det}({\text{fermions}})= \prod_{\alpha} \prod_{n=1}^\infty \bigg( (n + \ri {\alpha(\sigma_0)})(- n - 1 + \ri {\alpha(\sigma_0)}) \bigg)^{n(n+1)},
\end{equation}
and the quotient gives
\begin{equation}
\begin{array}{ll}
Z_{\text{1-loop}}^{\text{gauge}}[\sigma_0] &\displaystyle = \prod_{\alpha}\prod_{n=1}^\infty \frac{( n + \ri {\alpha(\sigma_0)})^{n(n+1)} ( - n - 1 + \ri {\alpha(\sigma_0)} )^{n(n+1)} }{ ((n+1)^2 + {{\alpha(\sigma_0)}}^2 )^{n(n+2)} } \\ \\
&\displaystyle = \prod_{\alpha}\prod_{n=1}^\infty \frac{( n + \ri {\alpha(\sigma_0)})^{n(n+1)} ( - n - 1 + \ri {\alpha(\sigma_0)} )^{n(n+1)} }{(n + \ri {\alpha(\sigma_0)})^{(n-1)(n+1)}(n+1 - i {\alpha(\sigma_0)})^{n(n+2)} },
\end{array}
\end{equation}
up to a $\sigma_0$-independent sign. We see there is partial cancellation between the numerator and the denominator, and this becomes:
\begin{equation}
\ba
Z_{\text{1-loop}}^{\text{gauge}}[\sigma_0] &= \prod_{\alpha}\prod_{n=1}^\infty \frac{( n + \ri {\alpha(\sigma_0)} )^{n+1} }{ (n - \ri {\alpha(\sigma_0)})^{n-1} }=\prod_{\alpha>0}\prod_{n=1}^\infty \frac{( n^2 + {\alpha(\sigma_0)}^2 )^{n+1} }{ (n^2 + {\alpha(\sigma_0)}^2)^{n-1} }\\
&= \prod_{\alpha>0}\prod_{n=1}^\infty ( n^2 + {\alpha(\sigma_0)}^2 )^2,
\ea
\end{equation}
where we used the fact that the roots split into positive roots $\alpha>0$ and negative roots $-\alpha$, $\alpha>0$. We finally obtain
\begin{equation}
Z_{\text{1-loop}}^{\text{gauge}}[\sigma_0] =  \left( \prod_{n=1}^\infty n^4 \right)  \prod_{\alpha>0} \prod_{n=1}^\infty \left( 1 + \frac{{\alpha(\sigma_0)}^2}{n^2} \right)^2.
\end{equation}
We can regularize this infinite product with the zeta function. This will lead to a finite, numerical result for the infinite product
\be
\prod_{n=1}^\infty n^4.
\ee
On the other hand, we can use the well-known formula
\be
{\sinh (\pi z) \over \pi z} =\prod_{n=1}\left( 1+{z^2 \over n^2} \right) 
\ee
to write
\begin{equation}
Z_{\text{1-loop}}^{\text{gauge}}[\sigma_0] \propto \prod_{\alpha>0} \left(\frac{ \sinh( \pi {\alpha(\sigma_0)})}{\pi {\alpha(\sigma_0)}} \right)^2,
\end{equation}
where the proportionality factor is independent of $\sigma_0$. We conclude that the localization of the vector multiplets leads to a total contribution 
to the partition function 
\be
\label{vectormm}
\int \rd \mu \, \prod_{\alpha>0} \left(2 \sinh\left(\alpha\left({\mu \over 2}\right)\right) \right)^2 \, \re^{-{1\over 2 g_s} \tr(\mu^2)}
\ee
where we defined the convenient coupling, 
\be
\label{convcoup}
g_s={2 \pi \ri \over k}
\ee
and we wrote 
\be
\label{mudef}
\sigma_0={\mu\over 2\pi},
\ee
where $\mu$ takes values in the Cartan subalgebra. 

\subsection{Localization in Chern--Simons--matter theories: matter sector}

Let us now consider the matter sector. We will follow the computation in \cite{hama}, which simplifies a little bit 
the original computation in \cite{kapustin}. As shown in (\ref{matterder}), the matter Lagrangian is in itself a total superderivative, so we can introduce a coupling $t$ in the form
\be
-t S_{\text{matter}}.
\ee
By the by now familiar localization argument, the partition function is independent of $t$, as long as $t>0$, and we can compute it for $t=1$ (which is the original case) or for 
$t\rightarrow \infty$. We can also restrict this Lagrangian to the localization locus of the gauge sector. The matter kinetic terms are then 
\be
\ba
 {\cal L}_\phi =&
 g^{\mu\nu}\partial_\mu\bar\phi\partial_\nu\phi +\bar\phi\sigma_0^2\phi
 +{2\ri(\Delta-1) \over r} \bar\phi\sigma_0 \phi
 +{\Delta (2-\Delta) \over r^2}\bar\phi\phi,
 \nonumber \\
 {\cal L}_\psi =&
 -\ri\bar\psi\gamma^\mu\partial_\mu\psi
 +\ri\bar\psi\sigma_0\psi
 -{\Delta-2 \over r}\bar\psi\psi.
\ea
\ee
The real part of the bosonic Lagrangian is positive definite, and it is minimized (and equal to zero) when 
\be
\phi=0. 
\ee
Like before, in the $t \rightarrow \infty$ limit, only quadratic terms in the matter fields contribute to the localization computation. In particular, there is no contribution from the superpotential terms involving the matter multiplets, like (\ref{superpot}). After using (\ref{lapL}) and (\ref{finaldirac}), we find that the operators governing the quadratic fluctuations around this fixed point are given by the operators
\begin{eqnarray}
 \CO_\phi &=&
 \frac1{r^2}\left\{ 4 {\bf L}^2-(\Delta-\ri r \sigma_0)(\Delta-2-\ri r \sigma_0) \right\},
 \nonumber \\
 \CO_\psi &=& \frac1r \left\{
  4 {\bf L} \cdot {\bf S} +\ri r\sigma_0+2-\Delta \right\}.
\end{eqnarray}
Their eigenvalues are, for the bosons, 
\be
  \lambda_\phi (n) =
 r^{-2}(n+2+\ri r \sigma_0-\Delta)(n-\ri r \sigma_0+\Delta), \qquad n=0, 1, 2, \cdots, 
\ee
with multiplicity $(n+1)^2$, and for the fermions
\be
 \lambda_\psi (n) =
 r^{-1}(n+1+\ri r\sigma_0-\Delta),\quad
 r^{-1}(-n+\ri r \sigma_0-\Delta), \qquad n=1, 2, \cdots, 
\ee
with multiplicity $n(n+1)$. We finally obtain, after setting $r=1$, 
\be
 {|\text{det}\Delta_\psi|  \over \text{det}\Delta_\phi} =\prod_{m>0} {(m+1+\ri r \sigma_0 -\Delta)^{m(m+1)} (m-\ri r \sigma_0 +\Delta)^{m(m+1)}  \over 
 (m+1+\ri r \sigma_0 -\Delta)^{m^2} (m-1-\ri r \sigma_0 +\Delta)^{m^2}},
 \ee
 and we conclude
 \be
 \label{onelmatter}
Z_{\text{1-loop}}^{\text{matter}}[\sigma_0] =\prod_{m>0}\Big(\frac{m+1-\Delta+\ri r \sigma_0}{m-1+\Delta-\ri r \sigma_0}\Big)^m.
 \ee
 As a check, notice that, when $\Delta=1/2$ and $\sigma_0=0$, we recover the quotient of determinants (\ref{freequot}) of the free theory. The quantity (\ref{onelmatter}) can be easily 
 computed by using $\zeta$-function regularization \cite{jafferis, hama}. Denote
 \be
 z = 1 - \Delta + \ri r \sigma_0
\ee
and
\be 
 \ell(z) = \log
Z_{\text{1-loop}}^{\text{matter}}[\sigma_0] . 
\ee
We can regularize this quantity as
\be
 \ell(z) = -\frac{\partial}{\partial s}\bigg{|}_{s=0}\,  \sum_{m=1}^\infty \left(
\frac{m}{(m+z)^s } - \frac{m}{(m-z)^s}\right). 
\ee
On the other hand, 
\be
\sum_{m=1}^\infty \left(
\frac{m}{(m+z)^s } - \frac{m}{(m-z)^s}\right)=\zeta_H(s-1,z) -
z\zeta_H(s,z)
-\zeta_H(s-1,-z)-z\zeta_H(s, -z),
\ee
 where 
 \be
 \zeta_H(s,z)=\sum_{m=0}^{\infty} {1\over (m+z)^s}
 \ee
is the Hurwitz zeta function. Using standard properties of this function (see for example \cite{noc}), one finally finds the regularized result
 \begin{equation} \ell(z) = - z
\log\left(1-\re^{2\pi \ri z}\right) + \frac{\ri}{2} \left(\pi z^2 +
\frac{1}{\pi} \textrm{Li}_2(\re^{2\pi \ri z}) \right) - \frac{\ri
\pi}{12}. 
\ee
As a check of this, notice that 
\be
\ell\left({1\over 2}\right)=-{1\over 2} \log 2
\ee
in agreement with (\ref{freelogz}). 

There is an important property of $\ell(z)$, namely when $\Delta=1/2$ (canonical dimension) one has
\be
\label{realpart}
{1\over 2}\left(\ell(z)+\ell(z^*)\right)=-{1\over 2} \log \left(2 \cosh (\pi r \sigma_0) \right). 
\ee
To prove this, we write 
\be
z={1\over 2}+\ri \theta, 
\ee
and we compute
\be
{1\over 2}\left(\ell(z)+\ell(z^*)\right)=-{1\over 2} \log \left( 2 \cosh (\pi \theta)\right) +{1\over 2} \pi \ri \theta^2+ {\ri \pi  \over 24} +{\ri \over 4 \pi} \left( {\rm Li}_2(-\re^{-2\pi \theta}) + {\rm Li}_2(-\re^{2\pi \theta})\right). 
\ee
After using the following property of the dilogarithm, 
\be
{\rm Li}_2(-x)+{\rm Li}_2(-x^{-1})=-{\pi^2 \over 6} -{1\over 2} \left( \log(x)\right)^2, 
\ee
we obtain (\ref{realpart}). 

When the matter is in a self-conjugate representation of the gauge group, the set of eigenvalues of $\sigma_0$ is invariant under change of sign, 
therefore we can calculate the contribution of such a multiplet by using (\ref{realpart}). We conclude that for such a matter multiplet,
\be
\label{matterloop}
Z_{\text{1-loop}}^{\text{matter}}[\mu] =\prod_{\Lambda} \left( 2 \cosh  {\Lambda (\mu) \over 2}  \right)^{-1/2},
\ee
where we set $r=1$ and we used the variable $\mu$ in the Cartan defined in (\ref{mudef}). The product is over the weights $\Lambda$ of the representation of the 
matter multiplet.  
For general representations and anomalous dimensions, one has to use the more complicated result above for $\ell(z)$.

\subsection{The Chern--Simons matrix model}

As a first application of the results of localization, let us consider pure supersymmetric Chern--Simons theory, defined by the action (\ref{susycs}). If we don't add matter to the theory, the fields $D$, $\sigma$ and $\lambda, \bar \lambda$ are auxiliary and they can be integrated out. In other words, supersymmetric Chern--Simons theory on 
$\IS^3$ should be equivalent to pure (bosonic) Chern--Simons theory. There is however an important difference: in super-Chern--Simons theories with at least 
$\CN=2$ supersymmetry, 
there is no renormalization of the coupling $k$ due to the extended supersymmetry \cite{kaolee}. The localization argument developed above says that the partition function 
of Chern--Simons theory on $\IS^3$ with gauge group $G$ should be proportional to the matrix model (\ref{vectormm}):
\be
\label{csprop}
Z_{\rm CS} (\IS^3) \propto \int \rd \mu \, \prod_{\alpha>0} \left(2 \sinh {\alpha\cdot \mu \over 2}  \right)^2 \, \re^{-{1\over 2 g_s} \mu^2},
\ee
where we regard $\mu$ as a weight and we use the standard Cartan--Killing inner product in the space of weights. For example, in the case of $G=U(N)$, 
if we write $\mu$ and the positive roots in terms of 
an orthonormal basis $e_i$ of the weight lattice, 
\be
\mu=\sum_{i=1}^N \mu_i e_i, \qquad \alpha_{ij} =e_i-e_j, \quad i<j,
\ee
we find
\be
\label{zcsprop}
Z_{\rm CS}(\IS^3) \propto \int \prod_{i=1}^N \rd \mu_i \, \prod_{i<j} \left( 2 \, \sinh {\mu_i -\mu_j\over 2} \right)^2 \, \re^{-{1\over 2 g_s} \sum_{i=1}^N \mu_i^2}.
\ee
The proportionality constant appearing in (\ref{csprop}) should be independent of the coupling constant $k$, and it is only a function of $N$. 
The matrix model (\ref{vectormm}) is a ``deformation" of the standard Gaussian matrix model. It has a Gaussian weight, but instead of displaying the standard 
Vandermonde interaction between eigenvalues (\ref{genvander}) it has a ``trigonometric" deformation involving the $\sinh$.  This interaction reduces to the standard one for small 
$\alpha\cdot \mu$, which corresponds in the $U(N)$ case to a small separation between eigenvalues. 

The matrix model (\ref{vectormm}), with a sinh kernel, was first introduced in \cite{mmcs}. 
It was later rederived using geometric localization techniques in \cite{bw}, and abelianization techniques 
  in \cite{bttwo}. As we have seen following \cite{kapustin}, it can be derived in an elegant and simple way by using supersymmetric localization. Actually, the matrix integral 
  appearing in the r.h.s. of (\ref{csprop}) can be calculated in a very simple way by using Weyl's denominator formula, as pointed out in for example \cite{akmv}. 
 This formula reads, 
 \be
 \sum_{w \in {\cal W}} \epsilon (w) \re^{w(\rho)\cdot u} =\prod_{\alpha>0}
2 \sinh {\alpha\cdot u \over 2}.
\label{wdf}
\end{equation}
In this formula, $\CW$ is the Weyl group of $G$, $\epsilon(w)$ is the signature of $w$, and $\rho$ is the Weyl vector, given by the sum of the fundamental weights. Using this formula, the matrix integral reduces to a sum of Gaussian integrals which can be 
calculated immediately, and one finds
\be
\left( {\rm det}(C)\right)^{1/2} \left(2 \pi g_s\right)^{r/2} \left|\CW \right| \re^{g_s \rho^2} \sum_{w\in \CW} \epsilon(w)  \re^{g_s \rho\cdot w(\rho)}, 
\ee
where $C$ is the inverse matrix of the inner product in the space of weights (for simply connected $G$, this is the Cartan matrix), and $r$ is the rank of $G$. Using again Weyl's denominator formula we find, 
\be
\sum_{w\in \CW} \epsilon(w)  \re^{g_s \rho\cdot w(\rho)}=\ri^{|\Delta_+|} \prod_{\alpha>0} 2 \sin\left( {\pi \alpha\cdot \rho\over k} \right)
\ee
where $|\Delta_+|$ is the number of positive roots of $G$. The matrix integral then gives, 
\be
\label{mires}
\left( {\rm det}(C)\right)^{1/2} \left(2 \pi\right)^{r} \left|\CW \right| {\ri ^{|\Delta_+|-r/2} \over k^{r/2}}  \re^{ {\pi \ri \over 6 k } d_G y } \prod_{\alpha>0} 2 \sin\left( {\pi \alpha\cdot \rho\over k} \right)
\ee
where we have used Freudenthal--de Vries formula
\be
\rho^2={1\over 12} d_G y. 
\ee
The result (\ref{mires}) 
is indeed proportional to the partition function of Chern--Simons theory on $\IS^3$, and we can use the result to fix the normalization, $N$-dependent factor in the 
matrix integral. Let us particularize for $G=U(N)$. In this case, one has to take as $C$ the identity matrix, and the value of $\rho^2$ is the same as for $SU(N)$. Then, the matrix integral is 
\be
\ri^{-{N^2\over 2}}(2\pi)^N N! \,  \re^{{\pi \ri \over 6 k} N(N^2-1)} k^{-N/2}  \prod_{j=1}^{N} \left[ 2 \sin \left( \pi j \over k\right) \right]^{N-j},
\ee
which is indeed proportional to (\ref{csun}) (after changing $k\rightarrow k-N$), up to an overall factor
\be
\ri^{-{N^2\over 2}} (2\pi)^N N! \,  \re^{{\pi \ri \over 6 k} N(N^2-1)}. 
\ee
The phase appearing here depends on $k$, and it has the right dependence on $k$, $N$ to be understood as a change of framing of $\IS^3$ in the result (\ref{csun}). We can now use 
the above result to fix the normalization in the matrix model describing supersymmetric Chern--Simons theory, and we find
\be
\label{normCS}
Z_{\rm CS} \left(\IS^3\right)={\ri^{-{N^2\over 2}} \over N!} \int \prod_{i=1}^N {\rd \mu_i \over 2 \pi} \, \prod_{i<j} \left( 2 \, \sinh {\mu_i -\mu_j\over 2} \right)^2 \, \re^{-{1\over 2 g_s} \sum_{i=1}^N \mu_i^2}.
\ee
We will refer to this model as the Chern--Simons matrix model. Later on we will study its large $N$ limit. 

\subsection{The ABJM matrix model}

Let us now consider the matrix model calculating the partition function on $\IS^3$ of ABJM theory, or rather its generalization \cite{abj} to the gauge group $U(N_1)\times U(N_2)$. The contribution of the vector multiplets gives in the integrand 
\be
 \prod_{1\le i<j\le N_1} \left( 2 \, \sinh {\mu_i -\mu_j\over 2} \right)^2 \, \re^{-{1\over 2 g_s} \sum_{i=1}^{N_1} \mu_i^2} 
  \prod_{1\le a<b \le N_2} \left( 2 \, \sinh {\nu_a-\nu_b\over 2} \right)^2 \, \re^{{1\over 2 g_s} \sum_{a=1}^{N_2} \nu_a^2},
  \ee
  where the opposite signs in the Gaussian exponents are due to the opposite signs in the levels. 
  Since there are four hypermultiplets in the bifundamental representation, we have an extra factor due to (\ref{matterloop}), 
  \be
  \prod_{i=1}^{N_1} \prod_{a=1}^{N_2} \left( 2 \, \cosh {\mu_i -\nu_a \over 2} \right)^{-2}.
  \ee
The normalization of the matrix model can be fixed by using the normalization for the Chern--Simons matrix model (\ref{normCS}), and by comparing to the perturbative one-loop 
result. In this way we find, 
\be
\label{abjmmm} 
\ba
Z_{\rm ABJM}(\IS^3)=& {\ri^{-{1\over 2} \left(N_1^2-N_2^2 \right) } \over N_1! N_2!} \int  \prod_{i=1}^{N_1}{ \rd \mu_i  \over 2\pi} \prod_{a=1}^{N_2} {\rd \nu_a \over 2\pi}  \prod_{1\le i<j\le N_1} \left( 2 \sinh  \left( {\mu_i -\mu_j \over 2}\right) \right)^2 \\
\times &  \prod_{1\le a<b \le N_2} \left( 2 \sinh  \left( {\nu_a-\nu_b \over 2}\right) \right)^2\prod_{i,a} \left( 2  \cosh  \left( {\mu_i -\nu_a \over 2}\right) \right)^{-2}\,  \re^{-{1\over 2g_s}\left(  \sum_i \mu_i^2 -\sum_a \nu_a^2\right)}.
\ea
\ee
This model is closely related to a matrix model that computes the partition function of Chern--Simons theory on lens spaces $L(p,1)$, in particular to the model with $p=2$. These models 
were introduced in \cite{mmcs}, and the case $p=2$ was extensively studied in \cite{akmv}. The matrix integral for $p=2$ is given by, 
\be
\label{lensmm}
\ba
Z_{\rm CS}(L(2,1))=& {\ri^{-{1\over 2} \left(N_1^2+ N_2^2 \right) } \over N_1! N_2!} \int  \prod_{i=1}^{N_1}{ \rd \mu_i  \over 2\pi} \prod_{a=1}^{N_2} {\rd \nu_a \over 2\pi}  \prod_{1\le i<j\le N_1} \left( 2 \sinh  \left( {\mu_i -\mu_j \over 2}\right) \right)^2 \\
\times &  \prod_{1\le a<b \le N_2} \left( 2 \sinh  \left( {\nu_a -\nu_b \over 2}\right) \right)^2\prod_{i,a} \left( 2  \cosh  \left( {\mu_i -\nu_a \over 2}\right) \right)^{2}\,  \re^{-{1\over 2g_s}\left(  \sum_i \mu_i^2 +\sum_a \nu_a^2\right)}.
\ea
\ee
We will refer to this matrix model as the lens space matrix model. It turns out \cite{mp} 
that the partition functions (\ref{abjmmm}) and (\ref{lensmm}) are related, order by order in perturbation theory, 
by the analytic continuation
\be
\label{changen2}
N_2 \rightarrow -N_2. 
\ee
We will show this explicitly in the analysis of the planar limit (indeed, we will study the planar limit of (\ref{lensmm}) rather than (\ref{abjmmm})). It can be shown 
that (\ref{abjmmm}) is the super-matrix model version of (\ref{lensmm}) \cite{mp,dt}, and this leads to the relation (\ref{changen2}) between both matrix models.

In order to derive the interpolating function for (\ref{introgoal}), we just have to compute the planar free energy of the matrix model (\ref{abjmmm}) for $N_1=N_2=N$, and 
for any value of the 't Hooft coupling $\lambda$. The calculation of exact planar free energies of matrix models is a classical problem which was first solved in \cite{bizp} for a 
simple class of matrix models. We will now review the standard techniques to do that, which we will then generalize to the matrix models appearing in ABJM theories.

\sectiono{Matrix models at large $N$}

In this section we will focus on 
conventional matrix models, and in the next section we will use the same techniques (and the same formulae) to analyze 
the matrix models appearing in supersymmetric Chern--Simons--matter theories. A more detailed treatment of matrix models in the 
large $N$ expansion, as well as a complete list of references, can be found in \cite{dfgzj,mmhouches, eo}.

\subsection{Saddle-point equations and one-cut solution}

Let us consider the matrix model partition function
\be
\label{parteigen}
Z={1 \over N!} {1 \over (2\pi)^N}\int \prod_{i=1}^N \rd\lambda_i \, \Delta^2(\lambda)
\re^{-{1\over  g_s} \sum_{i=1}^N V(\lambda_i)}.
\ee
$V(\lambda)$, called the {\it potential} of the matrix model, will be taken to be a polynomial
\be
V(\lambda)={1 \over 2}\lambda^2+  \sum_{p\ge 3} {g_p\over p} \lambda^p
\ee
where the $g_p$ are coupling constants of the model. In (\ref{parteigen}), 
\be
\Delta^2(\lambda)= \prod_{i<j} (\lambda_i-\lambda_j)^2
\ee
is the Vandermonde determinant (\ref{genvander}) for the group $U(N)$. The integral (\ref{parteigen}) is typically obtained as a reduction to eigenvalues of integrals over the space of $N\times N$ Hermitian matrices, see \cite{dfgzj,mmhouches} for more details. 
We want to study $Z$ in the so-called 't Hooft limit, in which 
\be
g_s \rightarrow 0, \quad N \rightarrow \infty,
\ee
but the `t Hooft parameter of the matrix model 
\be
t=g_s N
\ee
is fixed. In particular, we want to study the leading asymptotic behavior of the free energy
\be
F=\log Z
\ee
in this limit. Let us 
write the partition function (\ref{parteigen}) 
as follows:
\be
\label{inteff}
Z={1 \over N!} \int \prod_{i=1}^N {\rd\lambda_i \over 2 \pi}\re^{g_s^{-2} S_{\rm eff} (\lambda)}
\ee
where the effective action is given by
\be
S_{\rm eff}(\lambda)= -{t \over N} \sum_{i=1}^N V(\lambda_i) +
{2 t^2 \over N^2}\sum_{i< j}\log |\lambda_i -\lambda_j|.
\ee
We can now regard $g^2_s$ as a sort of $\hbar$, in such a way
that, as $g_s \rightarrow 0$ with $t$ fixed, the
integral (\ref{inteff}) will be dominated by a saddle-point configuration that
extremizes the effective action. Notice that, since a sum over $N$ eigenvalues is roughly of order $N$, in the 't Hooft limit the effective
action is of order ${\cal O}(1)$, and the free energy scales as 
\be
F(g_s, t) \approx g_s^{-2} F_0(t).
\ee
$F_0(t)$ is called the genus zero, or planar, free energy of the matrix model, and it is obtained by evaluating the effective action at the 
saddle point. This dominant contribution is just the first term in an asymptotic expansion around $g_s=0$, 
\be
F=\sum_{g=0}^{\infty} F_0(t) g_s^{2g-2}. 
\ee
In order to obtain the saddle-point equation, we just vary $S_{\rm eff} (\lambda)$ w.r.t. the eigenvalue
$\lambda_i$. We obtain the equation
\be
\label{sadd}
{1 \over 2 t} V'(\lambda_i) ={1\over N} \sum_{j \not= i} {1 \over \lambda_i - \lambda_j}, \quad i=1, \cdots, N.
\ee
This equation can be given a simple interpretation: we can regard the eigenvalues as
coordinates of a system of $N$ classical particles moving on the real line. 
(\ref{sadd}) says that these particles are subject to an effective potential
\be
\label{veffdiscrete}
V_{\rm eff}(\lambda_i)= V(\lambda_i) -{2 t \over N}\sum_{j \not=i} \log |\lambda_i -\lambda_j|
\ee
which involves a logarithmic Coulomb
repulsion between eigenvalues. For small 't Hooft parameter, the potential term dominates over
the Coulomb repulsion, and the particles tend to be at a critical point $x_{*}$ of the
potential: $V'(x_{*})=0$. As $t$ grows, the logarithmic Coulomb interaction will force the eigenvalues to repel each
other and to spread out away from the critical point. 

To encode this information about the 
equlibrium distribution of the particles, it is convenient to define an {\it eigenvalue
distribution} (for finite $N$) as
\be
\rho (\lambda)={1 \over N} \sum_{i=1}^N\langle  \delta (\lambda- \lambda_i) \rangle,
\label{rhofinite}
\ee
where the $\lambda_i$ solve (\ref{sadd}) in the saddle-point approximation.
In the large $N$ limit, it is reasonable to expect that this 
distribution becomes a continuous distribution $\rho_0(\lambda)$. As we will see in a moment, this distribution 
has a compact support. The simplest case occurs when $\rho_0(\lambda)$ vanishes outside a connected interval 
${\cal C}=[a,b]$. This is the so-called {\it one-cut solution}. Based on the considerations above, 
we expect $\cal C$ to be centered around a critical point $x_*$ of the potential. In particular, as $t \rightarrow 0$, the interval $\cal C$ should collapse
to the point $x_*$. 

We can now write the saddle-point equation in terms of continuum quantities, by using the rule
\be
{1\over N}
\sum_{i=1}^N f(\lambda_i)\rightarrow \int_{\cal C} f(\lambda)\rho_0(\lambda) \rd\lambda.
\ee
Notice that the distribution of
eigenvalues $\rho_0(\lambda)$ satisfies the normalization condition
\be
\label{normrho}
\int_{\cal C} \rho_0(\lambda) d\lambda =1.
\ee
The equation (\ref{sadd}) then becomes
\be
\label{saddrho}
{1 \over 2t}V' (\lambda)= {\rm P} \int_{\CC} {\rho_0(\lambda') \rd \lambda' \over \lambda -\lambda'}
\ee
where ${\rm P}$ denotes the principal value of the integral. The above equation
is an integral equation that allows one in principle to compute $\rho_0(\lambda)$, given
the potential $V(\lambda)$, as a function of the 't Hooft parameter $t$ and the
coupling constants. Once $\rho_0(\lambda)$ is known, one can easily compute $F_0(t)$, since the effective action in the continuum limit is a functional of $\rho_0$:
\be
\label{rhofunct}
S_{\rm eff}(\rho_0)=-t  \int_{\cal C} \rd\lambda \, \rho_0 (\lambda) V(\lambda) +
t^2 \int_{\cal C \times \cal C} \rd \lambda \, \rd\lambda'  \, \rho_0(\lambda)\rho_0(\lambda')\log |\lambda -\lambda'|.
\ee
The planar free energy is given by
\be
\label{plant}
F_0(t)= S_{\rm eff}(\rho_0).
\ee

We can obtain (\ref{sadd}) directly in the continuum formulation by computing the extremum of the functional 
\be
S(\rho_0, \Gamma) = S_{\rm eff}(\rho_0) + \Gamma \biggl(t \int_{\CC} \rd\lambda \, \rho_0 (\lambda)  -t \biggr)
\ee
with respect to $\rho_0$. Here, $\Gamma$ is a Lagrange multiplier that imposes the normalization condition of the density of eigenvalues (times $t$). This leads to
\be
V(\lambda)=2t  \int {\rm d} \lambda' \, \rho_0(\lambda') \log |\lambda -\lambda'| + \Gamma, 
\label{intsaddle}
\ee
which can be also obtained by integrating (\ref{saddrho}) with respect to $\lambda$. It is convenient to introduce the 
{\it effective potential on an eigenvalue} as 
\be
\label{veff}
V_{\rm eff}(\lambda) = V(\lambda)-2t \int {\rm d}\, \lambda' \rho_0(\lambda') \log |\lambda -\lambda'|. 
\ee
This is of course the continuum counterpart of (\ref{veffdiscrete}). 
In terms of this quantity, the saddle--point equation (\ref{intsaddle}) says that the effective potential is {\it constant} on the 
interval $\CC$:
\be
V_{\rm eff}(\lambda) =\Gamma, \qquad \lambda \in \CC. 
\ee
The Lagrange multiplier $\Gamma$ appears 
in this way as an integration constant that only depends on $t$ and the coupling constants. As in any other Lagrange minimization problem, the 
multiplier is obtained by taking minus the derivative of the target function w.r.t. the constraint, which in this case is $t$. We then find the very useful equation
\be
\label{derf0}
\partial_t F_0(t)=-\Gamma=-V_{\rm eff}(b). 
\ee
where $b$ is the endpoint of the cut $\CC$.

The density of eigenvalues is obtained as a solution to the saddle-point equation (\ref{saddrho}).
This equation is a singular
integral equation which has been studied in detail in other contexts of
physics (see, for example, \cite{georgia}). The way to solve it
is to introduce an auxiliary function called the {\it resolvent}. The
resolvent is defined, at finite $N$, as
\be
\label{resolvent}
\omega(p)={ 1\over N} \left \langle\sum_{i=1}^N {1 \over p -\lambda_i} \right\rangle,
\ee
and we will denote its large $N$ limit by $\omega_0(p)$, which is also called the genus zero resolvent. 
This can be written in terms of the eigenvalue density as
\be
\label{zeroresint}
\omega_0(p) =\int d \lambda {\rho_0 (\lambda)\over p -\lambda}.
\ee
The genus zero resolvent (\ref{zeroresint}) has three important properties. First of all, due to the 
normalization property of the eigenvalue distribution (\ref{normrho}), it
has the asymptotic behavior
\be
\label{asymres}
\omega_0(p) \sim {1 \over p}, \qquad p\rightarrow \infty.
\ee
Second, as a function of $p$ it is an analytic function
on the whole complex plane except on the interval $\cal C$, where it has a discontinuity as one crosses the
interval $\cal C$. This discontinuity can be computed by standard contour deformations. We have
\be
\label{omegaup}
\omega_0(p+\ri \epsilon)= \int_{\IR} \rd \lambda {\rho_0 (\lambda)\over p+ \ri \epsilon-\lambda}=
\int_{\IR-\ri \epsilon} \rd \lambda {\rho_0 (\lambda)\over p-\lambda}= {\rm P}\, \int  \rd \lambda {\rho_0 (\lambda)\over p-\lambda} +\int_{C_{\epsilon}} \rd \lambda {\rho_0 (\lambda)\over p-\lambda}, 
\ee
where $C_{\epsilon}$ is a contour around $\lambda=p$ in the lower half plane, and oriented counterclockwise. The last integral can be evaluated as a residue, and we finally obtain, 
\be
\omega_0(p+\ri \epsilon)= {\rm P}\int  \rd \lambda {\rho_0 (\lambda)\over p-\lambda} - \pi \ri \rho_0(p). 
\ee
Similarly 
\be
\label{omegadown}
\omega_0(p-\ri \epsilon)= 
\int_{\IR+\ri \epsilon} \rd \lambda {\rho_0 (\lambda)\over p-\lambda}= {\rm P}\, \int  \rd \lambda {\rho_0 (\lambda)\over p-\lambda} +\pi \ri \rho_0(p).
\ee
One then finds the key equation
\be
\label{rhow}
\rho_0(\lambda) =-{1 \over 2 \pi \ri} \bigl(\omega_0(\lambda+ \ri\epsilon) -\omega_0 (\lambda-\ri \epsilon)\bigr).
\ee

From these equations we deduce that, if the resolvent at genus zero is known, the planar eigenvalue distribution follows from
(\ref{rhow}), and one can compute the planar free energy. On the other hand, by using again (\ref{omegaup}) and (\ref{omegadown}) we can compute
\be
\omega_0(p+ \ri\epsilon) +\omega_0 (p-\ri \epsilon) =2  {\rm P}\, \int  \rd \lambda {\rho_0 (\lambda)\over p-\lambda} 
\ee
and we then find
\be
\label{wdisco}
\omega_0(p+ \ri \epsilon) +\omega_0 (p-\ri \epsilon)={1 \over t} V'(p),\qquad p \in \CC, 
\ee
which determines the resolvent in terms of the potential. In this way we have reduced the 
original problem of computing $F_0(t)$ to the Riemann-Hilbert problem of computing $\omega_0(\lambda)$. 
In order to solve (\ref{wdisco}), we write it as a sum of an analytic or regular part $\omega_r (p)$, and a singular part $\omega_s(p)$, 
\be
\omega_0(p)=\omega_r (p) + \omega_s(p),
\ee
where
\be
\omega_r(p)={1\over 2t} V'(p). 
\ee
It follows that the singular part satisfies 
\be
\omega_s(p+ \ri \epsilon) +\omega_s (p-\ri \epsilon)=0,\qquad p \in \CC. 
\ee
This is automatically satisfied if $\omega_s(p)$ has a square-root branch cut across $\CC$, and we find
\be
\omega_s(p)=-{1\over 2t} M(p){\sqrt{(p-a)(p-b)}}, 
\ee
where $a,b$ are the endpoints of $\CC$, and $M(p)$ is a polynomial, which is fully determined by the 
asymptotic condition (\ref{asymres}). There is in fact a closed expression for the
planar resolvent in terms of a contour integral \cite{migdal} which reads
\be
\label{solwo}
\omega_0(p) ={1 \over 2t} \oint_{\cal C} {\rd z \over 2 \pi \ri} { V'(z) \over p-z} \biggl( { (p-a)(p-b)\over
(z -a) (z -b)}\biggr)^{1 /2},
\ee
where $\CC$ denotes now a contour encircling the interval. The r.h.s. of (\ref{solwo}) behaves
like $c + d/p+{\cal O}(1/p^2)$. Requiring the asymptotic behavior (\ref{asymres})
imposes $c=0$ and $d=1$, and this leads to
\be
\label{endpo}
\ba
\oint_{\cal C} {\rd z \over 2\pi \ri} {V'(z) \over {\sqrt {(z-a)(z-b)}}}&=0, \\
\oint_{\cal C}{\rd z \over 2\pi \ri} {z V'(z) \over {\sqrt {(z-a)(z-b)}}}&=2t.
\ea
\ee
These equations are enough to determine the endpoints of the cuts, $a$ and $b$, as 
functions of the 't Hooft coupling $t$ and the coupling constants of the model. Equivalently, after deforming the contour in (\ref{solwo}) 
to infinity, we pick a pole at $z=p$, which gives the regular piece, and we find 
the equation
\be
\label{wpot}
\omega_0(p)={1 \over 2t} V'(p) -{1 \over 2 t}  M(p) {\sqrt {(p-a)(p-b)}},
\ee
where 
\be
\label{momentf}
M(p)=\oint_{\infty} {\rd z \over 2 \pi i} {V'(z)\over z-p} {1 \over{\sqrt {(z-a)(z-b)}}}.
\ee
\begin{figure}
\begin{center}
\includegraphics[height=3.5cm]{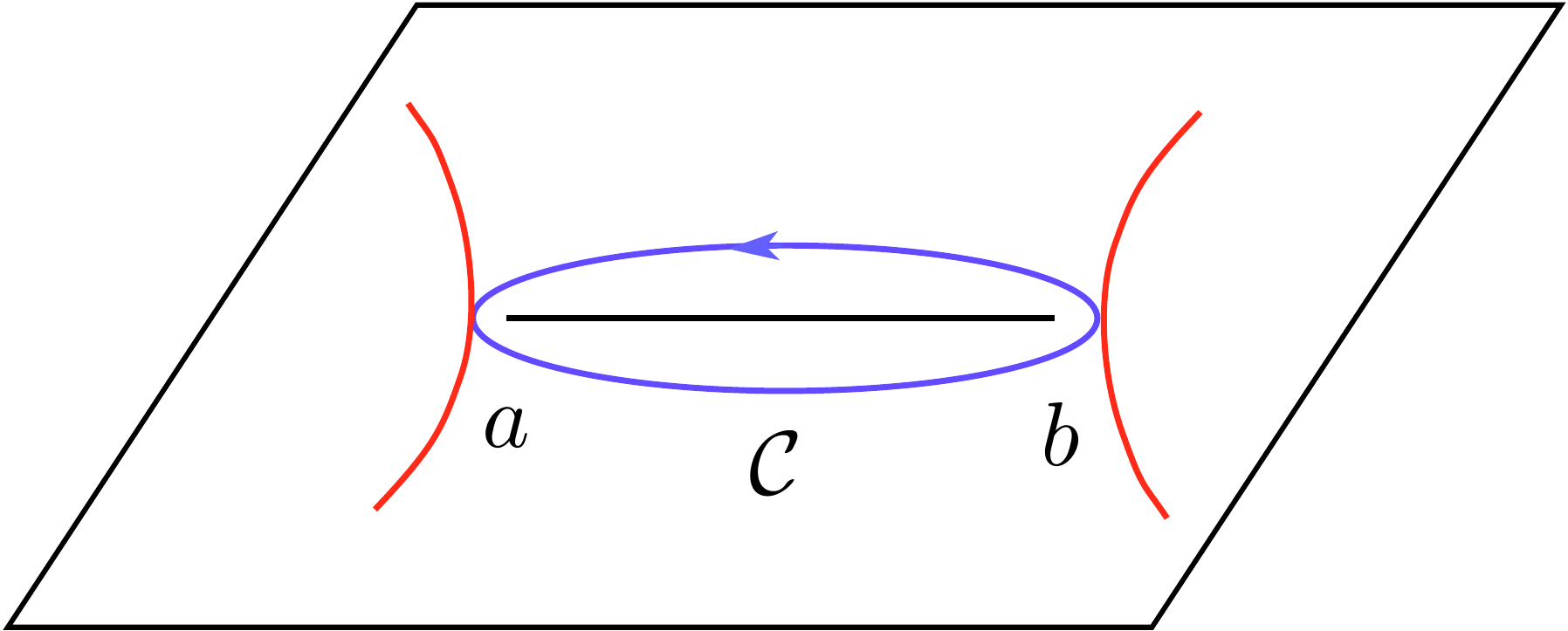}
\end{center}
\caption{The contour $\CC$ encircling the support of the density of eigenvalues, which can be regarded as a contour in the spectral 
curve $y(p)$.}
\label{onecut}
\end{figure}
A useful way to encode the solution to the matrix model is to define the {\it spectral curve} of the matrix model by
\be
\label{defymm}
y(p)= V'(p) -2t\, \omega_0(p)=M(p) {\sqrt {(p-a)(p-b)}}.
\ee
Notice that, up to a constant, 
\be
\label{ceffpot}
\int ^{\lambda}\rd p \, y(p)=V_{\rm eff}(\lambda).
\ee
If we regard $\omega_0(p)\rd p$ as a differential on the spectral curve, the 't Hooft parameter can be written as a contour integral
\be
\label{tper1}
t=\oint_{\CC}{\rd p \over 4 \pi \ri} 2t \omega_0(p).
\ee
This contour on the spectral curve (regarded as a complex curve) is represented in \figref{onecut}.

\begin{example} {\it The Gaussian matrix model}. 
Let us now apply these results to the simplest case, the Gaussian model with $V(z) =z^2/2$. 
We first look for the position of the endpoints from (\ref{endpo}). After deforming
the contour to infinity and changing $z\rightarrow 1/z$, the first equation
in (\ref{endpo}) becomes
\be
\oint_{0} {\rd z \over 2 \pi \ri} {1 \over z^2} {1 \over {\sqrt {(1-az)(1-bz)}}}=0,
\ee
where the contour is now around $z=0$. Therefore $a+b=0$, in accord with the symmetry of the potential.
Taking this into account, the second equation becomes:
\be
\oint_{0} {\rd z \over 2 \pi \ri} {1 \over z^3} {1 \over {\sqrt {1-a^2 z^2}}}=2t,\ee
and gives
\be
a=2 {\sqrt t}.
\ee
We see that the interval ${\cal C}=[-a,a]=[-2{\sqrt t} , 2{\sqrt t}  ]$ opens as the 't Hooft
parameter grows up, and as $t\rightarrow 0$ it collapses to the minimum of the potential 
at the origin, as expected. We immediately find from (\ref{wpot})
\be
\omega_0(p)={1\over 2 t} \Bigl(p -{\sqrt { p^2 -4t}}\Bigr),
\ee
and from the discontinuity equation we derive the density of eigenvalues
\be
\rho_0(\lambda)={1 \over 2 \pi t}{\sqrt { 4t -\lambda^2}}.
\ee
The graph of this function is a semicircle of radius $2{\sqrt t} $, and the above
eigenvalue distribution is the famous Wigner-Dyson semicircle law. Notice also that the equation 
(\ref{defymm}) is in this case
\be
y^2=p^2 - 4t.
\ee
This is the equation for a curve of genus zero, which resolves the singularity $y^2=p^2$. We then 
see that the opening of the cut as we turn on the 't Hooft parameter can be interpreted as 
a deformation of a geometric singularity. 
\end{example}

\subsection{Multi--cut solutions}

So far we have considered the so-called one-cut solution to the one-matrix model.
This is not, however, the most general solution, and we will now consider the so-called multi-cut solution, 
in the saddle-point approximation. Recall from our previous discussion that the cut appearing in the one-matrix model
was centered around a critical point of the potential. If the potential has many
critical points, one can have a saddle--point solution with various cuts, centered around different critical points. The most general
solution has then $n$ cuts (where $n$ is lower or equal than the number 
of critical points), and the support of the eigenvalue
distribution is a disjoint union of $n$ intervals
\be
{\cal C}=\bigcup_{i=1}^n \CC_i.
\ee
The total number of eigenvalues $N$ splits into $n$ integers $N_i$,
\be
N=N_1+\cdots + N_n, 
\ee
where $N_i$ is the number of eigenvalues in the interval $\CC_i$. We introduce the {\it filling fractions} 
\be
\label{fillingfr}
\epsilon_i={N_i \over N} = \int_{\CC_i} \rd \lambda\, \rho_0(\lambda), \qquad i=1, \cdots, n. 
\ee
Notice that 
\be
\label{epsconstraint}
\sum_{i=1}^n \epsilon_i=1. 
\ee
A closely related set of variables are the partial 't Hooft parameters 
\be
\label{pthooft}
t_i=t \epsilon_i=g_s N_i, \qquad i=1, \cdots, n. 
\ee
Notice that there are only $g=n-1$ independent filling fractions, but the partial 't Hooft parameters are all 
independent. 

The multi-cut solution is just a more general solution of the saddle-point equations that we derived above. It can be found by extremizing the functional 
(\ref{rhofunct}) with the condition that the partial 't Hooft parameters are {\it fixed}, 
\be
\label{tots}
S(\rho_0, \epsilon^I) =S_{\rm eff}(\rho_0) + \sum_{i=1}^n \Gamma_i \biggl( t \int_{\CC_i} \rd \lambda\, \rho_0(\lambda)-t_i \biggr), 
\ee
where $\Gamma_i$ are Lagrange multipliers. If we take the variation w.r.t. the density $\rho_0(\lambda)$ we find the equation
\be
\label{saddlemcut}
V(\lambda)=2t  \int_{\CC} {\rm d} \lambda' \rho_0(\lambda') \log |\lambda -\lambda'| + \Gamma_i, \qquad \lambda \in \CC_i 
\ee
which can be rewritten as 
\be
\label{cpot}
V_{\rm eff}(\lambda)=\Gamma_i,  \qquad \lambda \in \CC_i. 
\ee
The planar resolvent still solves (\ref{wdisco}), and the way to implement the multi--cut solution is to require $\omega_0(p)$
to have $2n$ branch points. Therefore we have
\be
\omega_0(p)={1 \over 2t} V'(p) -{1 \over 2 t} M(p) {\sqrt {\prod_{k=1}^{2n} (p-x_k)}},
\ee
which can be solved in a compact way by
\be
\label{solwmulti}
\omega_0(p) ={1 \over 2t} \oint_{\cal C} {\rd z \over 2 \pi \ri} { V'(z) \over p-z}
\left( \prod_{k=1}^{2n} { p-x_k\over
z-x_k}\right)^{1 /2}.
\ee
In order to satisfy the asymptotics (\ref{asymres}) the following conditions must hold:
\be
\label{splusone}
\delta_{\ell n}={1\over 2t} \oint_{\cal C} {\rd z \over 2 \pi \ri} {z^{\ell} V'(z)
\over \prod_{k=1}^{2n} (z-x_k)^{1\over 2}}, \qquad \ell=0,1, \cdots, n.
\ee
In contrast to the one-cut case, these are only $n+1$ conditions for the $2n$ variables
$x_k$ representing the endpoints of the cut. The remaining $n-1$ conditions are obtained by fixing the values of the filling fractions through (\ref{fillingfr}) 
(or, equivalently, by fixing the partial 't Hooft parameters). The multipliers in (\ref{tots}) are obtained, as before, by taking derivatives w.r.t. the constraints, and we find the equation 
\be
\label{pdF}
{\partial F_0 \over \partial t_i}-{\partial F_0 \over \partial t_{i+1}} =\Gamma_{i+1}-\Gamma_i, 
\ee
which generalizes (\ref{derf0}) to the multi--cut situation.

\FIGURE{
\includegraphics[height=3cm]{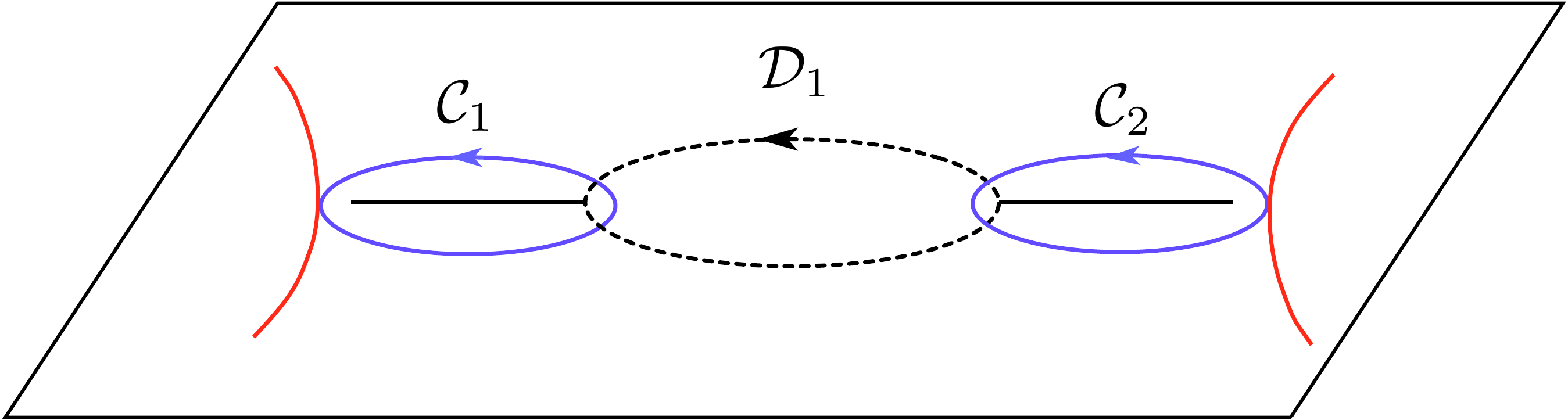}
\caption{A two-cut spectral curve, showing two contours $\CC_{1,2}$ around the cuts where $N_{1,2}$ eigenvalues sit. The ``dual" cycle ${\cal D}_1$ goes from $\CC_2$ to 
$\CC_1$.}
\label{twocuts}
}

We can write the multi-cut solution in a very elegant way by using contour integrals. 
First, the partial 't Hooft parameters are given by
\be
\label{tper2}
t_i={1\over 4 \pi \ri} \oint_{\CC_i} 2t \omega_0 (p)\rd p. 
\ee
We now introduce dual cycles ${\cal D}_i$ cycles, $i=1, \cdots, n-1$, going from the $\CC_{i+1}$ cycle to the $\CC_{i}$ cycle counterclockwise, see \figref{twocuts}. In terms of these, we can write (\ref{pdF}) as
\be
\label{dfy}
{\partial F_0 \over \partial t_i}-{\partial F_0 \over \partial t_{i+1}} =-{1\over 2} \oint_{{\cal D}_i} 2t \omega_0(p) \rd p.
\ee

\sectiono{Chern--Simons and ABJM matrix models at large $N$}
In this section we will solve the matrix models (\ref{normCS}) and (\ref{abjmmm}), in the planar limit, by using the saddle-point techniques of the last section. 

\subsection{Solving the Chern--Simons matrix model}

Let us first consider the Chern--Simons matrix model (\ref{normCS}). The analogue of the saddle-point equation (\ref{sadd}) is 
\be 
\label{EM}
\frac{1}{g_s} \mu_i=\sum_{j\neq i}\coth \left (\frac{\mu_i-\mu_j}{2}\right).
\ee
The form of the r.h.s. suggest to define the resolvent as
\be
\label{omegadef}
\omega (z)=g_s\left\langle \sum_{i=1}^N \coth \left (\frac{z-\mu_i}{2}\right ) \right\rangle.
\ee
The large $N$ limit of (\ref{EM}) gives then
\be
z={1\over 2} \left( \omega_0(z+\ri \epsilon) + \omega_0 (z-\ri \epsilon)\right).
\ee
The planar resolvent satisfies the boundary conditions
\be
\label{obounds}
\omega_0(z) \sim \pm t, \quad z\rightarrow \pm \infty,
\ee
where
\be
t=g_s N
\ee
is the appropriate 't Hooft parameter for this model. Let us define the exponentiated variable
\be
Z=\re^z. 
\ee
In terms of the $Z$ variable, the resolvent is given by
\be
\omega(z)\rd z =-t {\rd Z  \over Z}  +2  g_s\left\langle \sum_{i=1}^{N}{\rd Z \over Z-\re^{\mu_i}}\right\rangle.
\ee
From this resolvent it is possible to obtain the density of eigenvalues at the cuts. In the planar approximation, we have that
\be
\label{planarresCS}
\omega_0(z) =-t +2  t \int_{\CC} \rho_0 (\mu) {Z \over Z-\re^{\mu}}\rd \mu , 
\ee
where $\rho_0(\mu)$ is the density of eigenvalues, normalized in the standard way
\be
\int_{\CC} \rho_0(\mu) \rd \mu=1.
\ee
The discontinuity argument which gave us (\ref{rhow}) in the standard matrix model case tells us now that 
\be
\label{csdensity}
\rho_0(X) \rd X =-{1\over 4 \pi \ri t}{\rd X \over X} \left( \omega_0(X+\ri \epsilon) -\omega_0(X-\ri \epsilon)\right), \qquad X\in \CC.
\ee
Let us solve explicitly for $\omega_0$ by using analyticity arguments, following \cite{hy}. We first construct the function 
\be \label{gdef}
g(Z)= \re^{\omega_0/2}+Z\re^{-\omega_0/2}.
\ee
This function is regular everywhere on the complex $Z$ plane. Indeed, we have 
\be
g(Z+\ri \epsilon)= \re^{\omega_0(Z+\ri \epsilon) /2}+Z\re^{-\omega_0(Z+\ri \epsilon)/2}=Z\re^{-\omega_0(Z-\ri \epsilon)/2} +\re^{-\omega_0(Z-\ri \epsilon)/2}=g(Z-\ri \epsilon),
\ee
so it has no branch cut. The boundary conditions for this function, inherited from (\ref{obounds}), are
\be
\lim_{Z\rightarrow \infty} g(Z) =\re^{-t/2}Z, \qquad  \lim_{Z\rightarrow 0} g(Z) =\re^{-t/2}.
\ee
These conditions are solved by
\be
g(Z)=\re^{-t/2} (Z+1), 
\ee
and we can now regard (\ref{gdef}) as a quadratic equation that determines $\omega_0$:
\be
\label{cs-res}
\omega_0(Z)=2 \log \left[ \frac{1}{2}\left (g(Z)-\sqrt{g^2(Z)-4 Z}\right )\right].
\ee
From this resolvent we can determine immediately the density of eigenvalues, 
\be
\rho_0(x)={1\over \pi t} \tan^{-1} \left[ { {\sqrt{\re^t -\cosh^2 \left({x\over 2}\right)}} \over  \cosh \left({x\over 2}\right) }\right]
\ee
supported on the interval $[-A, A]$ with 
\be
A=2 \cosh^{-1}\left(\re^{t/2}\right). 
\ee
The result (\ref{cs-res}) can be also obtained directly from (\ref{solwo}), see \cite{mmhouches}.

\subsection{Solving the ABJM matrix model}
Let us now solve the matrix model we are interested in, namely (\ref{abjmmm}). The saddle point equations for the eigenvalues $\mu_i$, $\nu_a$ 
are 
\be
\ba
{\mu_i \over g_s} =& \sum_{j\ne i}^{N_1}\coth\frac{\mu_i-\mu_j}2-\sum_{a=1}^{N_2}\tanh\frac{\mu_i-\nu_a}2, \\
-{\nu_a \over g_s} =& \sum_{b\ne a}^{N_2}\coth\frac{\nu_a-\nu_b}2-\sum_{i=1}^{N_1}\tanh\frac{\nu_a-\mu_i}2.
 \label{saddleabjm}
\ea
\ee
We will solve instead the saddle-point equations
\be
\ba
\mu_i =&{t_1 \over N_1} \sum_{j\ne i}^{N_1}\coth\frac{\mu_i-\mu_j}2+{t_2 \over N_2} \sum_{a=1}^{N_2}\tanh\frac{\mu_i-\nu_a}2, \\
\nu_a =&{t_2 \over N_2}  \sum_{b\ne a}^{N_2}\coth\frac{\nu_a-\nu_b}2+{t_1 \over N_1}\sum_{i=1}^{N_1}\tanh\frac{\nu_a-\mu_i}2,
 \label{saddlealt}
\ea
\ee
where 
\be
t_i=g_s N_i 
\ee
are the partial 't Hooft parameters for this model. Clearly, from the solution to (\ref{saddlealt}) we can recover the solution to (\ref{saddleabjm}) by 
simply performing the analytic continuation 
\be
\label{chsign}
t_2 \rightarrow -t_2
\ee
in the solution. The equations (\ref{saddlealt}) are the saddle-point equations for the lens space matrix model (\ref{lensmm}), and the analytic continuation 
(\ref{chsign}) is just the planar version of the relation (\ref{changen2}) mentioned before. 

The total resolvent of the matrix model, 
$\omega(z)$, is defined as \cite{hy}
\be
\label{resolv}
\omega(z)
=g_s \left\langle \sum_{i=1}^{N_1} \coth \left( {z-\mu_i \over 2} \right) \right\rangle +g_s \left\langle\sum_{a=1}^{N_2}  \tanh \left( {z-\nu_a \over 2} \right)\right\rangle.
\ee
In terms of the $Z$ variable, it is given by
\be
\omega(z)\rd z =-t {\rd Z  \over Z}  +2  g_s\left\langle \sum_{i=1}^{N_1}{\rd Z \over Z-\re^{\mu_i}}\right\rangle+ 2 g_s \left\langle \sum_{a=1}^{N_2}{\rd Z \over Z+\re^{\nu_a}}\right\rangle, 
\ee
where $t=t_1 +t_2$, and it has the following expansion as $Z\rightarrow \infty$
\be
\label{Zex}
\omega(z) \rightarrow t + {2 g_s  \over Z}\left \langle \sum_{i=1}^{N_1} \re^{\mu_i} -\sum_{a=1}^{N_2} \re^{\nu_a}\right\rangle+\cdots
\ee
From the total resolvent it is possible to obtain the density of eigenvalues at the cuts. In the planar approximation, we have that
\be
\label{planarres}
\omega_0(z) =-t +2  t_1 \int_{\CC_1} \rho_1(\mu) {Z \over Z-\re^{\mu}}\rd \mu + 2  t_2 \int_{\CC_2} \rho_2 (\nu) {Z \over Z+\re^{\nu}}\rd \nu, 
\ee
where $\rho_1(\mu)$, $\rho_2(\nu)$ are the large $N$ densities of eigenvalues on the cuts $\CC_1$, $\CC_2$, respectively, normalized as
\be
\int_{\CC_1} \rho_1(\mu) \rd \mu=\int_{\CC_2} \rho_2 (\nu) \rd \nu=1.
\ee
The standard discontinuity argument tells us that 
\be
\label{densities}
\ba
\rho_1(X) \rd X &=-{1\over 4 \pi \ri t_1}{\rd X \over X} \left( \omega_0(X+\ri \epsilon) -\omega_0(X-\ri \epsilon)\right), \qquad X\in \CC_1,\\
\rho_2(Y) \rd Y &={1\over 4 \pi \ri t_2 } {\rd Y \over Y}\left( \omega_0(Y+\ri \epsilon) -\omega_0(Y-\ri \epsilon)\right), \qquad Y\in \CC_2.
\ea
\ee

Let us now find an explicit expression for the resolvent, following \cite{hy}. First, notice that it can be split in two pieces, 
\be
\omega(z)=\omega^{(1)}(z) + \omega^{(2)}(z + \ri \pi), 
\ee
where
\be
\ba
\omega^{(1)} (z)&=g_s \left\langle \sum_{i=1}^{N_1}\coth\left (\frac{z-\mu_i}{2}\right )\right\rangle,\\
\omega^{(2)} (z)&=g_s\left\langle \sum_{a=1}^{N_2} \coth\left (\frac{z-\nu_a}{2}\right )\right\rangle
\ea
\ee
are just the resolvents of the Chern--Simons matrix model (\ref{omegadef}). 
 In fact, it is easy to see that the matrix model (\ref{lensmm}) is equivalent to a Chern--Simons matrix model for $N$ variables $u_i$, $i=1, \cdots, N$, 
 where $N_1$ variables 
 \be
 u_i =\mu_i, \quad i=1, \cdots, N_1, 
 \ee
 are expanded around the point $z=0$, and $N_2=N-N_1$ 
 variables 
 \be
 u_{N_1+a}=\ri \pi + \nu_a, \quad  a=1, \cdots, N_2, 
 \ee
 are expanded around the point $z=\ri \pi$. At large $N_{1,2}$ it is natural to assume that the first set of eigenvalues will condense in a cut around $z=0$, and 
 the second set will condense in a cut around $z=\pi \ri$. It follows that $\omega^{(1)}(z)$ will have a discontinuity on an interval $[-A, A]$, while $\omega^{(2)} (z)$ will 
 have a discontinuity on an interval $[-B, B]$. When $g_s$ is real, these cuts occur in the real axis, and the two cuts in the total resolvent are separated by $\ri\pi$ 
 (see \figref{cuts}).

\FIGURE{
\includegraphics[height=4.5cm]{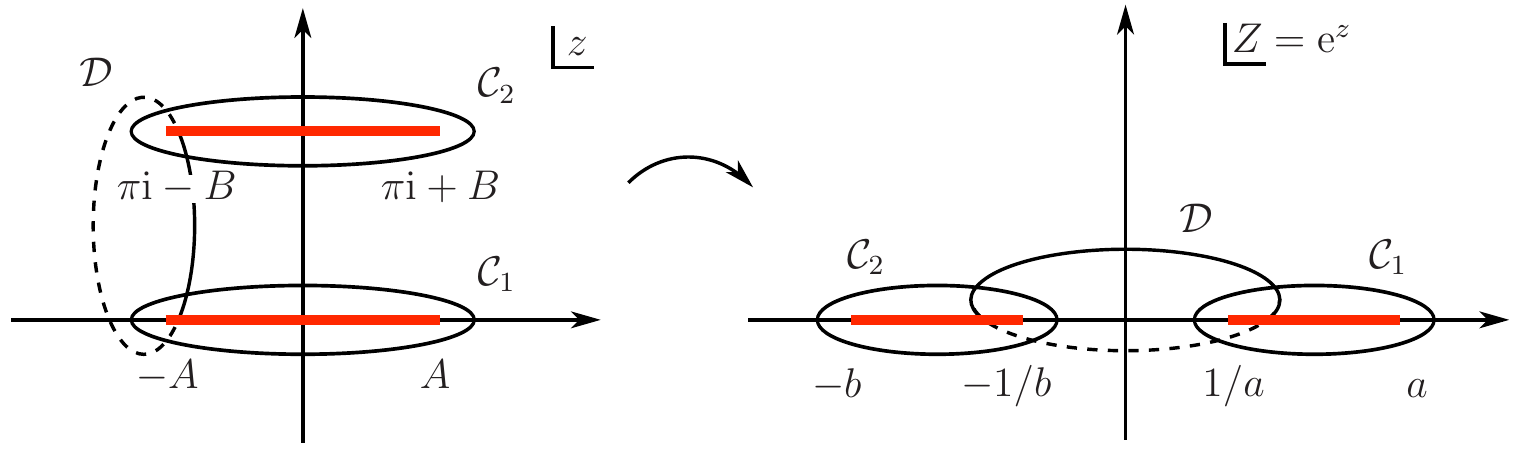} 
\caption{Cuts in the $z$-plane and in the $Z$-plane.}
\label{cuts}
}

The saddle-point equations (\ref{saddlealt}) become then, at large $N$, 
\be
\ba 
z=& \frac{1}{2}\left( \omega_0(z+\ri \epsilon)+\omega_{0}(z-\ri \epsilon)\right), \quad z \in [-A, A], \\
z=& \frac{1}{2}\left( \omega_0(z+\ri \pi +\ri \epsilon)+\omega_{0}(z+\ri \pi-\ri \epsilon)\right), \quad z \in [-B, B].
\ea
\ee
It follows that the function
\be\label{reg}
f(Z)=\re^t\left( \re^{\omega_0}+Z^2\re^{-\omega_0} \right)
\ee
is regular everywhere on the complex plane and has limiting behavior
\be \lim_{Z\rightarrow \infty} f(Z)=Z^{2}, \qquad 
 \lim_{Z\rightarrow 0} f(Z)=1.
\ee
The unique solution satisfying these conditions is 
\be \label{gp2}
f(Z)=Z^2-\zeta Z +1,
\ee
where $\zeta$ is a parameter to be determined. Solving now (\ref{reg}) as a quadratic equation for $\re^{\omega_0}$ yields,
\be \label{w/2}
\omega_0(Z)   = \log \biggl( {\re^{-t} \over 2  } \Bigl[ f(Z) -{\sqrt{ f^2(Z) -4 \re^{2t} Z^2}}  \Bigr]\biggr).
\ee
Notice that $\re^{\omega_0}$ has a square root branch cut involving the function 
\be
\label{sigmaz}
\sigma(Z)= f^2(Z) -4 \re^{2t} Z^2 =\left(Z-a\right) \left(Z-1/a\right) \left(Z+b\right) \left(Z+1/b\right) 
\ee
where $a^{\pm 1}, -b^{\pm1}$ are the endpoints of the cuts in the $Z=\re^z$ plane, see \figref{cuts}. We deduce that the parameter $\zeta$ is related to the positions of the endpoints of the cuts by the relation
\be
\label{zetadef}
\zeta={1\over 2}\left( a  +{1\over a} -b -{1\over b}\right), 
\ee
and we also find the constraint
\be
\label{const}
{1 \over 4}\left( a  +{1\over a} +b +{1\over b}\right)=\re^t.
\ee

Once the resolvent is known, we can obtain both the 't Hooft parameters and the derivative of the genus zero free energy in terms of 
period integrals. The 't~Hooft parameters 
are given by
\be
\label{tperiods}
t_i ={1\over 4\pi \ri} \oint_{\CC_i} \omega_0 (z)  \rd z , \qquad i=1,2.
\ee
This is the same equation that (\ref{tper2}), since the resolvent (\ref{resolv}) has an extra factor $2t$ as compared to (\ref{resolvent}). 
It is easy to see, by using the same techniques that we used in section 7, that the planar free energy $F_0$ satisfies the analogue of (\ref{dfy}) after 
taking into account this extra factor, 
\be
\label{pder}
\CI\equiv
{\partial F_0 \over \partial s} -\pi\ri t
=-\frac{1}{2}\oint_{{\cal D}} \omega_0(z) \rd z, 
\ee
where 
\be
s={1\over 2} (t_1 -t_2)
\ee
and the ${\cal D}$ cycle encloses, in the $Z$ plane, the interval between $-1/b$ 
and $1/a$ (see \figref{cuts}). The extra term $-\pi \ri t$ in (\ref{pder}) is due to the fact that $\omega_0(z)\rd z$ has a nonzero residue at $z=0$.

The above period integrals are hard to compute, but their derivatives can be easily found by adpating a trick from \cite{brinit}. One obtains, 
\be
\label{derperzeta}
\frac{\partial t_{1,2}}{\partial \zeta}=-
\frac{1}{4\pi \ri}\oint\limits_{\mathcal{C}_{1,2}}\frac{\rd Z}{\sqrt{(Z^2-
\zeta Z+1)^2-4\re^{2t} Z^2}}=\pm \frac{\sqrt{ab}}{\pi(1+ab)}\,K(k),
\ee
where $K(k)$ is the complete elliptic integral of the first kind, and its modulus is given by 
\be
\label{modulus}
k^2={(a^2-1)(b^2-1) \over (1+a b)^2}=1-\left(\frac{a + b}{1 + ab}\right)^2. 
\ee
Likewise for the period integral in (\ref{pder}) we find 
\begin{equation}
 \frac{\partial \CI}{\partial \zeta}=
 -2\,\frac{\sqrt{ab}}{1+ab}\,K(k'), \label{I-derivs}
\end{equation}
where the complementary modulus $k'={\sqrt{1-k^2}}$ is given by
\begin{equation}
 k'=\frac{a + b}{1 + ab}.
\end{equation} 

\subsection{ABJM theory and exact interpolating functions}

We will now analyze the exact planar results for the ABJM matrix model in the simplest case, namely the one corresponding to the original ABJM theory with 
gauge groups of the same rank $N$. If we recall the definition of $g_s$ (\ref{convcoup}) in terms of the level $k$, we find 
\be
t_1=-t_2=2\pi \ri {N \over k}=2 \pi \ri \lambda, 
\ee
where $\lambda$ is the 't Hooft coupling of ABJM theory defined in (\ref{abjmthooft}). 
If we think about $t_{1,2}$ as ``moduli" parametrizing the space of complex 't Hooft couplings, 
the ABJM theory of \cite{abjm} corresponds to a real, one-dimensional submanifold in this moduli space. We will call this submanifold the ABJM 
slice. In this slice, the total 
't Hooft parameter of the matrix model vanishes: $t=0$. The theory has only one parameter, $\lambda$, which should be related to the only parameter (\ref{zetadef})
appearing in the resolvent. It follows from 
(\ref{zetadef}) and (\ref{const}) that
\be
a+{1\over a}=2+\zeta, \qquad b+{1\over b}=2-\zeta, 
\ee
The derivative (\ref{derperzeta}) can be expressed in a simpler way by using appropriate transformations of the elliptic integral $K(k)$. Let us consider the elliptic moduli
\be
\label{emoduli}
k_1={1-k'\over 1+k'}, \qquad 
k_2={\rm i} {k_1\over k_1'}. 
\ee
One has that (see for example \cite{gr}, 8.126 and 8.128)
\be
K(k)=(1+k_1)K(k_1) ={1+k_1 \over k_1'} K(k_2),
\ee
and we can write
\be
\frac{\sqrt{ab}}{\pi(1+ab)}K(k) =\frac{\sqrt{ab}}{\pi(1+ab)} {1+k_1\over k_1'} K(k_2)={1\over \pi} {\sqrt{ {a b \over (a+b)(1+a b)}}} K(k_2). 
\ee
Notice that
\be
k_1={(a-1)(b-1) \over (a+1)(b+1)}, \qquad k_2^2=-{(a-1)^2(b-1)^2 \over 4 (a+b)(1+ab)}
\ee
In the ABJM slice we have
\be
k_2^2={\zeta^2 \over 16}, \qquad {\sqrt{ {a b \over (a+b)(1+a b)}}} ={1\over 2}
\ee
and
\be
\label{lamkapex}
 \frac{\rd \lambda}{\rd \zeta} ={1 \over 4\pi^2 \ri} K\left( {\zeta \over 4}\right).
\ee
Since $K(k)$ depends only on $k^2$, t follows from this equation that, if we want $\lambda$ to be real (as it should be in the ABJM theory), 
$\zeta$ has to be pure imaginary, and we can 
write
\be
\zeta=\ri \kappa, \qquad \kappa \in \IR
\ee
so that (\ref{lamkapex}) becomes
\be
 \frac{\rd \lambda}{\rd \kappa} ={1 \over 4\pi^2} K\left( {\ri \kappa \over 4}\right).
 \ee
 This can be integrated explicitly in terms of a hypergeometric function \cite{mp}
 \be
 \label{lamkap}
 \lambda(\kappa)={\kappa \over 8 \pi}   {~}_3F_2\left(\frac{1}{2},\frac{1}{2},\frac{1}{2};1,\frac{3}{2};-\frac{\kappa^2
   }{16}\right),
   \ee
where we have used that $\lambda=0$ when $\kappa=0$ (in this limit, the cut $[a,1/a]$ collapses to zero size, and the period $t_1$ vanishes). (\ref{lamkap}) gives 
the relation between $\lambda$ and $\zeta$ (or $\kappa$). This closed expression makes it possible 
to perform an analytic continuation for $\kappa \gg 1$, where $\lambda$ behaves 
as
 \be
 \label{lamas}
 \lambda(\kappa) ={\log ^2(\kappa)\over 2 \pi
   ^2}+\frac{1}{24}+\CO\left( {1\over \kappa^2}\right). 
   \ee
This suggests to define the shifted coupling
\be
\label{shift-lam}
\hat \lambda=\lambda-{1\over 24}.
\ee
Notice from (\ref{qshift}) 
that this shift is precisely the one needed in order for $\hat\lambda$ to be identified with $Q/k$ at leading order in the string coupling constant. 
The relationship (\ref{lamas}) is immediately inverted to 
   \be
   \kappa=\re^{\pi {\sqrt{2\lambda}}}\left(1 +\CO\left({1\over {\sqrt{\lambda}}},  \re^{-2\pi {\sqrt{2\lambda}}} \right) \right).
   \ee

Let us now consider the genus zero free energy. Its second derivative w.r.t. $s$, evaluated at $t=0$, can be calculated as 
\be
{\partial^2 F_0 \over \partial s^2}\Bigl|_{t=0}={\partial \CI \over \partial \zeta}\Bigl|_{t=0}\cdot \left( {\rd t_1 \over \rd \zeta}\right)^{-1}. 
\ee
Like before, we will use the transformation properties of the elliptic integral $K(k')$ to write (\ref{I-derivs}) in a more convenient way. From (\ref{emoduli}) we deduce
\be
k_1'={2 {\sqrt{k'}} \over 1+k'}, \qquad k_2'={1\over k_1'}, 
\ee
and we have, using again \cite{gr}, 8.126 and 8.128, 
\be
K(k')={1\over 1+k'}K(k_1')={k_2' \over 1+k'}\left( K(k_2') + \ri K(k_2) \right). 
\ee
In the ABJM slice we find, 
\be
{\partial \CI \over \partial \zeta}\biggl|_{t=0}=-{1\over 2} \left[K'\left( {\ri \kappa \over 4}\right) + \ri K\left( {\ri \kappa \over 4}\right)\right], 
\ee
so that 
\be
{\partial^2 F_0 \over \partial s^2}\biggl|_{t=0}=-\pi { K'\left( {\ri \kappa \over 4}\right)  \over K'\left( {\ri \kappa \over 4}\right)} -\pi \ri. 
\ee
We conclude that, in the ABJM theory, where $s=2\pi \ri \lambda$, 
\be
\label{seconder}
\partial_{\lambda}^2 F_0(\lambda) =4 \pi^3  {K'\left({\ri \kappa \over 4}\right)\over K \left({\ri \kappa \over 4}\right)} + 4 \pi^3 \ri.
\ee
A further integration leads to the following expression in terms of a Meijer function
\be
\label{comf}
 \partial_\lambda F_0 (\lambda)={\kappa \over 4} G^{2,3}_{3,3} \left( \begin{array}{ccc} {1\over 2}, & {1\over 2},& {1\over 2} \\ 0, & 0,&-{1\over 2} \end{array} \biggl| -{\kappa^2\over 16}\right)+{ \pi^2 \ri \kappa \over 2} 
  {~}_3F_2\left(\frac{1}{2},\frac{1}{2},\frac{1}{2};1,\frac{3}{2};-\frac{\kappa^2.
   }{16}\right).
\ee
This is, indeed, the exact interpolating function we were looking for! To see this, we can expand it at weak coupling as follows:
\be
\label{Gcomp}
 \partial_\lambda F_0 (\lambda)=-8 \pi ^2 \lambda  \left(\log \left(\frac{\pi  \lambda }{2}\right)-1\right)+\frac{16 \pi ^4 \lambda ^3}{9} +\CO\left(\lambda^5\right). 
 \ee
 After integrating w.r.t. $\lambda$ and multiplying by 
 \be
 g_s^{-2}=-{k^2 \over 4 \pi^2}, 
 \ee
 we find that the first term exactly reproduces the weak-coupling answer (\ref{fabjmweak}). The comparison with the weak coupling expansion also fixes the integration constant, 
 \be
 \label{correctweak}
  F_0(\lambda)=\int_0^{\lambda} \rd \lambda' \,  \partial_{\lambda'} F_0 (\lambda').
\ee
To study the strong-coupling behavior, we can now analytically continue the r.h.s. of (\ref{comf}) to $\kappa=\infty$, 
and we obtain
\be
\label{slice}
\partial_\lambda F_0 (\lambda)=2\pi^2 \log \kappa +{4 \pi^2 \over \kappa^2} \, {}_4 F_3 \left( 1, 1, {3\over 2}, {3\over 2}; 2,2,2; -{16 \over \kappa^2} \right). 
\ee
After integrating w.r.t. $\lambda$ and using the shifted coupling $\hat \lambda$ defined in (\ref{shift-lam}), we find, 
\be
\label{prepotf}
F_0 (\hat \lambda)= {4\pi^3 \sqrt{2}  \over 3} \hat \lambda^{3/2}
+\sum_{\ell\ge1}  \re^{- 2\pi \ell  {\sqrt{2\hat \lambda}}} f_{\ell}\left({1\over \pi {\sqrt{2 \hat \lambda}}} \right)
\ee
where $f_{\ell}(x)$ is a polynomial in $x$ of degree $2 \ell-3$ (for $\ell\ge 2$). If we multiply by $g_s^{-2}$, we find that the leading term agrees precisely with the 
prediction from the AdS dual in (\ref{introgoal}). The series of exponentially small corrections in (\ref{prepotf}) were interpreted in \cite{dmp} as coming from 
worldsheet instantons of type IIA theory wrapping the $\IC\IP^1$ cycle in $\IC\IP^3$. This is a novel type of correction in AdS$_4$ dualities which is not present in 
AdS$_5$ spaces, see \cite{sorokin} for a preliminary investigation of these effects.

\FIGURE{
\includegraphics[height=4cm]{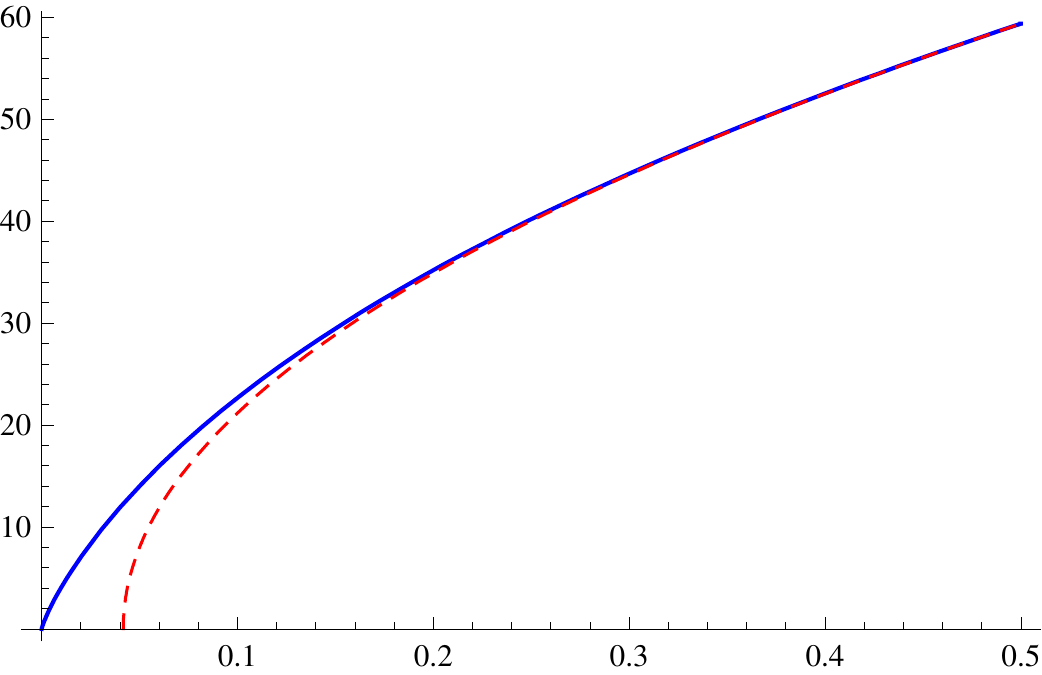} \qquad 
\includegraphics[height=4cm]{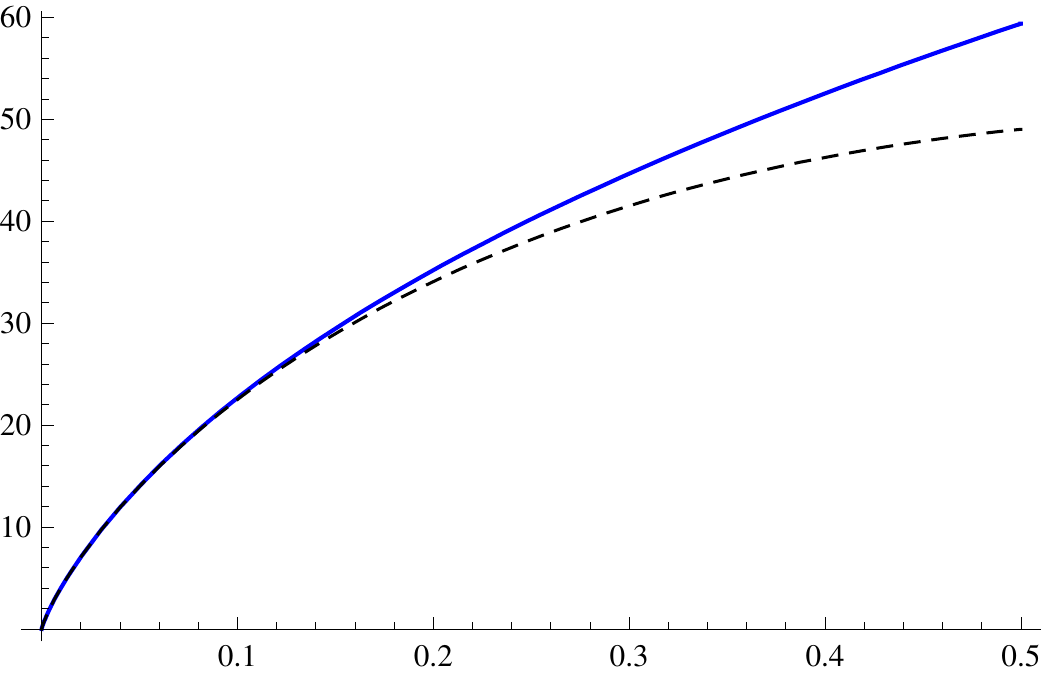}
\caption{Comparison of the exact result for $\partial_\lambda F_0(\lambda)$ given in (\ref{comf}), plotted as a solid blue line, and the weakly coupled and 
strongly coupled results. In the figure on the left, the red dashed line is the supergravity result given by the first term in (\ref{prepotf}), while in the figure on the right, the 
black dashed line is the Gaussian result given by the first two terms in (\ref{Gcomp}).}
\label{prepotfig}
}

\sectiono{Further developments and results}

In these lectures our goal has been to find the interpolating function for (\ref{introgoal}) in ABJM theory, but many other results can be obtained
with the ideas and techniques reviewed here. In this last section we will give a brief summary of more recent results along these lines. 

\subsection{Wilson loops and higher genus corrections}

One of the main results of the localization techniques developed in \cite{pestun}, in the context of four-dimensional gauge theories, was a proof of the conjecture of \cite{esz,dg}. According to this conjecture, the vacuum expectation 
value of $1/2$ BPS Wilson loops can be calculated by a matrix model. Similarly, the original motivation 
of \cite{kapustin} was to provide a matrix model calculation of Wilson loops in ABJM theory, generalizing \cite{pestun}. One can indeed show that 
both the vev of the 1/6 BPS Wilson loops of \cite{drukker,rey, cw}, as well as of the vev of the 1/2 BPS Wilson loop of \cite{dt}, can be computed 
as correlators in the matrix model (\ref{abjmmm}). 
Using the resolvents that we derived in these lectures, one obtains exact interpolating functions for the vevs of these Wilson loops \cite{mp,dmp}, which at 
strong coupling are in agreement with the AdS predictions. 

When $N_1=N_2$, i.e. for the original ABJM theory, the matrix model (\ref{abjmmm}) can be explicitly solved to {\it all} 
orders in the $1/N$ expansion \cite{dmp} by using techniques from topological string theory (see \cite{hkr} and references therein). This is a remarkable fact, since it amounts to solving for the free energy of the dual type IIA superstring 
theory at {\it all} genera, and it makes it possible to address interesting nonperturbative questions \cite{dmpnp}. 

\subsection{More general Chern--Simons--matter theories}

In these lectures we have only considered the ABJM theory, but there are many other supersymmetric Chern--Simons--matter theories with proposed large $N$ AdS duals. According to 
the AdS$_4$/CFT$_3$ correspondence, the free energy on the three-sphere of these theories should behave, in the planar limit and strong coupling, as (\ref{fvol}). This probes the volume of the compactification manifold in M-theory, which can be a non-trivial function of other parameters of the model, and suggests a strategy to make precision tests of the 
AdS/CFT correspondence: compute the planar free energies of the matrix models 
associated to more general Chern--Simons--matter theories, and see if they display the behavior (\ref{fvol}) at strong 't Hooft coupling. 

However, these matrix models become increasingly harder to solve as we move away from the ABJM theory, even in the planar limit, and exact results for the resolvent are only known for the Gaiotto--Tomasiello theories \cite{suyama} and for theories with matter in the fundamental \cite{cmp}. Even when the exact resolvent is known, 
extracting the strong coupling behavior might be difficult. For this reason, 
it is important to have techniques that make possible to find the strong coupling limit of the free energy, even without having an exact expression. In \cite{cmp} a geometric technique based on 
ideas of tropical geometry was proposed to derive in an easy way the strong coupling behavior from the resolvent, and the behavior (\ref{fvol}) was tested for a one-parameter family of 
tri-Sasaki-Einstein manifolds $X_7$ dual to $\CN=3$ theories with fundamental 
matter. Moreover, in \cite{hklebanov}, a very powerful method was introduced which makes possible to analyze the 
matrix models describing Chern--Simons--matter theories {\it directly} in the strong coupling limit, without the need to solve first for the resolvent. This allowed to test (\ref{fvol}) for a 
large class of $\CN=3$ theories. 

\subsection{Extremization of $Z$}

Perhaps the most interesting spinoff of these developments, from the QFT point of view, is the idea that the free energy of QFTs on the three-sphere should play the role, 
in three dimensions, of Zamolodchikov's $c$-function in two dimensions, and of the anomaly $a$-coefficient in four dimensions: it has been conjectured that this quantity
decreases along RG trajectories and it is stationary at RG fixed points. This has been dubbed the ``F-theorem" in \cite{jk}. The first evidence for this result comes from the extension of the localization computation of \cite{kapustin} to theories with 
$\CN=2$ supersymmetry \cite{jafferis, hama}. In this case, the anomalous dimensions of the matter fields $\Delta$ are not canonical, and the partition function is itself a function 
of them. It was shown in \cite{jafferis} that the partition function was extremized as a function of these anomalous dimensions, 
giving an efficient way to calculate them. 
This result is similar to the four-dimensional result that the R-charges extremize the anomaly $a$-coefficient \cite{intri}, and it leads naturally to the ``F-theorem" conjecture. 
The extremization property proved in \cite{jafferis} has been applied and tested in various situations \cite{ms,ckk,jk,amariti}, although the validity of the F-theorem is still conjectural 
and possible counterexamples have been discussed in \cite{jk,niarchos}.

\subsection{Other extensions and applications}

The localization techniques of \cite{kapustin} have been used as well to test field theory dualities in three-dimensional supersymmetric gauge theories \cite{kwyone,kwytwo,jy, kthree, wy}, by showing that their partition functions agree as functions of the various parameters. They have also been used to compute supersymmetric indices in $\IS^2\times \IS^1$ \cite{kim,imamura,ksv} 
and to study Chern--Simons--matter theories in other three-manifolds \cite{hhl,gang,kallen}. 

In conclusion, the localization ideas of \cite{pestun, kapustin} give us new ways of computing exact quantities in gauge theories where matrix model techniques play 
a crucial r\^ole, and we anticipate many interesting developments coming from this line of research. 

\section*{Acknowledgements}
These notes are based on lectures given at the University of Edinburgh in march 2011. My first thanks goes to Jos\'e Figueroa-O'Farrill, Patricia Ritter and Joan Sim\'on for the invitation and their warm hospitality, as well as to the participants for their questions and comments. I would also like to thank my collaborators in this subject for the enjoyable work: Ricardo Couso, Nadav Drukker and Pavel Putrov. I'm specially grateful to Nadav Drukker and Pavel Putrov for a careful reading of the manuscript. Finally, I would like to thank Kostas Skenderis for his very useful 
remarks on holographic renormalization. 
This work is supported in part by the Fonds National Suisse.

 \appendix

\sectiono{Harmonic analysis on $\IS^3$}

\subsection{Maurer--Cartan forms}

We will first introduce some results and conventions for the Lie algebra and the Maurer--Cartan forms.
The basis of a Lie algebra ${\bf g}$ satisfies
\be
[T_a, T_b]=f_{abc} T_{c}.
\ee
If $g\in G$ is a generic element of $G$, one defines the {\it Maurer--Cartan forms} $\omega_a$ through the 
equation
\be
g^{-1} \rd g = \sum_{a} T_{a} \omega_a,
\ee
and they satisfy
\be
\label{mc}
\rd \omega_a+{1\over 2} f_{abc } \omega_b\wedge \omega_c=0.
\ee
This is due to the identity
\be
\rd\left( g^{-1} \rd g\right)+ g^{-1} \rd g \wedge g^{-1} \rd g=0. 
\ee

Let us now specialize to $SU(2)$. A basis for the Lie algebra is given by:
\be
T_a={\ri\over 2} \sigma_a.
\ee
Explicitly 
\be
T_1={\ri\over 2}  \left( \begin{matrix} 0&  1 \\
                                 1 &   0 \end{matrix} 
                              \right), \qquad T_2={\ri\over 2}  \left( \begin{matrix} 0&  -\ri \\
                                 \ri &   0 \end{matrix} 
                              \right), \qquad T_3={\ri\over 2}  \left( \begin{matrix} 1&  0 \\
                                 0 &   -1 \end{matrix} 
                              \right).
                              \ee
                              The structure constants are 
                              \be
 f_{abc}=-\epsilon_{abc}.
 \ee
 Any element of $SU(2)$ can be written in the form
 \be
 g= \left( \begin{matrix} \alpha&  \beta \\
                                -\bar \beta &   \bar \alpha \end{matrix} \right), \qquad |\alpha|^2 + |\beta|^2=1.
                                \ee
We parametrize this element as (see for example \cite{VK})
\be
|\alpha|=\cos \, {t_1\over 2}, \qquad |\beta|=    \sin \, {t_1\over 2}, \qquad {\rm Arg}\, \alpha={t_2 + t_3\over 2} , \qquad    
 {\rm Arg}\, \beta={t_2 - t_3 + \pi \over 2},
 \ee
 where $t_i$ are the Euler angles and span the range
 \be
0\le   t_1 <\pi, \qquad   0\le t_2 < 2\pi, \qquad   -2\pi\le   t_3 <2\pi.                  
\ee
The general element of $SU(2)$ will then be given by
\be
\ba
g&=
u(t_1, t_2. t_3)=  \left( \begin{matrix} \cos(t_1/2) \re^{\ri(t_2 + t_3)/2} &  \ri \sin(t_1/2) \re^{\ri(t_2 - t_3)/2} \\
                                 \ri \sin(t_1/2) \re^{\ri(-t_2 + t_3)/2}  &   \cos(t_1/2) \re^{-\ri(t_2 + t_3)/2}  \end{matrix} 
                              \right)\\
                              &=u(t_2,0,0) u(0,t_1,0) u(0,0,t_3).
                              \ea
                              \ee
 We then have
 \be
 \Omega=g^{-1} \rd g= {\ri\over 2}      \left( \begin{matrix} \rd t_3 + \cos\, t_1 \rd t_2& 
\re^{-\ri t_3} (\rd t_1 +  \ri \rd t_2  \sin\, t_1) \\ \re^{\ri t_3} (\rd t_1 -  \ri \rd t_2  \sin\, t_1)  & -\rd t_3 -\cos\, t_1 \rd t_2
 \end{matrix} 
                              \right).
                              \ee
 Therefore, 
 \be
 \ba
 \omega_1=&  \cos\, t_3 \rd t_1  +  \sin\, t_3 \sin\,t_1 \rd t_2 ,\\
  \omega_2=&\sin\, t_3 \rd t_1  - \cos\, t_3 \sin\, t_1 \rd t_2,\\
 \omega_3=&    \cos\, t_1 \rd t_2 +\rd t_3,
 \ea
 \ee
 and one checks explicitly 
 \be
\rd \omega_a = {1\over 2} \epsilon_{abc}\, \omega_b\wedge \omega_c, 
 \ee
 as it should be according to (\ref{mc}).

 \subsection{Metric and spin connection}
 
 The metric on $SU(2)=\IS^3$ is induced from the 
 metric on $\IC^2$
 \be
 \rd s^2= r^2\biggl( \rd |\alpha|^2 + |\alpha|^2 \rd{\rm Arg}\alpha^2 + \rd|\beta|^2 + |\beta|^2 \rd{\rm Arg}\beta^2   \biggr),
 \ee
 where $r$ is the radius of the three-sphere. A simple calculation leads to 
 \be
 \rd s^2= {r^2 \over 4} \biggl( \rd t_1^2 + \rd t_2^2 + \rd t_3^2 + 2 \cos\, t_1 \, \rd t_2 \rd t_3   \biggr),
 \ee
 with inverse metric 
 \be
 G^{-1}={4 \over r^2} \left( \begin{matrix}   1& 0& 0\\ 0& \csc^2 \, t_1& -\cot\, t_1 \, \csc\, t_1 \\0 &  -\cot\, t_1 \, \csc\, t_1&  \csc^2 \, t_1
 \end{matrix} 
                              \right)
                              \ee
and volume element
 \be
({\rm det}\, G)^{1/2}= {r^3 \sin\, t_1\over 8}.
\ee
The volume of $\IS^3$ is then
\be
\int_{SU(2)}   ({\rm det}\, G)^{1/2}  \rd t_1\, \rd t_2 \, \rd t_3=2\pi^2 \, r^3
\ee
which is the standard result. The only nonzero Christoffel symbols of this metric are
\be
\Gamma_{23}^1={1\over 2} \sin t_1, \qquad \Gamma_{13}^2=\Gamma_{12}^3=-{1\over 2 \sin t_1}, \qquad \Gamma_{13}^3=\Gamma_{12}^2={1\over 2} \cot t_1. 
\ee

We can use the Maurer--Cartan forms to analyze the differential geometry of $\IS^3$. The dreibein of $\IS^3$ is proportional 
 to $\omega_a$, and we have
 \be
 e^a_\mu={r\over 2} \left( \omega_a \right)_\mu. 
 \ee
 In terms of forms, we have
 \be
 e^a=e^a_\mu \rd x^\mu ={r\over 2} \omega_a. 
 \ee
 Indeed, one can explicitly check that
 \be
 e^a_\mu e^b_\nu \eta_{ab}=G_{\mu \nu}. 
 \ee
 The inverse vierbein is defined by
\be
E^\mu_a=\eta_{ab} G^{\mu \nu}e^b_\mu, 
\ee
which can be used to define left-invariant vector fields
 \be
 \ell_a =E^\mu_a {\partial \over \partial x^\mu}.
 \ee
 Let us give their explicit expression in components:
 \be
 \ba
 \ell_1&={2\over r} \left( \cos t_3 {\partial \over \partial t_1} + {\sin t_3 \over \sin t_1} {\partial \over \partial t_2} -\sin t_3 \cot t_1 {\partial \over \partial t_3}\right),\\
 \ell_2&={2\over r} \left( \sin t_3 {\partial \over \partial t_1} - {\cos t_3 \over \sin t_1} {\partial \over \partial t_2} +\cos t_3 \cot t_1 {\partial \over \partial t_3}\right),\\  
 \ell_3&= {2\over r}  {\partial \over \partial t_3}.\ea
 \ee
 Of course, they obey
 \be
 e^a(\ell_b) =\delta^a_b, 
 \ee
as well as the following commutation relations
 \be
 [\ell_a, \ell_b]=-{2\over r} \epsilon_{abc} \ell_c. 
 \ee
This can be checked by direct computation. If we now introduce the operators $L_a$ through
\be
\ell_a={2\ri \over r} L_a. 
\ee
we see that they satisfy the standard commutation 
relations of the $SU(2)$ angular momentum operators: 
\be
[L_a, L_b]=\ri \epsilon_{abc} L_c. 
\ee

 The spin connection $\omega^a_{~b}$ is characterized by
 \be
 \label{defspin}
 \rd e^a + \omega^a_{~b} \wedge e^b=0.
 \ee
 %
 %
 %
Imposing no torsion one finds the explicit expression, 
\be
 \omega^a_{~b \mu}=-E_b^\nu \left(  \partial_{\mu} e^a_{\nu} -\Gamma^\lambda_{\mu \nu} e^a_\lambda\right), 
 \ee
 or, equivalently, 
 \be
 \partial_{\mu} e^a_{\nu} =\Gamma^\lambda_{\mu \nu} e^a_\lambda -e^b_\nu \omega^a_{b\mu}. 
 \ee
In our case we find
 \be
 \label{spincon}
 \omega^a_{~b}={1\over r} \epsilon^a_{~bc}e^c.
 \ee

\subsection{Laplace--Beltrami operator and scalar spherical harmonics}

The scalar Laplacian on $\IS^3$ can be calculated in coordinates from the general formula
\be
-\Delta^{0} \phi={1\over {\sqrt {{\rm det}\, G}}} \sum_{m,n}{\partial \over \partial x^m}  \biggl( {\sqrt{ {\rm det}\, G}} G^{mn}{\partial  \phi\over 
\partial x^n} \biggr),
\ee
or equivalently
\be
\label{deltag}
-\Delta^{0}=G^{\mu \nu} \partial_\mu \partial_\nu -G^{\mu \nu}\Gamma^{\rho}_{\mu \nu} \partial_\rho.
\ee
In this case it reads
\be
-\Delta^{0}={4 \over r^2} \biggl( {\partial^2 \over \partial t_1^2} + \cot\, t_1 {\partial \over \partial t_1} + \csc^2\,t_1{\partial^2 \over \partial t_2^2}
+\csc^2\,t_1{\partial^2 \over \partial t_2^3}-2\csc\, t_1\cot \, t_1  {\partial^2 \over \partial t_2 \partial t_3} \biggr).
\ee
It is easy to check that it can be written, in terms of left-invariant vector fields, as
\be
-\Delta^{0}=\sum_a \ell_a^2.
\ee
To see this, we write
\be
\sum_a \ell_a^2 =\sum_a  E^\mu_a  E^\nu_a \partial_\mu \partial_\nu+\sum_a E^\mu_a {\partial E^\nu_a \over \partial x^\mu} {\partial \over \partial x^\nu}. 
\ee
The first term reproduces the first term in (\ref{deltag}). We now use the identity
\be
\partial_\mu E_b^\nu=E_c^\nu \omega^c_{b\mu} -\Gamma^\nu_{\mu \lambda} E_b^\lambda. 
\ee
After contraction with $E^\mu_a$ and use of the explicit form of the spin connection, we see that only the second term survives, which is indeed the second term in (\ref{deltag}).

The Peter--Weyl theorem says that any square-integrable function on $\IS^3\simeq SU(2)$ can be 
written as a linear combination of 
\be
\label{sshar}
S_j^{m n}, \qquad m,n=1, \cdots, d_j
\ee
where 
\be
S_j: SU(2)  \rightarrow M_{d_j \times d_j}
\ee
is the representation of spin $j$ and dimension $d_j$, and $M_{d_j \times d_j}$ are the inversible square matrices 
of rank $d_j$. The function $S_j^{mn}$ 
is just the $(m,n)$-th entry of the matrix. The eigenvalues of the Laplacian might be calculated immediately by noticing that, in terms of the $SU(2)$ angular 
momentum operators, it reads
\be
\label{lapL}
\Delta^0 ={4\over r^2} {\bf L}^2, 
\ee
and since the possible eigenvalues of ${\bf L}^2$ are 
\be
j(j+1), \qquad j=0, {1\over 2}, \cdots, 
\ee
we conclude that the eigenvalues of the Laplacian are of the form 
\be
\label{eigenlaplace}
\lambda_j={4\over r^2} j (j+1), \qquad j=0, \, \, {1\over 2}, \cdots
\ee
Notice that the dependence on $r$ is the expected one from dimensional analysis. The degeneracy of these eigenvalues is 
\be
d_j^2=(2j+1)^2 
\ee
which is the dimension of the matrix $M_{d_j \times d_j}$.

\subsection{Vector spherical harmonics}

The space of one-forms on $\IS^3$ can be decomposed in two different sets. One set is spanned by gradients of $S_j^{mn}$, and it 
is proportional to 
\be
\label{gradient}
S_j^{mq} (T_a)^{qn}_j \omega_a. 
\ee
The other set is spanned by 
the so-called {\it vector spherical harmonics},
\be
V_{j  \pm }^{m n}, \qquad \epsilon=\pm1, \quad m=1, \cdots, d_{j  \pm {1\over 2}}, \quad n=1, \cdots, d_{j  \mp {1\over 2}},
\ee
see Appendix B of \cite{aharony} for a useful summary of their properties. The $\epsilon=\pm 1$ corresponds to two linear combinations of the $\omega_a$ which are independent from the one appearing in (\ref{gradient}). 
The vector spherical harmonics are in the representation 
\be
\left( j \pm {1\over 2} , j \mp {1\over 2} \right) 
\ee
of $SU(2) \times SU(2)$. We will write them, as in \cite{aharony}, as $V^{\alpha}$, where 
\be
\alpha= (j, m, m', \epsilon),
\ee
and we will regard them as one-forms. They satisfy the properties
\be
\label{dvh}
\rd^{\dagger} V^{\alpha}=0, \qquad 
*\rd V^{\alpha} = -\epsilon(2j +1) V^{\alpha}.
\ee
It follows that
\be
\label{lapvsh}
*\rd*\rd V^{\alpha}=-\Delta^{1} V^{\alpha}=(2j+1)^2 V^{\alpha}.
\ee
Their degeneracy is 
\be
2 d_{j  +{1\over 2}} d_{j  - {1\over 2}} =4 j (2j+2).
\ee

\subsection{Spinors}
Using the dreibein, we define the ``locally inertial" gamma matrices as
\be
\gamma_a =E_a^\mu \gamma_\mu, 
\ee
which satisfy the relations
\be
\{\gamma_a, \gamma_b \}=2\delta_{ab}, \qquad [\gamma_a , \gamma_b]=2\ri \epsilon_{abc} \gamma_c. 
\ee
 The standard definition of a covariant derivative acting on a spinor is 
 \be
 \nabla_\mu=\partial_\mu+{1\over 4} \omega_\mu^{ab} \gamma_a \gamma_b=\partial_\mu+{1\over 8} \omega_\mu^{ab} [\gamma_a, \gamma_b].
 \ee
 Using the  commutation relations of the gamma matrices $\gamma_a$ and the explicit expression for the spin connection (\ref{spincon}) we find 
\be
\ba
\nabla_\mu=&\partial_\mu+{\ri \over 4 r}  \epsilon_{abc} \epsilon_{abd} e_\mu^c \gamma_d =\partial_\mu+{\ri \over 2}  e_\mu^c \gamma_c\\
=&\partial_\mu+{\ri \over 2r}  \gamma_\mu.
\ea
\ee
It follows that the Dirac operator is 
\be
-\ri \Ds =-\ri \gamma^\mu \partial_\mu +{3\over 2r}=-\ri \gamma^a E_a^\mu\partial_\mu +{3\over 2r}=-\ri \gamma^a \ell_a +{3\over 2r}. 
\ee
Let us now introduce the spin operators
\be
S_a ={1\over 2} \gamma_a, 
\ee
which satisfy the $SU(2)$ algebra
\be
[S_a, S_b]=\ri \epsilon_{abc} S_c. 
\ee
In terms of the $S_a$ and the $SU(2)$ operators $L_a$, the Dirac operator reads
\be
\label{finaldirac}
-\ri \Ds ={1\over r} \left( 4 {\bf L}\cdot {\bf S} +{3\over 2} \right).
\ee
The calculation of the spectrum of this operator is as in standard Quantum Mechanics: we introduce the total angular momentum 
\be
{\bf J}={\bf L} + {\bf S}, 
\ee
so that 
\be
4 {\bf L}\cdot {\bf S}=2 \left( {\bf J}^2-{\bf L}^2 - {\bf S}^2\right). 
\ee
Since ${\bf S}$ corresponds to spin $s=1/2$, and ${\bf L}$ to $j$, the possible eigenvalues of ${\bf J}$ are $j \pm 1/2$, and we conclude that the 
eigenvalues of (\ref{finaldirac}) are (we set $r=1$)
\be
2 \left( \left(j\pm {1\over 2}\right) \left(j\pm {1\over 2} +1\right) -j(j+1)\right)=\begin{cases} 2j+{3\over 2} &\text{for $+$}\\
-2j -{1\over 2} &\text{for $-$}, \end{cases},
\ee
with degeneracies
\be
d_{j\pm{1\over 2}}=\left( 2 \left( j\pm {1\over 2} \right) \right) (2j+1)=\begin{cases} 2(j+1)(2j+1) &\text{for $+$}\\
2j(2j+1) &\text{for $-$}. \end{cases}
\ee
These can be written in a more compact form as
\be
\label{spinoreigen}
\lambda_n^{\pm}=\pm \left(n+{1\over 2}\right), \quad d_n^{\pm} = n(n+1), \qquad n=1,2, \cdots
\ee

\end{document}